\documentclass{elsart}

    \usepackage{latexsym,bm,amsmath,amssymb,amsfonts}
    \usepackage{epsfig,graphics,graphicx}
    \usepackage{makeidx}
    \usepackage{citesort}
    \usepackage{slashed}
    \usepackage{float,multicol}
\newcommand{\avg}[1]{\left\langle #1 \right\rangle}

    \newcommand{\mcal}{\mathcal}
    \newcommand{\rmK}{{\rm K}}
    \newcommand{\erfc}{{\rm Erfc}}
    \newcommand{\f}{\frac}
    \newcommand{\rmd}{{\rm d}}   %ELS%
    \newcommand{\dif}{{\rm d}}
    \newcommand{\abar}{\bar{\alpha}_s}
    \newcommand{\atpi}{\frac{\bar{\alpha}_s}{2\pi}}
    \newcommand{\del}{\partial}
    \newcommand{\lan}{\left\langle}
    \newcommand{\ran}{\right\rangle}
    \newcommand{\cal}{\mathcal}
    \newcommand{\rme}{{\rm e}}
    \newcommand{\tr}{{\rm tr}}

    \newcommand{\x}{\bm x}
    \newcommand{\y}{\bm y}
    
    \newcommand{\z}{\bm z}

    \newcommand{\nn}{\nonumber\\}

    \long\def\comment#1{ }
    
    \newcommand{\beq}{\vspace{-.4cm}\begin{eqnarray}}
    \newcommand{\eeq}{\vspace{-.5cm}\end{eqnarray}}
    \newcommand{\be}{\vspace{-.3cm}\begin{eqnarray}}
    \newcommand{\ee}{\vspace{-.4cm}\end{eqnarray}}
\def\simge{\mathrel{%
   \rlap{\raise 0.511ex \hbox{$>$}}{\lower 0.511ex \hbox{$\sim$}}}}
\def\simle{\mathrel{
   \rlap{\raise 0.511ex \hbox{$<$}}{\lower 0.511ex \hbox{$\sim$}}}}

\begin{document}

\begin{flushright}
~\vspace{-1.25cm}\\
{\small\sf SACLAY--T06/001\\ECT$^*$--06--01\\RBRC--581}
\end{flushright}
\vspace{2.cm}

\begin{frontmatter}

\parbox[]{16.0cm}{ \begin{center}
\title{Diffusive scaling and the high--energy limit\\
of deep inelastic scattering in QCD at large $N_c$}

\author{Y.~Hatta$^{\rm a}$},
\author{E.~Iancu$^{\rm b,}$\thanksref{th2}},
\author{C.~Marquet$^{\rm b}$},
\author{G.~Soyez$^{\rm b}$\thanksref{th3}}
\author{and D.N.~Triantafyllopoulos$^{\rm c}$}

\address{$^{\rm a}$ RIKEN BNL Research Center, Brookhaven National Laboratory,
Upton, NY 11973, USA}

\address{$^{\rm b}$ Service de Physique Theorique de Saclay, CEA/DSM/SPhT,
F-91191 Gif-sur-Yvette, France}

\address{$^{\rm c}$ ECT$^*$, Villa Tambosi, Strada delle Tabarelle
286, I-38050 Villazzano (TN), Italy}

\thanks[th2]{Membre du Centre National de la Recherche Scientifique
(CNRS), France.}
\thanks[th3]{On leave from the Fundamental Theoretical Physics group
of the University of Li\`ege.}

\date{\today}
%\vspace{0.8cm}
\begin{abstract}
Within the limits of the large--$N_c$ approximation (with $N_c$
the number of colors), we establish the high--energy behaviour of
the diffractive and inclusive cross--sections for deep inelastic
scattering at fixed impact parameter. We demonstrate that for
sufficiently high energies and up to very large values of $Q^2$,
well above the proton average saturation momentum $\langle
Q_s^2\rangle$, the cross--sections are dominated by dense
fluctuations in the target wavefunction, that is, by the
relatively rare gluon configurations which are at saturation on
the resolution scale $Q^2$ of the virtual photon. This has
important physical consequences, like the emergence of a new, {\em
diffusive}, scaling, which replaces the `geometric scaling'
property characteristic of the mean field approximation. To
establish this, we shall rely on a dipole version of the
Good--Walker formula for diffraction (that we shall derive here in
the context of DIS), together with the high--energy estimates for
the dipole scattering amplitudes which follow from the recently
established evolution equations with Pomeron loops and include the
relevant fluctuations. We also find that, as a consequence of
fluctuations, the diffractive cross--section at high energy is
dominated by the elastic scattering of the quark--antiquark
component of the virtual photon, up to relatively large
virtualities $Q^2\gg \langle Q_s^2\rangle$.

\end{abstract}
\end{center}}

\end{frontmatter}

\newpage
\tableofcontents
\newpage

\section{Introduction}
\setcounter{equation}{0}
\label{SECT_INTRO}

In this paper, we shall present the first analysis of the physical
consequences of the fluctuations in the high--energy evolution in
QCD for the phenomenology of lepton--hadron deep inelastic
scattering (DIS). By ``fluctuations'' we mean in general the
correlations in the gluon distribution of an energetic hadron (the
target in DIS) which are generated through the radiative processes
contributing to its evolution with increasing energy. More
precisely, we shall be interested here in the {\em gluon--number
fluctuations}, which have been recently shown
\cite{IM032,MS04,IMM04,IT04} to play an essential role in the
evolution of the elastic $S$--matrix for dipole--hadron scattering
towards the unitarity limit. The ``dipole'' is a quark--antiquark
pair in a colorless state, so like the $q\bar q$ excitation
through which the virtual photon couples to the hadronic target in
DIS. As demonstrated in Refs. \cite{MS04,IMM04,IT04}, both the
energy dependence of the characteristic  scale for the onset of
unitarization
--- the target {\em saturation momentum} $Q_s$ --- and the
functional form of the scattering amplitudes at high energy, are
strongly modified, and to a large extent even determined, by such
fluctuations. Motivated by these observations, and by the
importance played by dipole configurations for the DIS processes
at high energy, we shall provide a detailed analysis of the
influence of gluon number fluctuations on the inclusive and
diffractive DIS cross--sections. As we shall discover, the effects
of fluctuations are in fact overwhelming: For sufficiently high
energy, and within a wide kinematical range around the (average)
saturation momentum, they  wash out the corresponding predictions
of the mean field approximation, and replace them with a new kind
of universal behaviour, which is relatively simple.

A priori, one may expect the gluon--number fluctuations not to be
important for the problem of the high--energy evolution, since the
latter is characterized by a high--density environment (especially
in the saturation regime at transverse momenta below $Q_s$), where
the gluon occupation numbers are large. And as a matter of facts,
such fluctuations have been left out by the modern approaches to
non--linear evolution in QCD, namely, the Balitsky equations for
the eikonal scattering amplitudes \cite{B} and the (functional)
JIMWLK equation \cite{JKLW,W,CGC} for the evolution of the `color
glass condensate' \cite{MV,CGC}. These formalisms do include
non--trivial correlations --- as obvious from the fact that they
generate infinite hierarchies of equations ---, but these are
purely {\em color} correlations (like the exchange of a single
gluon between two dipole amplitudes), and as such they die away
when $N_c\to\infty$ (with $N_c$ the number of colors). In this
limit, the Balitsky--JIMWLK hierarchy reduces to a single
non--linear equation\footnote{More precisely, they reduce to a
simplified hierarchy which is equivalent to the BK equation for
uncorrelated initial conditions, and which does not generate new
correlations in the course of the evolution; see the discussion in
Ref. \cite{IT04}.}
--- the Balitsky--Kovchegov (BK) equation \cite{B,K} ---, which
is often referred to as {\em the} mean field approximation (MFA).
But in so far as the gluon--number fluctuations of interest here
are concerned, the Balitsky--JIMWLK formalism as a whole is a kind
of MFA --- the generalization of the BK equation to finite $N_c$.

Anticipated by early studies \cite{Salam95} (mostly numerical)
within Mueller's ``color dipole picture'' \cite{AM94}, the
importance of the gluon--number fluctuations for the high--energy
evolution in QCD has been fully appreciated only recently, over
the last couple of years \cite{IM032,MS04,IMM04,IT04}. Rare
fluctuations associated with gluon splitting have been shown to
control the formation of higher--point correlations in the dilute
tail of the gluon distribution at relatively large transverse
momenta (above $Q_s$). Once initiated through fluctuations, such
correlations are rapidly amplified by the BFKL evolution
\cite{BFKL}, and then play a crucial role in the growth of the
gluon distribution and its eventual saturation: they are at the
origin of the saturation and unitarization effects included in the
Balitsky--JIMWLK equations. Thus, while they are correctly
describing the role of the higher--point correlations in providing
saturation, the Balitsky--JIMWLK equations fail to also include
the physical {\em source} for such correlations, namely, the
gluon--number fluctuations in the dilute regime.

Soon after this failure has been first realized \cite{IT04}, new
equations have been proposed \cite{IT04,IT05,MSW05} which correct
the Balitsky--JIMWLK hierarchy by including the proper source
terms (the relevant fluctuations) within the limits of the
large--$N_c$ approximation. The latter is necessary since the
dynamics leading to fluctuations in the dilute tail is described
in the framework of the dipole picture \cite{AM94,IM031}. There is
currently an intense, ongoing, effort aiming at generalizing these
equations to finite $N_c$
\cite{KL05,KL3,KL4,BREM,MMSW05,HIMS05,Balit05,KL5}, but in spite
of some interesting discoveries --- like the recognition
\cite{KL3,BIIT05} of a powerful self--duality of the high--energy
evolution, and the construction of an effective Hamiltonian which
is explicitly self--dual \cite{BREM,Balit05}
---, the general evolution equations valid for arbitrary $N_c$ are
not yet known. This is the main reason for our restriction to the
large--$N_c$ approximation in what follows.

The new equations in Refs. \cite{IT04,IT05} form an infinite
hierarchy for the $N$--dipole amplitudes $\langle T^{(N)}\rangle$
with $N\ge 1$ which describe the scattering between a system of
$N$ color dipoles and an energetic hadronic target. For reasons to
be briefly mentioned in Sect. \ref{SECT_FACT}, we shall refer to
these equations as the {\em `Pomeron loop equations'}. They are
currently under active investigation
\cite{LL05,BIIT05,Stasto05,GS05,EGBM05,IST05,SX05,ArmMil06}, and
although their general solutions are not known, a lot of
information is already available about their {\em asymptotic
behaviour} at high energy, via the correspondence
\cite{MP03,IMM04,IT04} between high--energy QCD and statistical
physics (see the discussion below and also Sect. \ref{SECT_PL} for
more details). This is the information that we shall use in our
analysis of deep inelastic scattering throughout this paper.

Previous studies of DIS at high energy, or small ``Bjorken--$x$''
\cite{LW94,BW93,BP96,W97,BMH97,BJW99,KM99,GBW99,KST00,Gotsman00,KL00,K01,KW01,BGBK,LevinHERA,IIM03,MS03,CM04,FS04,GBM05,MariaDiff05}
(see also the review papers \cite{WM99,Hebecker,Ingelman}), have
shown that {\em diffractive} DIS represents a privileged framework
to study both {\em gluon saturation} --- since the diffractive
cross--section $\sigma_{\rm diff}$ appears to be dominated by
relatively large $q\bar q$ configurations, with size $r \sim
1/Q_s$ ($Q_s$ is the target saturation momentum, and increases
with the energy as $Q_s^2(x)\sim 1/x^\lambda$)
--- and {\em fluctuations} --- since, as known
since long \cite{GW,PM78},  the inelastic part of $\sigma_{\rm
diff}$ is a measure of the dispersion of the virtual photon
wavefunction over its various Fock space components (like $q\bar
q$, $q\bar q g$, etc.). Note, however, that the ``fluctuations''
which appear in the standard discussion of diffraction
\cite{GW,PM78} refer to the wavefunction of the {\em projectile}
--- in DIS, the virtual photon and its various hadronic excitations ---,
and not to that of the {\em target} (the proton, or more generally
some hadron $h$). In such previous studies, the target has always
been treated in the spirit of the {\em mean field approximation},
like an optical potential or some fixed gluon distribution, off
which scatter the partonic components of the projectile. However,
our main emphasis here will be precisely on the gluon--number
fluctuations {\em  in the target wavefunction}, which are the
relevant fluctuations for a study of the high--energy limit, and
which influence {\em all} the DIS observables, and not only those
associated with diffraction. Indeed, as we shall see in Sect.
\ref{SECT_DIPOLE}, the calculation of both inclusive and
diffractive cross--sections in DIS at high energy and large $N_c$
naturally involves the amplitudes for dipole--target scattering,
as determined by the Pomeron loop equations alluded to above.

Let us clarify here what we shall mean by the {\em `high--energy
regime of DIS'\,} in what follows. This is the regime where, when
increasing the energy (or decreasing $x$), the resolution scale
$Q^2$ of the virtual photon is simultaneously increased (although
not as fast as the total energy, of course !), in such a way that
$Q^2$ remains comparable, or even larger, than the target
saturation momentum $Q_s^2(x)$. Indeed, the {\em strict}
high--energy limit, in which the energy is increased at a {\em
fixed} value for $Q^2$ (and for a fixed impact parameter $b$), is
less interesting, since conceptually clear: for sufficiently high
energy, the cross--section  reaches its {\em unitarity} (or `black
disk') {\em limit}, in which the projectile is completely absorbed
by the target. (This behaviour will be manifest in the theoretical
formalism that we shall use, which preserves unitarity.)  On the
other hand, the behaviour at relatively large $Q^2$, such that
$Q^2 \gg Q_s^2(x)$, is considerably more interesting, since in
that region the scattering is weak, yet --- as we shall discover
--- it is strongly influenced by the physics of saturation, within
a large window in $Q^2$ whose width is increasing with $1/x$.
Moreover, with increasing energy, the effects of {\em
fluctuations} in the evolution of the target gluon distribution
become more and more pronounced, and the interplay between
fluctuations and saturation leads to a rich structure in DIS at
high energy and $Q^2 \simge Q_s^2(x)$, that we now explain.

\begin{figure}[t]
    \centerline{\epsfxsize=12cm\epsfbox{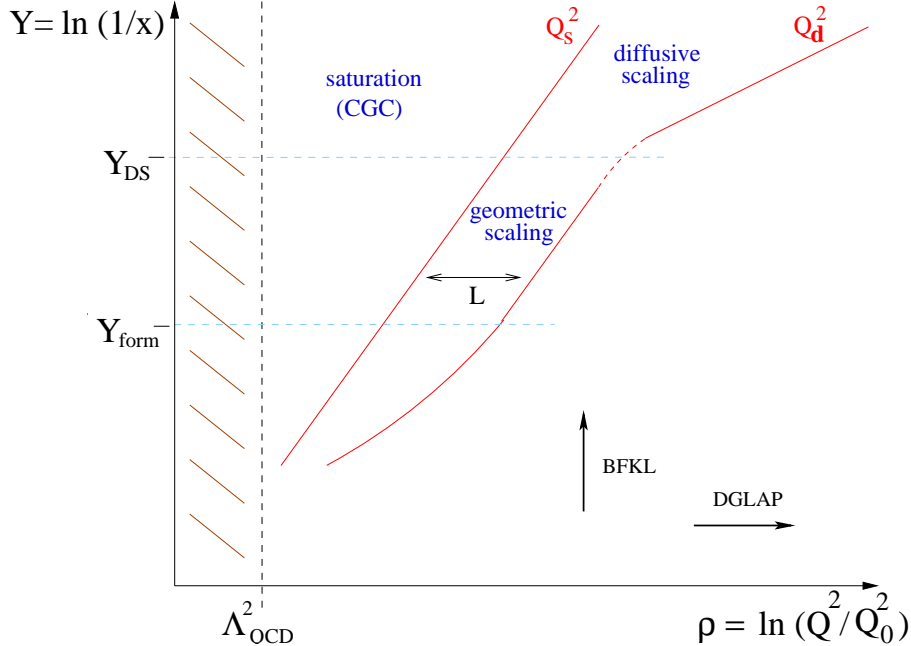}}
    \caption{\sl ``Phases'' of the hadronic wavefunction in the
kinematical plane $Y$--$\,\ln(Q^2/Q_0^2)$. When $Y \lesssim Y_{\rm
DS}$ the gluon distribution scales with momentum in a `geometric'
way, i.e.~it is a function of $Q^2/Q_s^2$. When $Y \gtrsim Y_{\rm
DS}$ it scales in a `diffusive' way, i.e.~it becomes a function of
$\,\ln(Q^2/Q_s^2)/\sqrt{Y}$.
    \label{Diag}}
    \end{figure}

To that aim, let us refer to Fig. \ref{Diag}, which represents the
kinematical plane $Y$--$\,\rho$ (with $Y\equiv\ln (1/x)$ and
$\rho\equiv \ln(Q^2/Q_0^2)$, where $Q_0^2$ is a fixed reference
scale) for DIS scattering, and also the physical phase--space for
the QCD evolution of the hadronic target. Within the latter
interpretation, $Y$ is the target rapidity and $Q^2$ can be
assimilated to either the transverse momentum of a gluon inside
the target ($k_\perp\sim Q$), or the transverse size of a dipole
within the projectile ($r\sim 1/Q$). The straight line denoted as
``$Q_s^2$'' in Fig. \ref{Diag} represents the {\em average}
saturation line in the target: $\langle\rho_s\rangle_Y \equiv
\ln(\langle Q_s^2\rangle/Q_0^2) \simeq \lambda Y$, with $\lambda$
the `saturation exponent'. This is an {\em `average'} since, in
the presence of fluctuations,  the saturation momentum becomes a
random quantity: the stochastic evolution of the target gives rise
to a {\em statistical ensemble} of gluon configurations, and the
value of $Q_s^2$ can vary from one configuration to another. The
{\em dispersion} $\sigma^2\equiv \lan\rho_s^2\ran -
\langle\rho_s\rangle^2$ which characterizes this variation will
play a fundamental role in what follows. This stochastic process
turns out to be a random walk in $\rho_s\equiv \ln( Q_s^2/Q_0^2)$,
hence the dispersion rises linearly with $Y$ \cite{IMM04}\,:
$\sigma^2(Y) \simeq D_{\rm fr} Y$, with $D_{\rm fr}$ the `front
diffusion coefficient' (see Sect. \ref{SECT_PL}). The second line
in Fig. \ref{Diag}, denoted as ``$Q_d^2$'', will be shortly
explained.

The most interesting region for us here is the {\em scaling
window} in between these two lines. At lower values of $Q^2$, to
the left of the saturation line, the gluons form a color glass
condensate with large occupation numbers $\sim 1/\alpha_s$, the
dipole amplitude reaches the unitarity limit $\langle T\rangle =
1$, and the mean field description is well justified. At much
larger $Q^2$, on the right of the scaling region, one finds the
standard perturbative regime, where the gluon system is very
dilute, the dipole amplitude shows `color transparency', $\langle
T(r)\rangle\propto r^2$, and neither saturation, nor fluctuations,
do play any role. But in the scaling region at intermediate $Q^2$,
the gluon occupation numbers are relatively low and the scattering
is weak, $\langle T\rangle \ll 1$, yet the dynamics is generally
influenced by both saturation and fluctuations, in proportions
which depend upon the actual value of $Y$.

The crucial scale for the present discussion is the rapidity
$Y_{\rm DS}$ which separates between an {\em intermediate} energy
regime at $Y < Y_{\rm DS}$ , where $\sigma^2\ll 1$ and the
mean--field picture roughly applies (with some important
limitations, though, as we shall discuss in Sect. \ref{SECT_PL}),
and a {\em high--energy} regime at $Y > Y_{\rm DS}$ , where
$\sigma^2\gg 1$ and the target expectation values are dominated by
{\em rare fluctuations} up to very large values of $\rho$.
(Parametric estimates for $Y_{\rm DS}$ and the other scales which
appear in Fig. \ref{Diag} will be given in Sect. \ref{SECT_PL}.)

Specifically, the lower part of the scaling region at $Y < Y_{\rm
DS}$ is the window for {\em geometric scaling}
\cite{geometric,SCALING,MT02,DT02,MP03}.  In that region, the
dipole amplitude and the gluon occupation number `scale' as
functions of the dimensionless variable $Q^2/\langle
Q_s^2\rangle$; e.g.,
 \be\label{TBK}
 \langle T(r)\rangle_Y \approx
(r^2 \langle Q_s^2\rangle)^{\gamma_0}\equiv \exp[-\gamma_0
 (\rho-\langle\rho_s\rangle)],\qquad\mbox{with}\qquad
 \gamma_0\simeq 0.63\,.\ee
This behaviour is particularly interesting as it may explain a
similar scaling observed in the HERA data for DIS at $x\le 0.01$
and up to rather large $Q^2$ ($Q^2\simle 400$ GeV$^2$)
\cite{geometric}.

But in the high--energy regime at $Y > Y_{\rm DS}$, the geometric
scaling is progressively washed out by fluctuations \cite{MS04}
and eventually replaced (when $Y\gg Y_{\rm DS}$) by a new type of
scaling \cite{IMM04,IT04}, for which we propose the name of {\em
diffusive scaling\,}: the dipole amplitude scales as a function of
the dimensionless variable $(\rho-\langle\rho_s\rangle)/\sigma$.
This scaling extends up to very large values of $\rho$, namely, it
holds so long as $\rho-\langle\rho_s\rangle \ll \sigma^2$. The
borderline of this domain at $\rho_d(Y)\equiv \langle\rho_s\rangle
+ \sigma^2\simeq (\lambda+D_{\rm fr})Y$ represents the upper part
of the line ``$Q_d^2$'' in Fig. \ref{Diag}. The functional form of
the dipole amplitude is known within this entire domain
\cite{IT04} (see Sect. \ref{SECT_PL}). In particular, deeply
within the scaling window in Fig. \ref{Diag}, this is roughly a
Gaussian:
 \be\label{TAS}
 \langle T(\rho)\rangle_Y \approx
  \frac{\sigma}{\rho-\langle\rho_s\rangle}\,
    \,{\exp}\left\{-\frac{
    (\rho-\langle\rho_s\rangle)^2}{\sigma^2}\right\}
     \qquad \mbox{for} \qquad \sigma \,\ll \,
     \rho-\langle\rho_s\rangle\,  \ll\, \sigma^2\,.
     \ee
The physics behind this asymptotic form is remarkably simple and
{\em universal} (i.e., insensitive to both the initial condition
at low energy and the details of the evolution with increasing
energy). It relies on just two basic facts: \texttt{(i)}
saturation exists --- meaning that, in the event--by--event
description, $T(\rho)=1$ for $\rho$ smaller than some $\rho_s$,
and \texttt{(ii)} the saturation scale  $\rho_s$ is a random
variable distributed according to a Gaussian law, with expectation
value $\langle\rho_s\rangle$ and dispersion $\sigma^2$ which both
increase linearly with $Y$. Then, the scaling behaviour and the
Gaussian shape manifest in Eq.~(\ref{TAS}) are a direct reflection
of the corresponding properties of the probability distribution
for $\rho_s$. These properties transmit so directly from the
probability law to the average amplitude since, within the range
indicated in Eq.~(\ref{TAS}), the expectation value $\langle
T(\rho)\rangle$ is small, $\langle T(\rho)\rangle\ll 1$, yet it is
dominated by gluon configurations which are {\em at saturation} at
the scale $\rho$ of interest (i.e., for which $\rho_s
\simge\rho\gg \langle\rho_s\rangle$, and thus $T=1$) : Indeed,
although relatively {\em rare}, such configurations yield
contributions of order one, whereas the respective contributions
of the {\em typical} configurations, for which $\rho_s \sim
\langle\rho_s\rangle$, are exponentially suppressed (cf.
Eq.~(\ref{TBK})).

We thus arrive at the important conclusion that, even though the
scattering is weak {\em on the average}, meaning that the target
looks {\em typically dilute} when probed on the scale $r\sim 1/Q$,
the average amplitude is nevertheless controlled by ``black spots''
within the target wavefunction, i.e., by rare fluctuations with
unusually large gluon density. At sufficiently high energy and up to
extremely large values of $Q^2$ (much larger than $\langle
Q_s^2\rangle$), the physical quantities are dominated by saturated
gluon configurations. This conclusion will be further substantiated
by our analysis of the DIS cross--sections at high energy.

Specifically, by using estimates like that in Eq.~(\ref{TAS}) for
the dipole amplitudes $\langle T^{(N)}\rangle$ (the general
formulae of this type will be presented in Sect. \ref{SECT_PL}),
together with appropriate, `dipole', factorization fomual\ae\ for
the inclusive and the diffractive DIS cross--sections (that we
shall discuss in Sect. \ref{SECT_DIPOLE}), we shall analytically
compute in Sect. \ref{SECT_FLUCT} the dominant behaviour of the
respective cross--sections in the {\em high--energy regime}
--- by which we shall more precisely mean the diffusive
scaling regime at $\sigma^2\gg 1$ and $\rho-\langle\rho_s\rangle
\ll \sigma^2$ (see Fig. \ref{Diag}). The main conclusion of this
analysis can be concisely formulated as follows: All the physical
properties (like the diffusive scaling or the Gaussian decrease in
Eq.~(\ref{TAS})) that have been previously noticed at the level of
the dipole amplitudes translate literally to the DIS
cross--sections. This is so since, as we shall demonstrate in
Sect. \ref{SECT_FLUCT}, the various convolutions which relate the
DIS cross--sections to the dipole amplitudes --- namely, the
convolutions with the BFKL kernel which enter the high--energy
evolution of the projectile, and those with the wavefunction of
the virtual photon
--- are dominated by {\em small} dipole sizes $r\sim 1/Q$ (within the
kinematic regime of interest here).

This last point is perhaps surprising, especially in relation with
the diffractive sector, where the cross--section at large $Q^2$
($Q^2\gg Q_s^2$) was known to be dominated by relatively large
dipoles, with $r\sim 1/Q_s$ \cite{GLR,LW94,BW93,GBW99}. This
standard picture is correct indeed, but at high energy it applies
only in the event--by--event description, and not also on the
average. Indeed, even for $Q^2\gg \langle Q_s^2\rangle$, the
statistical ensemble representing the target does still contain
gluon configurations (`black spots') for which $Q_s^2\sim Q^2\,$;
although rare, such configurations will dominate the (inclusive and
diffractive) cross--sections, because the photon wavefunction
strongly favors the small dipole sizes $r\sim 1/Q$.

Another surprise of our analysis is that, in the whole
`high--energy regime' defined as above --- i.e., up to rather
large values of $Q^2$, such that $Q^2 \gg \langle Q_s^2\rangle$
---, the diffractive cross--section is dominated by its {\em
elastic} component, that is, by the elastic scattering between the
{\em onium} as a whole and the hadronic target. (The `onium'
represents the wavefunction of the projectile at the time of
scattering, as produced via the BFKL evolution of the original
$q\bar q$ pair; see Sect. \ref{SECT_DIPOLE} for details.) Whereas
expected in the `black disk' regime at $Q^2 \simle \langle
Q_s^2\rangle$ \cite{GW,PM78,Hebecker}, this property may look
surprising in the high--$Q^2$ regime at $Q^2 \gg \langle
Q_s^2\rangle$ (e.g., it contradicts the corresponding expectations
of the MFA \cite{GBW99,BGBK,MS03}), but it is in fact natural in
view of our present wisdom: even for such high values of $Q^2$, the
cross--section is dominated by `black spots' which are already in
the `black disk' regime.

The dominance of the elastic scattering has interesting
consequences for the dependence of the diffractive cross--section
upon the rapidity gap $Y_{\rm gap}$  : in the high--energy regime,
the differential cross--section per unit rapidity gap is strongly
peaked near $Y_{\rm gap}=Y$. Alternatively
--- and this is the language that we shall prefer in this paper
--- the cross--section integrated over all the values of the
rapidity gap $Y_{\rm gap}$ from a {\em minimal} value $Y_{\rm
gap}^{\rm min}$ up to $Y$ is independent of the lower limit
$Y_{\rm gap}^{\rm min}$, so long as the latter is not too close to
$Y$. This second language will be more convenient for us here
since the approximations that we shall employ
--- a ``leading logarithmic approximation'' with respect to $\ln
(1/\beta)\equiv Y-Y_{\rm gap}$ --- will allow us to compute the
integrated cross--section, but not also to control the details of
the differential cross--section near $\beta=1$.

Let us mention here some limitations of our approach (besides the
high--energy and large--$N_c$ approximations already discussed).
We shall not be able to describe the impact parameter dependence
of the scattering amplitudes; that is, we shall follow their
evolution with increasing $Y$ at a fixed impact parameter $b$.
There are two reasons for that: First, the asymptotic solutions to
the Pomeron loop equations are known only after a
`coarse--graining' in impact parameter space, which eliminates the
dependence upon $b$ \cite{IT04}. Second, more generally, we do not
expect perturbation theory to properly describe the
$b$--dependence of the scattering amplitudes, especially near the
edge of the hadron, where the gluon density is low and the
dynamics is sensitive to confinement. Because of that, in
perturbation theory one can study the `blackening' of the hadron
disk, but not also the radial expansion of the `black disk', i.e.,
the problem of the Froissart bound \cite{KW02,FB}.

Furthermore, in our study of diffraction, we shall not be able to
describe processes with arbitrary large diffractive mass $M_X$, or
arbitrary small values of $\beta$ (see kinematics in Fig.
\ref{kinem}). Indeed, the formalism for diffraction that we shall
develop in Sect. 2 can accommodate saturation effects in the
wavefunction of the target (the hadron), but not also in that of
the projectile (the virtual photon). This implies an upper limit
on the rapidity $\ln (1/\beta)$ of the projectile, which is
however parametrically large: $\ln (1/\beta)\ll
(1/\abar)\ln(1/\alpha_s^2)$. Hence, in the case  of diffraction,
the `high--energy limit' that we shall consider is the limit in
which the total rapidity $Y$ and the rapidity gap $Y_{\rm gap}$
can be arbitrarily large, but such that their difference $Y-Y_{\rm
gap}=\ln (1/\beta)$ remains finite and constrained as indicated
above.

This paper is organized as follows: In the first half of it, which
covers Sects. 2 and 3, we shall develop the formalism which is
required for a calculation of the DIS cross--sections in the
high--energy regime and for large $N_c$, and in the second half
(i.e., Sect. 4), we shall explicitly compute these
cross--sections, and compare to the respective mean--field
results.

Specifically, in Sect. 2 we shall develop a factorization scheme
for the diffractive dipole--hadron scattering at high energy (the
`dipole' being, of course, the $q\bar q$ pair produced via the
dissociation of the virtual photon in DIS).
%By the ``dipole'' we mean, of course, the $q\bar q$ pair produced
%via the dissociation of the virtual photon, but the electromagnetic
%vertex describing this dissociation will be left aside and
%reintroduced only in Sect. 4.
Our final formula turns out to be an interesting combination
between the `color dipole picture'  by Mueller \cite{AM94} and an
early formula for hadron--hadron diffraction by Good and Walker
\cite{GW}: Within the frame in which the target rapidity coincides
with the minimal rapidity gap $Y_{\rm gap}^{\rm min}$, the
projectile is described as an {\em onium}  --- a collection of
dipoles produced via the BFKL evolution of the original $q\bar q$
pair over a rapidity interval $Y-Y_{\rm gap}^{\rm min}$
---, and the diffractive cross--section is expressed as the onium
expectation value of the {\em elastic} cross--sections for all the
dipole configurations within the onium. Whereas the emergence of
such a formula is not unexpected (as the dipoles fulfill the main
requirement of the analysis in Ref. \cite{GW}, namely, they are
eigenstates of the $S$--matrix operator), our explicit derivation
has the merit to clarify the importance of properly choosing the
Lorentz frame in order for this formula to be valid. The relation
to previous approaches \cite{BP96,KW01,MS03}, in particular to a
non--linear evolution equation proposed by Kovchegov and Levin
\cite{KL00}, will be also discussed.

The aforementioned factorization formula involves the average
dipole amplitudes $\langle T^{(N)}\rangle$ evaluated at the target
rapidity $Y_{\rm gap}$, which in our formalism can be arbitrarily
large. In Sect. 3 we shall describe the calculation of these
amplitudes in the high--energy regime, with the purpose of
justifying expressions like Eq.~(\ref{TAS}) and completing the
physical picture developed before, in relation with Fig.
\ref{Diag}. The presentation in Sect. 3 is based on recent
developments in Refs. \cite{IMM04,IT05}, but it also contains some
new elements and conceptual clarifications, like a parametric
estimate for $Y_{\rm DS}$, the concept of diffusive scaling, and
the recognition of an interesting `rigidity' property for the
average dipole amplitude in the presence of fluctuations.

Sect. 4 is our main section, in which we establish the dominant
behaviour of the DIS cross--sections in the high--energy regime.
Remarkably, it turns out that this behaviour can be computed from
the scattering amplitude $\langle T(r)\rangle_Y$ for a {\em
single} dipole --- the original $q\bar q$ fluctuation of the
virtual photon. This is {\em a priori} clear for the {\em
inclusive} cross--section, but it is true for the {\em
diffractive} one (at high energy), because of the dominance of the
elastic scattering, as aforementioned. Our calculation will be
organized as follows: First, we shall consider the $q\bar q$
component alone and compute the respective contributions to
$\sigma_{\rm tot}$ and $\sigma_{\rm diff}$. Then, we consider the
additional contributions associated with the $q\bar q g$ component
and demonstrate that, in the regime of diffusive scaling, the
inelastic contribution is parametrically suppressed as compared to
the elastic one. To better emphasize the effects of fluctuations,
we shall compare our results to the corresponding predictions of
the MFA.

Sect. 5 summarizes our results and presents our conclusions and
some perspectives.

%As we shall argue there, a main effect of the fluctuations is to
%extend some of the salient features of saturation up to very high
%values of $Q^2$,  within the weak scattering regime.

\section{A dipole picture for deep inelastic scattering at high energy}
\setcounter{equation}{0}\label{SECT_DIPOLE}

\begin{figure}[t]
    \centerline{\epsfxsize=15cm\epsfbox{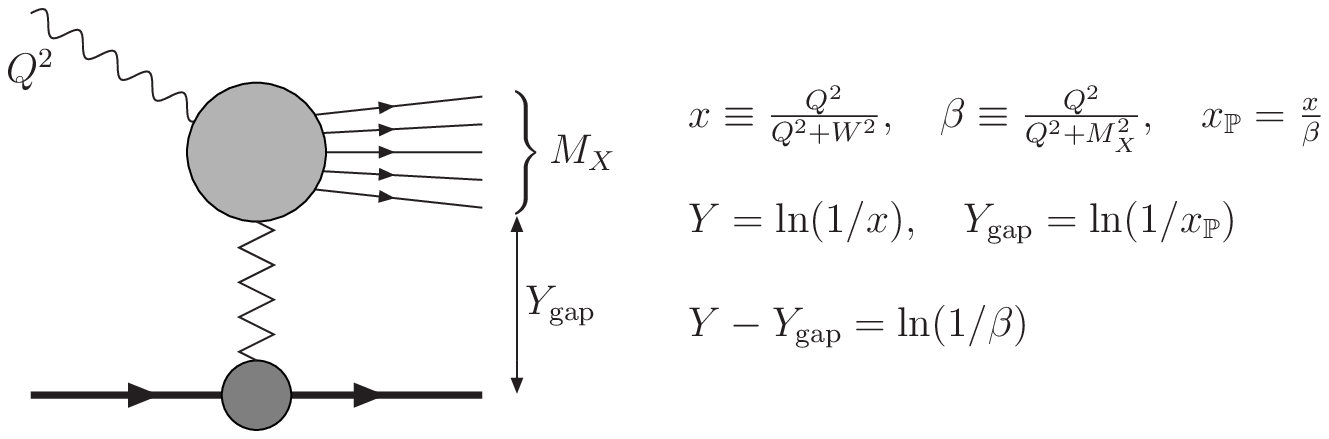}}
    \caption{\sl Kinematics for diffractive DIS at high energy, or
     small Bjorken--$x$: $Q^2$ is the virtuality of $\gamma^*$;
     $W$ is the center--of--mass energy of the $\gamma^* h$ system
     (with $W^2\gg Q^2$);
   $M_X^2$ is the invariant mass squared of the diffractive system.
    \label{kinem}\bigskip}
    \end{figure}

As explained in the Introduction, we would like to provide a
theoretical description for diffractive deep inelastic scattering
at high energy and in the multi--color limit $N_c\to\infty$. By
`diffractive events' we shall understand the DIS events $\gamma^*
h \,\to\, X p$ which contain an elastically scattered hadron $h$
in their final state together with a `diffractive hadronic state'
$X$ which is separated by a {\em rapidity gap} $Y_{\rm gap}$ from
the scattered hadron.  (We use here standard notations in the
context of DIS, as summarized in Fig. \ref{kinem}. See, e.g.,
Refs. \cite{Hebecker,WM99} for review papers on the phenomenology
of diffraction and more details about the kinematics.) In the high
energy regime at $W^2\gg Q^2,\,M^2_X$, the existence of a gap
$Y_{\rm gap}\simeq \ln [W^2/(Q^2+M^2_X)]$ follows automatically
from the condition that the proton undergoes elastic scattering
together with the relevant kinematics.  We shall be mainly
interested here in the {\em high--energy limit}, defined as the
limit $W^2\to\infty$ at fixed values for $Q^2$ and $M^2_X$. This
means that the total rapidity difference $Y= \ln (1/x)$ between
the projectile ($\gamma^*$) and the target ($h$), as well as the
rapidity gap $Y_{\rm gap}=\ln (1/x_\mathbb P)$ can increase
arbitrarily large, but such that their difference $Y-Y_{\rm
gap}=\ln (1/\beta)$ remains finite, and not too large (see
Eq.~(\ref{Ymaxproj}) below for the precise condition). Note that
our definitions for either diffraction or its high energy limit
are not the most general possible ones (e.g., one could consider
diffractive processes in which the target too breaks up after the
collision), but they do cover interesting physical situations and,
besides, they are constrained by the limitations of the subsequent
theoretical developments.

As also mentioned in the Introduction, we shall directly compute
the diffractive cross--section integrated over the rapidity gap
from some minimal value $Y_{\rm gap}^{\rm min}$ up to $Y$. For
convenience, from now on, we shall reserve the simpler notation
$Y_{\rm gap}$ for this {\em minimal} rapidity gap, as this
quantity will appear very often, and we shall measure the actual
rapidity by the corresponding $\beta$--variable, cf.  Fig.
\ref{kinem}. We have, clearly, the following relation
 \be\label{sigmadiffint}
 \frac{\rmd\sigma^{\gamma}_{\rm diff}}
 {\rmd^2 {b}}\,(Y,Y_{\rm gap},Q^2)=
\frac{\rmd\sigma^{\gamma}_{\rm el}}
 {\rmd^2 {b}}\,(Y,Q^2)\,+
 \int\limits_0^{Y-Y_{\rm gap}}
 \rmd \ln (1/\beta)\
 \frac{\rmd\sigma^{\gamma}_{\rm diff}}
 {\rmd^2 {b} \ \rmd \ln (1/\beta)}(Y,\ln (1/\beta),Q^2)\nn
 \ee
between the integrated quantity in the  l.h.s., that we shall
explicitly compute in what follows, and the differential
cross--section per unit rapidity --- which appears under the
integral sign in the r.h.s. and is the quantity usually considered
in phenomenological studies of diffraction. As also indicated by
the notations in Eq.~(\ref{sigmadiffint}), we shall consider
cross--sections at fixed impact parameter $\bm{b}$. The first,
``elastic'', term in the r.h.s. of the above equation is simply
the boundary term for the integration at $\beta=1$, or $Y_{\rm
gap}=Y$ :
  \be\label{sigmael}
 \frac{\rmd\sigma^{\gamma}_{\rm el}}
 {\rmd^2 {b}}\,(Y,Q^2)\,\equiv\,
 \frac{\rmd\sigma^{\gamma}_{\rm diff}}
 {\rmd^2 {b}}\,(Y,Y,Q^2)\,,\ee
and the reason why we refer to it as ``elastic'' is because,
within the `leading--logarithmic approximation' in $\ln (1/\beta)$
that we shall use, it corresponds to the elastic scattering
between the dissociation products of the virtual photon and the
target  (see Sect. 2.1 below for details and Appendix B for
results going beyond the leading--logarithmic approximation
alluded to above). Note however that the scattering is not elastic
from the viewpoint of the virtual photon itself (see the
discussion at the end of Sect. 2.1 and also Fig. \ref{DipEl}).

Let us briefly describe here the physical picture underlying our
subsequent developments. In a convenient `dipole' frame in which
the hadron $h$  carries most of the total energy, the scattering
between the virtual photon $\gamma^*$ and the target proceeds as
follows: Long before the scattering (say, at time $t_0\to
-\infty$), the virtual photon dissociates into a quark--antiquark
pair in a color singlet state (a `color dipole'), which then
evolves through soft gluon radiation until it meets the hadron (at
time $t=0$) and scatters off the color fields therein. At high
energy, the color dipole and the accompanying soft gluons are
eigenstates of the $S$--matrix operator --- the collision acts
merely as a color rotation on these states ---, and the original
picture of diffraction by Good and Walker \cite{GW} can be taken
over: The diffractive process $\gamma^* h \,\to\, X h$ consists in
the {\em elastic} scattering between the various Fock space
components of the projectile and the target. More precisely, we
shall argue below that this simple picture holds only in a {\em
well--tuned frame}, in which the rapidity $Y_0$ of the target
coincides with the (minimal) rapidity gap $Y_{\rm gap}$ in the
final state\footnote{In practice, the identification of $Y_0$ with
$Y_{\rm gap}$ must hold within the accuracy of the leading
logarithmic approximation: $Y_0$ and $Y_{\rm gap}$ can differ by
an amount ${\rm d}Y$ such that $\bar\alpha_s{\rm d}Y \ll 1$.}.
Such a special choice of the frame is necessary in order to avoid
the explicit treatment of the {\em final state interactions},
i.e., the gluon emissions and absorptions which take place in the
wavefunction of the projectile after the time of scattering, and
whose detailed description goes beyond the dipole picture.

As implicit in the above picture, we shall restrict ourselves to
the {\em leading logarithmic approximation} at high energy, in
which the evolution consists in the emission of `small--$x$
gluons', that is, gluons which carry only a small fraction $x\ll
1$ of the longitudinal momentum of their parent parton. So long as
the energy is not too high, the evolution remains linear and is
described by the BFKL formalism \cite{BFKL}, which leads to a
rapid growth of the gluon occupation numbers. But for sufficiently
high energy, {\em saturation effects} which tame this growth start
to be important \cite{GLR,MQ85,MV}, and introduce non--linear
effects in the evolution equations
\cite{B,K,JKLW,CGC,W,IT04,IT05}. In the physical situations that
we shall consider, the projectile will be always dilute and thus
evolve linearly. On the other hand, we shall allow for arbitrarily
high energies (and thus for saturation effects) on the side of the
target.

The high--energy evolution will be further simplified through
approximations valid at large $N_c$. On the projectile side, the
combination of the BFKL evolution with the large--$N_c$
approximation leads to Mueller's `color dipole picture'
\cite{AM94} : When $N_c\to\infty$, a gluon can be effectively
replaced with a pointlike quark--antiquark pair in a color octet
state, and a soft gluon emission from a color dipole can be
described as the splitting of the original dipole into two new
dipoles with a common leg. In this picture, the original $q\bar q$
pair produced by the dissociation of the virtual photon evolves
through successive dipole splittings and becomes an {\em onium}
--- i.e., a collection of dipoles --- at the time of scattering.
Strictly speaking, the dipole picture provides only the {\em norm}
of the onium wavefunction, that is, the probability distribution
for the dipole configurations within the onium \cite{AM94,IM031}.
But, as we shall see, this knowledge is indeed sufficient to
compute diffraction within the present formalism. On the target
side, the large--$N_c$ approximation turns out to be essential for
the successful construction of evolution equations for the
amplitudes describing the scattering between the projectile
dipoles and the target gluons \cite{IT04,IT05,MSW05}.

\subsection{Dipole factorization for the diffractive
cross--section} \label{SECT_FACT}

To develop our formalism, let us ignore for a while the
electromagnetic process $\gamma^* \to q\bar q$ (at high energy,
this process factorizes out \cite{AM90,NZ91} and can be easily
reintroduced later; see the beginning of Sect. \ref{SECT_QQ}), and
focus on the onium--hadron (${\cal O}h$) scattering. The quantity
that we would like to compute is the probability $P_{\rm
diff}(\x,\y; Y, Y_{\rm gap})$ for the diffractive process ${\cal
O} h \,\to\, X h$ in which the hadron $h$ undergoes elastic
scattering while the onium ${\cal O}$ dissociates into some
arbitrary hadronic state $X$ which is separated from the outgoing
hadron by a rapidity gap equal to, or larger than, $Y_{\rm gap}$
(with $Y_{\rm gap}\le Y$, of course). In this definition, $\x$ and
$\y$ are the transverse coordinates of the quark and,
respectively, the antiquark in the original $q\bar q$ pair, the
one which evolves into the onium.

\begin{figure}[t]
    \centerline{\epsfxsize=14cm\epsfbox{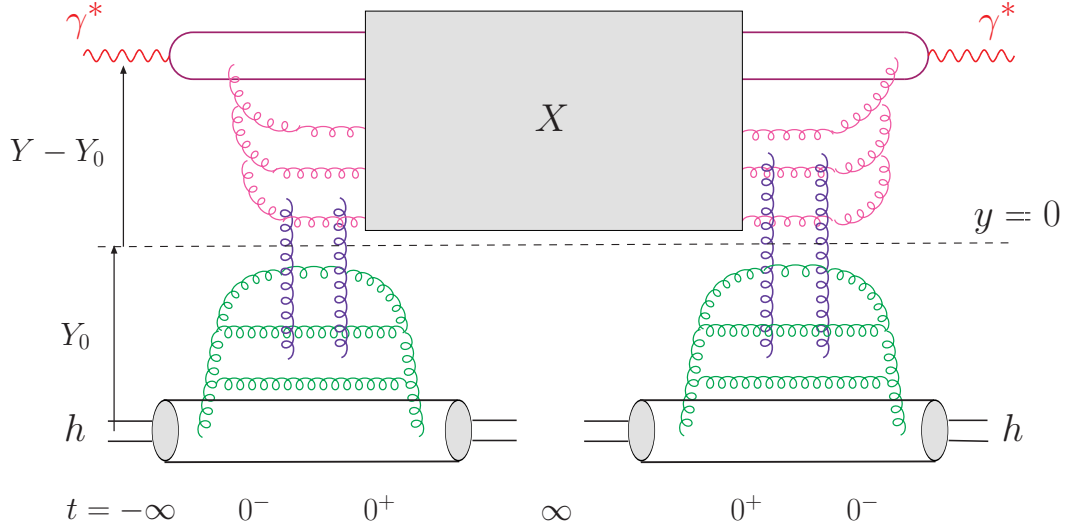}}
    \caption{\sl Typical diagram contributing to the diffractive process
$\gamma^* h \to X h$ in the frame where the target $Y_0$ coincides
with the rapidity gap. For the projectile, we illustrate the gluon
dynamics before and at the time of scattering. The final hadronic
state $X$ can be formed with an arbitrary number of gluons
produced via `final state interactions' (see the discussion in
Sect. \ref{SECT_PDIFF}). The gluons in the target recombine back
before the final state, so that the hadron emerges intact from the
collision. For simplicity, we exhibit only two--gluon exchanges.
    \label{GenGluon} \bigskip}
    \end{figure}

\begin{figure}[t]
    \centerline{\epsfxsize=14cm\epsfbox{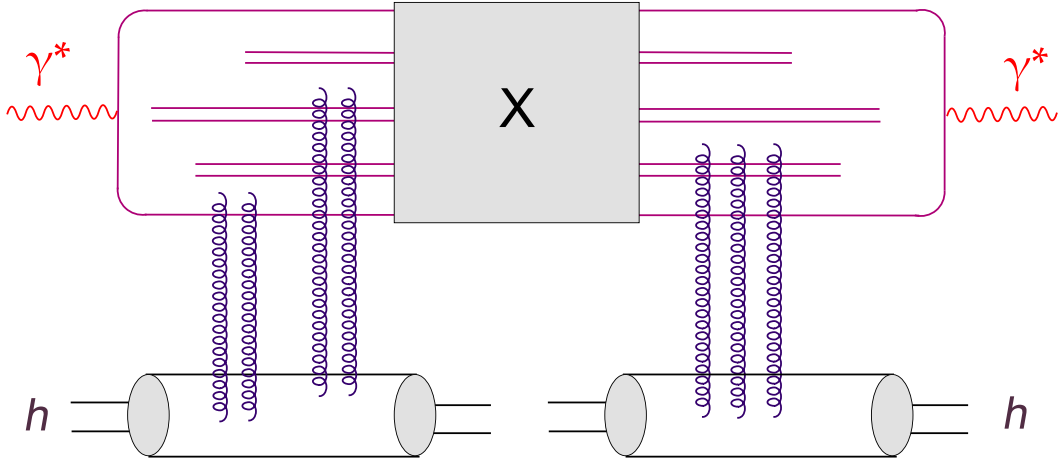}}
    \caption{\sl
    \label{GenDip}The same as in Fig.~3 but in the large--$N_c$ limit.
    Each gluon in
the wavefunction of the virtual photon has been replaced with a
pointlike quark--antiquark pair in a color octet state. The gluons
inside the target wavefunction are not shown explicitly anymore.
The relative simplicity of the $q\bar q$ representation allows us
to also exhibit some multiple gluon exchanges, corresponding to
unitarity corrections.\bigskip}
    \end{figure}

The probability $P_{\rm diff}(\x,\y; Y, Y_{\rm gap})$ is, of
course, frame--independent, but in what follows we shall derive an
explicit expression for it by working in the specific frame in
which $Y_{\rm gap}$ coincides with the rapidity $Y_0$ of the
target (and where the projectile has therefore a
rapidity\footnote{Within the leading--log approximation w.r.t.
$\ln (1/\beta)=Y-Y_0$, we can ignore the difference in rapidity
between the quark and the antiquark components of the original
$q\bar q$ pair and treat that pair as a `particle' with
unambiguous rapidity $Y-Y_0$. In Appendix B, we shall go beyond
this leading--log approximation for the case where $\beta$ is
close to one.} $Y-Y_0=Y-Y_{\rm gap}$). To avoid redundant
notations, we shall replace everywhere $Y_{\rm gap}\to Y_0$ and
write, e.g., $P_{\rm diff}(\x,\y; Y, Y_0)$. The final formula that
we shall arrive at (see Sect. \ref{SECT_PDIFF} for a derivation,
and Figs. \ref{GenGluon} and \ref{GenDip} for graphical
representations) reads
 \be\label{Pdiff}
 P_{\rm diff}(\x,\y; Y,Y_0)\,=\,
\sum_{N=1}^{\infty} \, \int \rmd \Gamma_N \,
 P_N(\{\bm{z}_i\}; Y-Y_0)
 \ \big|\big\langle 1\,-\, S(1)S(2)\cdots S(N)\big\rangle_{Y_0}\big|^2
  \, , \ee
with notations to be explained now:

\texttt{i)} $P_N(\{\bm{z}_i\};Y-Y_0)$ is the probability density
to produce a given configuration of $N$ dipoles after a rapidity
evolution $Y-Y_0$ starting with an original dipole $(\x,\y)$. The
configuration is specified by $N-1$ transverse coordinates
$\{\bm{z}_i\}=\{\bm{z}_1, \bm{z}_2,...\bm{z}_{N-1}\}$, which
physically represents the coordinates of the $N-1$ emitted gluons,
and in terms of which the coordinates of the $N$ dipoles are
$(\bm{z}_0,\bm{z}_1)$,
$(\bm{z}_1,\bm{z}_2)$,...,$(\bm{z}_{N-1},\bm{z}_N)$, with
$\bm{z}_0 \equiv \bm{x}$ and $\bm{z}_N \equiv \bm{y}$. Also,
$\sum_{N} \int\rmd \Gamma_N$ with $\rmd{\Gamma}_N\,=\,{\rm
d}^2\bm{z}_1{\rm d}^2\bm{z}_2\dots{\rm d}^2\bm{z}_{N-1}$
represents the sum over all the configurations. The dipole
probabilities are obtained by solving appropriate evolution
equations (see below) with the following initial conditions :
 \be\label{P0}
 P_1(Y=0)=1, \qquad P_{N>1}(Y=0)=0. \ee
The evolution of the probabilities is such that the correct
normalization condition
 \be\label{Pnotm}
 \sum_{N=1}^{\infty} \, \int \rmd \Gamma_N \,
 P_N(\{\bm{z}_i\}; Y)\,=\,1\ee
is satisfied at any $Y$.

\texttt{ii)} $S(i)\equiv S(\bm{z}_{i-1},\bm{z}_i)$ is the
$S$--matrix for the scattering between the $i$th dipole in the
projectile and a given configuration of color fields in the
target. (Recall that, in the high--energy scattering, a dipole is
an eigenstate of the interaction \cite{AM90,NZ91} and the internal
configuration of the target is frozen during the duration of the
collision; see also Eqs.~(\ref{Sdipole})--(\ref{CGCaverage})
below.) We shall also need the scattering amplitude corresponding
to a single dipole, defined as $T(i)\equiv 1-S(i)$. Then:
 \be\label{STN} {\cal S}_N\,\equiv\,\prod_{i=1}^{N}
 S(i)\,\quad\mbox{and}\quad
 {\cal A}_N\,\equiv\,1 - {\cal S}_N\,= \,1 -
 \prod_{i=1}^{N} \,[1 - T(i)]\,
 ,
 \ee
are respectively the $S$--matrix and the scattering amplitude for
a given configuration of $N$ dipoles. The brackets in $\langle
{\cal A}_N \rangle_{Y_0}$ denote the {\em target average}, that
is, the average over the ensemble of color fields in the target.
The target wavefunction, and thus the corresponding expectation
values, depend upon the rapidity interval $Y_0$ available for its
internal evolution. Note that, in a given event (i.e., for a given
configuration of the color fields in the target), the $N$ dipoles
scatter independently from each other --- the total $S$--matrix
${\cal S}_N$ is simply the product of $N$ factors corresponding to
the individual dipoles ---, but correlations are generally
introduced by the average over the target (because the color
fields there have non--trivial correlations). Thus, the
expectation value $\langle {\cal S}_N \rangle_{Y_0}$ is {\em not}
factorizing anymore. Such target correlations will play an
important role in the subsequent discussion in this paper.

\texttt{iii)} The quantity $|\langle 1- S(1)S(2)\cdots
S(N)\rangle_{Y_0}|^2\equiv |\langle {\cal A}_N \rangle_{Y_0}|^2$
is recognized as the probability for the {\em elastic} scattering
between a given set of $N$ dipoles and the target. Thus, as
anticipated, the diffractive probability (\ref{Pdiff}) represents
the {\em projectile average} of the elastic probabilities for all
the possible (dipole) configurations in the projectile. Thus
defined, the diffraction includes, but it does not reduces to, the
truly elastic collision, in which the onium {\em as a whole}
scatters elastically off the target. The difference appears
because, whereas the individual dipoles (or gluons) {\em are}
eigenstates of the interaction, this is not true also for their
superposition (the onium), since the various states in this
superposition interact differently with the target.

The {\em elastic probability} $P_{\rm el}$ for onium--target
scattering is rather computed as
 \be\label{Pel}
 P_{\rm el}(\x,\y; Y)\,=\,|1 - {\cal S}(\x,\y; Y)|^2\,\equiv\,
 | {\cal A}(\x,\y; Y)|^2
 \,,\ee
where ${\cal S}(\x,\y; Y)$ is the diagonal $S$--matrix element,
${\cal S}\equiv \langle \Psi_{in}| S|\Psi_{in} \rangle =\langle
\Psi_{in}| \Psi_{out} \rangle$, which measures the overlap between
the final state emerging from the collision and the initial state
prior to it. Therefore, $P_{\rm survival}\equiv |{\cal S}|^2$ is
the probability that the original state (for the ensemble
target+projectile) survives intact after the collision, whereas
 \be\label{Ptotinel}
 P_{\rm inel}(\x,\y; Y)\,\equiv\,1 - | {\cal S}(\x,\y; Y)
 |^2 \,,\ee
is the probability for some inelastic process to occur. In what
follows, the amplitude ${\cal A} \equiv 1 - {\cal S}$ for the
elastic scattering will be succinctly referred to as the {\em
forward amplitude}. Within the present framework, this quantity
can be computed as
 \be\label{Aforward}
 {\cal A}(\x,\y; Y)\,=\,
\sum_{N=1}^{\infty} \, \int \rmd \Gamma_N \,
 P_N(\{\bm{z}_i\}; Y-Y_0)
 \ \big\langle 1\,-\, S(1)S(2)\cdots S(N)\big\rangle_{Y_0}
  \, , \ee
which is perhaps a more familiar formula (various versions of it
can be found in the literature \cite{K,LL04,IST05}), and will be
also derived below.

As indicated by its notation, the forward amplitude ${\cal A}$ is
independent of the rapidity divider $Y_0$. This is {\em a priori}
true on physical grounds --- since, in the computation of ${\cal
A}$, $Y_0$ plays no dynamical role, but merely specifies the
Lorentz frame --- and is also verified by our explicit formula
(\ref{Aforward}), within its accuracy limits (see the discussion
below). In particular, one can choose to compute ${\cal A}$ in the
frame where $Y_0\simeq Y$. In that frame, the projectile is just
an elementary dipole $(\bm{x},\bm{y})$, without additional gluons,
and therefore ${\cal A}(\x,\y; Y) = 1 - \langle S(\x,\y)
\rangle_{Y} \equiv \langle T(\x,\y) \rangle_{Y}$, in agreement
with Eqs.~(\ref{Aforward}) and (\ref{P0}). Similarly, the elastic
probability (\ref{Pel}) can be computed as the elastic scattering
of the $q\bar q$ pair alone: according to Eqs.~(\ref{Pdiff}),
(\ref{Pel}), and (\ref{Aforward}) we can write
 \be\label{Pel1}
 P_{\rm el}(\x,\y; Y)\,=\,|\langle T(\x,\y) \rangle_{Y}|^2\,
 =\,P_{\rm diff}(\x,\y; Y,Y)\,,\ee
where the second equality reflects the physically obvious fact
that an elastic scattering is the same as a diffractive event
having $Y_0=Y$. This explains the identification performed in
Eq.~(\ref{sigmael}).

Via the optical theorem, the forward amplitude (\ref{Aforward})
also determines the total (or `inclusive')
cross--section\footnote{Note that our definitions for the
scattering amplitudes  $T$ and ${\cal A}$ differ by a factor of
$i$ from the usual definitions in the textbooks. With our
conventions, these amplitudes are predominantly {\em real} at high
energy.} :
 \be\label{Ptot}
 P_{\rm tot}(\x,\y; Y)%&\,=\,&2\,(1 - {\rm Re} \,{\cal S})
  \,=\,2\, {\rm Re}\,{\cal A}(\x,\y; Y)\,=\,
 P_{\rm el}+ P_{\rm inel}
 \,.\ee
At this point, a word of caution is necessary, concerning a slight
abuse in our terminology: The various ``probabilities'' introduced
so far are strictly speaking {\em differential cross--sections}
for onium--hadron scattering at fixed impact parameter; e.g.,
 \be\label{sigmab}
 \frac{\rmd\sigma_{\rm tot}}
 {\rmd^2 {b}}\,(\bm{r}, \bm{b}, Y)\,=\,P_{\rm tot}(\x,\y; Y),
 \qquad
 \frac{\rmd\sigma_{\rm diff}}
 {\rmd^2 {b}}\,(\bm{r}, \bm{b}, Y, Y_0)\,=\,
 P_{\rm diff}(\x,\y; Y, Y_0),
 \ee
(${\bm r}\equiv \x-\y$ and  $\bm{b}\equiv (\x+\y)/2$ are the
transverse size and the impact parameter of the original $q\bar q$
pair). As such, these quantities are certainly {\em proportional}
to the corresponding scattering probabilities, but they are not
necessarily bound to be smaller than one. Rather, they are
constrained by the unitarity of the $S$--matrix, which in the high
energy regime (where the scattering amplitudes are predominantly
real) requires $0\le\langle {\cal A}_N \rangle \le 1$ for any $N$.
This condition, together with the above formul\ae , implies the
standard inequality $P_{\rm diff} \le \frac{1}{2}P_{\rm tot}$
\cite{PM78}, together with a series of upper bounds like ${\cal
A}\le 1$, $P_{\rm diff} \le 1$, and $P_{\rm tot}\le 2$. Within the
formalism that we shall use below to compute the dipole
amplitudes, all these constraints are correctly respected, and the
various upper bounds are {\em saturated} in the high--energy
limit.

Before we turn to a derivation of Eqs.~(\ref{Pdiff}) and
(\ref{Aforward}) in the next subsection, let us specify their
range of validity in rapidity, and explain how to compute the
dipole probabilities $P_N(Y-Y_0)$ and the target--averaged matrix
elements $\langle {\cal A}_N \rangle_{Y_0}$ which enter these
formul\ae .

The rapidity $Y-Y_0$ of the projectile should be small enough for
the saturation effects to remain negligible. This in turn requires
\cite{AM94,IM031}
 \be\label{Ymaxproj}
  Y-Y_0\,\ll \,
 \frac{1}{\bar\alpha_s}\,\ln \frac{N_c^2}{\bar\alpha_s^2}\,\,\ee
where $\bar\alpha_s\equiv \alpha_s N_c/\pi$ should be treated as a
fixed quantity in the large--$N_c$ limit. (Recall that the typical
rapidity interval necessary for the emission of one small--$x$
gluon is ${\rm d}Y\sim 1/\bar\alpha_s$.) For rapidities satisfying
this constraint, the probabilities $P_N(Y-Y_0)$ can be obtained by
solving a `Master equation' \cite{IM031}, actually, a set of
coupled, linear, equations for the evolution with $Y$, whose
structure makes it clear that the dipole evolution in the dilute
regime is a Markovian stochastic process (see also Refs.
\cite{LL03,IT04}). Alternatively, the generating functional for
$P_N$, to be introduced in Eq.~(\ref{Zdip}) below, obey the
non--linear evolution equation (\ref{Zevol}), originally derived
by Mueller \cite{AM94}. For latter convenience, let us display
here the expressions for $P_N$ generated after only one step in
the evolution, that is, for $Y-Y_0={\rm d}Y$ and to linear order
in the small quantity $\bar\alpha_s {\rm d}Y$ :
 \be\label{PdY}
P_1({\rm d}Y)&\,=\,&1 - {\rm d}Y\,\frac{\bar\alpha_s}{2\pi} \int
{\rm d}{\bm z} \,{\cal M}({\bm{x}},{\bm{y}},{\bm z}),\nn
P_2({\bm{z}}|{\rm d}Y)&\,=\,&{\rm d
 }Y\,\frac{\bar\alpha_s}{2\pi}\,{\cal M}({\bm{x}},{\bm{y}},{\bm
 z}),\ee
and $P_N({\rm d}Y)=0$ for $N\ge 3$. In these equations,
\beq\label{dipkernel}
    \mathcal{M}(\bm{x},\bm{y},\bm{z})\,\equiv\,
    \frac{(\bm{x}-\bm{y})^2}
    {(\bm{x}-\bm{z})^2 (\bm{y}-\bm{z})^2}\, ,
    \eeq
is known as the {\em `dipole kernel'} \cite{AM94} : $(\abar/2\pi)
\mathcal{M}(\bm{x},\bm{y},\bm{z})$ is the differential probability
for an elementary dipole $(\bm{x},\bm{y})$ to split into two
dipoles $(\bm{x},\bm{z})$ and $(\bm{z},\bm{y})$ per unit rapidity.
Note that the integral over ${\bm z}$ in the formula for $P_1$ has
logarithmic singularities at $\bm{z}=\bm{x}$ and $\bm{z}=\bm{y}$.
Such singularities are expected at the level of the dipole
probabilities, but they cancel out in the calculation of physical
quantities, as it can be checked on the examples of
Eqs.~(\ref{Pdiff}) and (\ref{Aforward}).

Consider now the matrix elements for dipole--target scattering: At
high energy, the $S$--matrix $S(\bm{x},\bm{y})$ corresponding to a
single dipole $(\bm{x},\bm{y})$ can be computed in the eikonal
approximation as
 \be\label{Sdipole} S(\x,\y) \,=\,
\frac{1}{N_c}\,\tr \big(V^\dag({\x}) V({\y})\big)\,,\qquad
  V^\dag(\x)={\mbox P}\exp \left\{
ig\int dx^- A_a^+(x^-,\x)\,t^a \right\}\,,
 \ee
where the {\em Wilson lines} $V^\dag({\x})$ and $V({\y})$
represent the color rotations suffered by the quark and,
respectively, the antiquark after their scattering off the color
field $A^+_a$ in the target. (The $t^a$'s are the generators of
the SU$(N_c)$ algebra in the fundamental representation and the
symbol P denotes path--ordering in $x^-$.) Note that, in our
conventions, the projectile propagates in the negative $z$ (or
positive $x^-$) direction, so it couples to the $A^+$ component of
the color field in the target. At high energy and in a suitable
gauge, this is the only non--trivial component, and the average
over the target wavefunction amounts to a functional average over
$A^+$ (the `color glass' average \cite{MV,CGC}) :
 \be\label{CGCaverage}
\langle S(\x,\y) \rangle_{Y_0}\,=\, \int\,{\rm D}[A^+]\,
\,W_{Y_0}[A^+]\,\,\frac{1}{N_c}\, {\rm tr}\big(V^\dagger({\x})\,
 V({\y})\big), \ee
where the `color glass weight function' $W_{Y_0}[A^+]$ (a
functional probability density) can be interpreted as the squared
wavefunction of the target and depends upon the respective
rapidity $Y_0$. The expectation value $\langle {\cal S}_N
\rangle_{Y_0}$ for the scattering of $N$ dipoles is defined
similarly. Note that, physically, the non--linear effects in $A^+$
included via the Wilson lines in
Eqs.~(\ref{Sdipole})--(\ref{CGCaverage}) describe {\em multiple
scattering} to all orders.

In the high--gluon density regime where the gluon--number
fluctuations become negligible, the evolution of the weight
function $W_{Y_0}$ with increasing $Y_0$ is described by the
JIMWLK equation \cite{JKLW,CGC,W} --- a functional, non--linear,
equation of the Fokker--Plank type. Via equations like
(\ref{CGCaverage}), the JIMWLK equation generates an hierarchy of
ordinary evolution equations for the dipole amplitudes $\langle
{\cal S}_N \rangle_{Y_0}$, originally derived by Balitsky
\cite{B}. The non--linear effects encoded in the JIMWLK equation
describe gluon saturation in the target wavefunction and translate
into unitarity corrections in the Balitsky equations.

\begin{figure}[t]
    \centerline{\epsfxsize=12cm\epsfbox{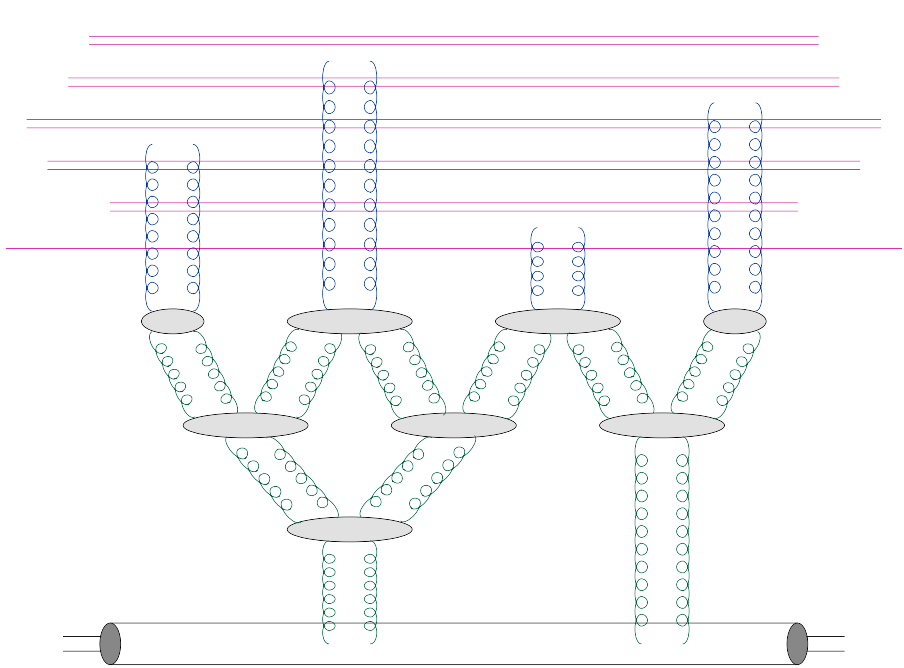}}
    \caption{\sl
    \label{Ploops} Pomeron loops in the forward amplitude for onium-hadron
    scattering. }
    \end{figure}

However, as pointed out in the Introduction, the gluon--number
fluctuations missed by the Balitsky--JIMWLK equations are in fact
essential for the physics at high energy, in that they represent
the source for the higher--point correlations responsible for
saturation. So far, the combined effects of fluctuations and
saturation on the dipole scattering at high energy have been taken
into account only in the large--$N_c$ limit
\cite{MS04,IMM04,IT04,MSW05}, where a new set of evolution
equations has been derived \cite{IT04,IT05} --- the `Pomeron loop
equations' alluded to in the Introduction (see also Refs.
\cite{LL05,BIIT05,Braun05}). In what follows, we shall assume that
the dipole amplitudes $\langle {\cal S}_N \rangle_{Y_0}$ obey
these new equations, which within the limits of the large--$N_c$
approximation are valid up to arbitrarily high energy. In addition
to the standard BFKL terms \cite{BFKL}, these equations include
{\em non--linear} terms
--- which correspond to gluon saturation in the target and ensure
the unitarization of the scattering amplitudes at high energy ---
and {\em source} terms --- which correspond to dipole splitting in
the dilute part of the target wavefunction and encode the
relevant, gluon--number, fluctuations at large $N_c$. The
perturbation theory for these equations can be organized in terms
of ``BFKL pomerons'' (the Green's function for the BFKL equation
\cite{BFKL}) which interact via ``triple Pomeron vertices''
\cite{BW93,BV99,Braun05} for Pomeron splitting and merging.
Accordingly, the solutions to these equations naturally encompass
the {\em Pomeron loops}. In Fig. \ref{Ploops}, we illustrate some
Pomeron loops effects in the calculation of the forward scattering
amplitude (\ref{Aforward}).

 To summarize, all the ingredients appearing in Eqs.~(\ref{Pdiff})
and (\ref{Aforward}) can be computed, at least in principle, by
solving evolution equations which are explicitly known. These
equations are valid for large $N_c$ and up to rapidities $Y$ and
$Y_0$ which can be arbitrarily large, but such that their
difference $Y-Y_0$ obeys the condition (\ref{Ymaxproj}). Within
this condition, the expression (\ref{Aforward}) for the forward
amplitude is independent of the rapidity divider $Y_0$, as it
should. Indeed, it has been demonstrated in Refs.
\cite{LL04,IST05} that the $Y_0$--dependence in the r.h.s. of
Eq.~(\ref{Aforward}) cancels out {\em exactly} when using the
Master equation for the dipole probabilities $P_N(Y-Y_0)$ together
with the (large--$N_c$ version of the) Balitsky--JIMWLK equations
for the dipole amplitudes $\langle {\cal S}_N \rangle_{Y_0}$. Now,
as argued before, the correct equations at high energy are not the
Balitsky--JIMWLK equations, but rather the Pomeron loop equations
of Refs. \cite{IT04,IT05}. The latter include the effects of gluon
number fluctuations, which are additional sources of
$Y_0$--dependence. In a more general calculation of ${\cal
A}(\x,\y; Y)$ which would be valid in {\em any} frame, such
additional dependencies would be compensated by recombination
effects in the wavefunction of the projectile. However, in any
`dipole frame' which satisfies the condition (\ref{Ymaxproj}) such
recombination effects are truly negligible, so
Eq.~(\ref{Aforward}) is indeed independent of $Y_0$, up to
higher--order corrections\footnote{This does not mean that the
physical consequences of the gluon--number fluctuations are also
negligible (these will be discussed in the next sections). It is
only the induced $Y_0$--dependence which is small indeed.}.
Incidentally, the above argument also shows that our formula for
the diffractive probability, Eq.~(\ref{Pdiff}), is {\em not}
independent of $Y_0$, as expected on physical grounds.

For later convenience, let us also introduce here the {\em
inelastic diffraction}, a process which, as shown in Ref.
\cite{PM78} in the context of hadron--hadron collisions, is a
direct probe of `parton' (here, dipole) fluctuations in the
wavefunction of the {\em projectile}. The corresponding
probability is obtained by simply subtracting out the (totally)
elastic component (\ref{Pel}) from the diffractive probability
(\ref{Pdiff}):
  \be\label{Pinel}
 P_{\rm diff}^{\rm\, inel}(\x,\y; Y,Y_0)\,=\,P_{\rm diff}(\x,\y; Y,Y_0)
 \,-\,P_{\rm el}(\x,\y; Y)\,.\ee
This formula takes on a particularly suggestive form after
noticing that Eqs.~(\ref{Pdiff}) and (\ref{Aforward}) can be
rewritten as (with simplified notations)
 \be\label{Pdifffor}
P_{\rm diff}(\x,\y; Y,Y_0)\,=\,\big\langle\, \big|\langle \,{\cal
A}\, \rangle_{\rm target}\big|^2 \big\rangle_{\rm proj}\,,\qquad
{\cal A}(\x,\y; Y)\,=\,\big\langle\, \langle \,{\cal A}\,
 \rangle_{\rm target}\, \big\rangle_{\rm proj}\,.\ee
In these formul\ae , the target--averaged amplitude ${\cal A}_{\rm
target}\equiv \langle \,{\cal A}\, \rangle_{\rm target}$ is still
an `operator' from the point of view of the projectile, in the
sense of depending upon a fixed configuration of dipoles. Also, at
high energy, the amplitude is predominantly real, so one can
ignore the modulus sign in the previous equations. We thus finally
arrive at \cite{PM78}
 \be\label{Pinel1}
 P_{\rm diff}^{\rm\, inel}(\x,\y; Y,Y_0)\,=\,\big\langle\,
 {\cal A}_{\rm target}^2\big\rangle_{\rm proj}\,-\,
 \big\langle\,
 {\cal A}_{\rm target}\big\rangle_{\rm proj}^2\,,\ee
which clearly exhibits the fact that $P_{\rm diff}^{\rm\, inel}$
is a measure of the dispersion of the dipole distribution within
the wavefunction of the projectile. The dipole picture that we
employ here provides an explicit realization for the projectile,
and thus allows one to compute this dispersion and any other
quantity pertinent to the distribution of dipoles. However, our
main emphasis in what follows will be not on the effects of
fluctuations in the {\em projectile} (these are already well
understood within the dipole picture; see, e.g., Refs.
\cite{AM94,Salam95,IM031,LL03,LL04,IT04}), but rather on the
physical consequences of the gluon number fluctuations in the {\em
target}, as encoded in the evolution equations with Pomeron loops
\cite{IT04,IT05}. These fluctuations affect separately all the
quantities introduced above (${\cal A}$, $P_{\rm diff}$, $P_{\rm
tot}$, etc.), because of their influence on the dipole scattering
amplitudes $\langle {\cal S}_N \rangle_{Y_0}$ which enter the
corresponding formul\ae .

Let us conclude this subsection with a warning against the abusive
interpretation of the notion of `inelastic diffraction' in the
context of deep inelastic scattering. As we shall see at the
beginning of Sect. 4, the diffractive probability (\ref{Pdiff})
determines the `integrated' cross--section for DIS diffraction,
i.e., the quantity in the l.h.s. of Eq.~(\ref{sigmadiffint}). In
view of this, and of the identification in Eq.~(\ref{Pel1}), it
becomes clear that the separation of $\sigma_{\rm diff}$ in
between an `elastic' plus an `integral' piece, as shown in the
r.h.s. of Eq.~(\ref{sigmadiffint}), corresponds to the
decomposition of $P_{\rm diff}$ in between its elastic and
inelastic components, cf. Eq.~(\ref{Pinel}): $P_{\rm diff}= P_{\rm
el}+P_{\rm diff}^{\rm\, inel}$. However, in DIS, and unlike in
hadron--hadron collisions, these two pieces of $\sigma_{\rm diff}$
cannot be {\em separately} measured. Indeed, even a process in
which the onium scatters elastically (see Fig. \ref{DipEl}) still
appears as {\em inelastic} at the level of the DIS experiment,
because the elastically scattered $q\bar q$ pair does not
recombine back into a virtual photon in the final state, but
rather emerges as a hadronic state. Hence, the only experimentally
relevant quantity is the `total' diffractive cross--section, as
determined by $P_{\rm diff}$.

\subsection{Justifying the dipole factorization}
\label{SECT_PDIFF}

To justify the above formul\ae\ for dipole factorization (in
particular, those pertinent to diffraction), we shall use the
light--cone wavefunction formalism, as developed in the
light--cone gauge $A_a^-=0$ (which is best suited for a study of
the eikonal scattering of the projectile within our conventions).
Our presentation will be rather sketchy, as similar techniques and
manipulations can be widely found in the literature, and it will
focus on the non--trivial aspects of the argument only.

Let us denote by $|\Psi({-\infty})\rangle$,  $|
\Psi({0^-})\rangle$, $| \Psi({0^+})\rangle$, and
$|\Psi({\infty})\rangle$ the wavefunction of the complete system
target plus projectile (including their hadronic descendants in
the case of the final state) at times\footnote{Strictly speaking,
the role of the `time variables' in the present formalism is
played by the light--cone coordinates --- $x^-$ in the case of the
projectile and, respectively, $x^+$ in the case of the target; for
simplicity, we shall use the more intuitive notation $t$.}
$t=-\infty$ (long before the scattering), $t=0^-$ (just before the
scattering), $t=0^+$ (immediately after the scattering), and,
respectively, $t=\infty$ (the final hadronic state which is
measured by the detector). We have:
 \be\label{Psiin}
 |\Psi({0^-})\rangle\,=\,|{\cal O}(Y-Y_0)\rangle\,\otimes\,
 |h(Y_0)\rangle\,,\ee
where the two factors in the r.h.s. are the states of the
projectile and, respectively, the target as produced after a
rapidity evolution $Y-Y_0$ and, respectively, $Y_0$.

In the case of the projectile, the initial state for this
evolution is the elementary color dipole $(\x,\y)$ produced by the
dissociation of the virtual photon: $|{\cal O}(0)\rangle = |
(\x,\y) \rangle$ (the color indices are kept implicit; see, e.g.,
Refs. \cite{KW01,CM04} for details). The evolved state at $t=0^-$
is then a superposition of partonic states containing the original
$q\bar q$ pair $(\x,\y)$ together with an arbitrary number of soft
gluons in a given spatial and color configuration. The detailed
structure of such individual Fock states turns out not to be
necessary for the present purposes. Rather, we shall simply denote
by $| N \rangle$ a generic state containing the $q\bar q$ pair
$(\x,\y)$ together with $N-1$ small--$x$ gluons. We thus write,
quite generically:
 \be |{\cal O}(Y-Y_0)\rangle\,=\,\sum_N \,c_N(Y-Y_0)\, | N \rangle\,,\ee
where the sum over $N$ should be really understood as a sum over
the number of gluons, an integral over their transverse
coordinates, and a sum over their polarizations and color
configurations (we refer again to Refs. \cite{KW01,CM04} for more
precise notations).

Concerning the target, its precise initial state at $t=-\infty$ is
unimportant here. Rather, all that we need to assume is that at
the time of scattering the target can be described as a
superposition of states with a given color field (eigenstates of
the gauge field operator $A\equiv A^+$) with coefficients
$\Phi[A]$ which depend upon the rapidity $Y_0$ :
 \be |h(Y_0)\rangle\,=\,\int D[A]\,\Phi[A](Y_0)\,\,|A
 \rangle\,.\ee

Consider now the onium--hadron collision which takes place at time
$t=0$. As anticipated, at high energy the partonic Fock--space
states composing the onium are eigenstates of the collision
operator: the transverse positions and the spins, or
polarizations, of the `partons' (quarks and gluons) are not
changed by the scattering, while their color orientations undergo
a field--dependent precession described by Wilson lines. We can
write, schematically,
 \be\label{Psiout}  |\Psi({0^+})\rangle\,=\,\int D[A]\,\Phi[A](Y_0)\,\,
 \sum_N \,c_N(Y-Y_0)\, \big(\prod_{i=1}^N\,S_i[A]\big)\,
 | N \rangle\,\otimes\,|A \rangle\,,\ee
where $S_i[A]$ is the $S$--matrix for the individual parton $i$,
and is a Wilson line (with generally open color indices).

For all the processes that we are interested in, the hadronic
target emerges intact from the collision, i.e., it undergoes
elastic scattering, so we need only the projection of the outgoing
state on the target state prior to the collision:
 \be\label{Psidiff} |\Psi_{\rm diff}({0^+})\rangle &\,\equiv\,&
 |h(Y_0)\rangle \,\langle h(Y_0)|\Psi({0^+})\rangle\nn
 &\,=\,&\int D[A]\, \big|\Phi[A](Y_0)\big|^2\,\,
 \sum_N \,c_N(Y-Y_0)\, \big(\prod_{i=1}^N\,S_i[A]\big)\,
 | N \rangle\,\otimes\,|h(Y_0)\rangle \nn
 &\,\equiv\,& \sum_N \,c_N(Y-Y_0)\, \big\langle \prod_{i=1}^N\,S_i[A]
 \big\rangle_{Y_0}\,
 | N \rangle\,\otimes\,|h(Y_0)\rangle
 \,,\ee
where the brackets in the third line refer to the average over the
target wavefunction. With the identification $W_{Y_0}[A]\equiv
|\Phi[A](Y_0)|^2$, this target averaging is recognized as the
color glass averaging in Eq.~(\ref{CGCaverage}).

Eq.~(\ref{Psidiff}) makes it manifest that, in the calculation of
diffractive processes, the average over the target wavefunction is
to be performed already at the level of the {\em amplitude}
\cite{KM99}, rather than at the level of the {\em squared}
amplitude (which is the quantity defining a probability; see
Eq.~(\ref{Pdiff0}) below). This peculiarity is, of course, related
to our restriction to processes in which the target undergoes
elastic scattering, and would not be true anymore for more
general, diffractive, processes in which the target too is allowed
to break up after the collision (see also the discussion at the
end of this subsection).

Note also that the target averaging in Eq.~(\ref{Psidiff})
automatically implies a color projection of the outgoing
projectile onto color singlet states: Since the post--collisional
state of the target (which is the same as its initial state
$|h(Y_0)\rangle$) is a color singlet, so must be also the
corresponding state of the projectile. In mathematical terms, the
weight function $W_{Y_0}[A]$ must be gauge--invariant, so the
operation of averaging $\prod_{i=1}^N S_i[A]$ with this weight
function must close all the Wilson lines into gauge--invariant
color traces. Under this average and for large $N_c$, the product
$ \prod_{i=1}^N S_i[A]$ naturally reduces to the product of $N$
dipolar factors, where is each of them is like the one appearing
in Eq.~(\ref{Sdipole}).

\begin{figure}[t]
    \centerline{\epsfxsize=14cm\epsfbox{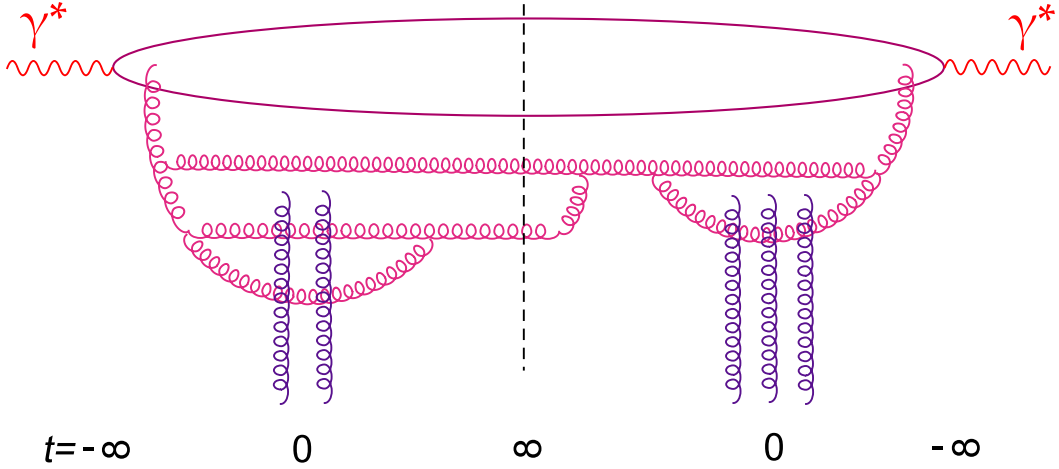}}
    \caption{\sl
    A particular diagram contributing to the diffractive process
$\gamma^* h \to X h$. The dashed vertical line at $t=\infty$
indicates the final (partonic) state. In the direct amplitude, a
three--gluon state (together with the original $q\bar{q}$ pair)
describes a component of the virtual photon wavefunction which
interacts with the target hadron (not shown here) at time $t=0$.
After the scattering, final state interactions (gluon
recombination) occur and at $t=\infty$ only two gluons are
``measured''. Similarly, in the complex conjugate amplitude, the
wavefunction contains two soft gluons at the time of scattering;
after the collision, these gluons recombine with each other, and
finally an additional soft gluon is emitted before the time of
measurement. The rapidity gap associated with this particular
process is larger than $Y_0$. \label{GluonFS}\bigskip}
    \end{figure}

\begin{figure}[t]
    \centerline{\epsfxsize=14cm\epsfbox{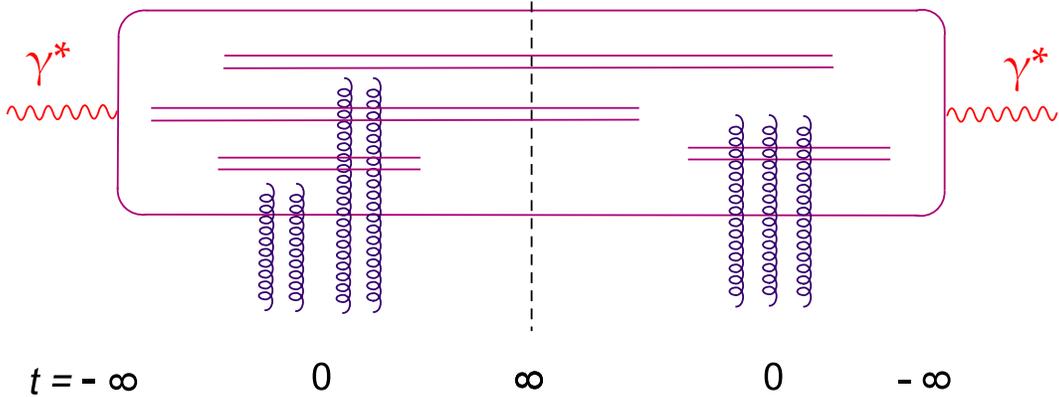}}
    \caption{\sl The same as in Fig.~6 in the large--$N_c$ limit; each gluon in
the wavefunction of the virtual photon has been replaced with a
pointlike quark--antiquark pair in a color octet state.
    \label{DipFS}}
    \end{figure}

\begin{figure}[t]
    \centerline{\epsfxsize=14cm\epsfbox{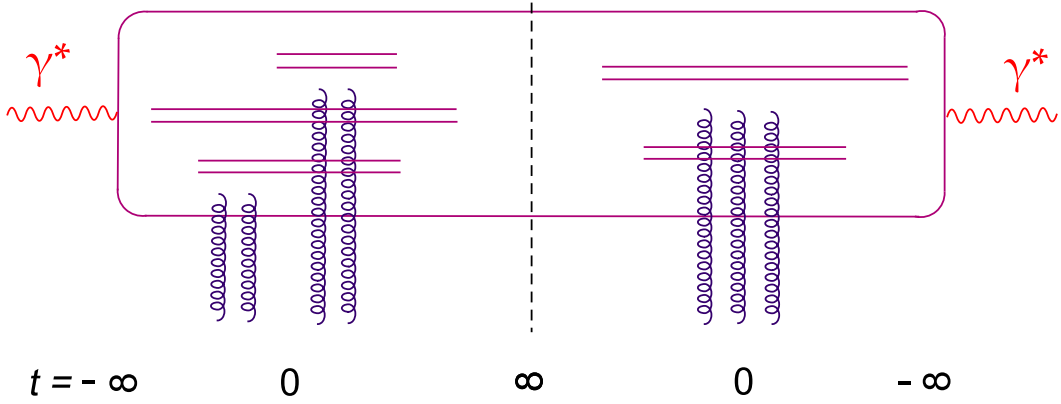}}
    \caption{\sl Another diagram contributing to the diffractive process
$\gamma^* h \to X h$. Both the amplitude and the c.c.~amplitude
correspond to contributions to the forward amplitude
(\ref{Aforward}) for dipole--hadron scattering. Hence, this
diagram is also an illustration of elastic dipole--hadron
scattering, Eq.~(\ref{Pel}). (Notice that this is not an elastic
scattering for $\gamma^*$ as well, since the virtual photon does
not appear in the final state.) Here the rapidity gap takes its
maximum value $Y$.
    \label{DipEl}\bigskip}
    \end{figure}

The final state at the time of detection is obtained by letting
the `diffractive' state in Eq.~(\ref{Psidiff}) evolve from $t=0$
up to $t=\infty$ under the action of the unitary evolution
operator $U(\infty,0)$ :
  \be\label{Psifin}
  |\Psi_{\rm diff}({\infty})\rangle\,=\,U(\infty,0)\,
  |\Psi_{\rm diff}({0^+})\rangle\,.\ee
It is this late time evolution which is responsible for the `final
state interactions' alluded to in the previous subsection: under
the action of $U(\infty,0)$, some of the soft gluons contained in
the diffractive state at $t=0^+$ may recombine with each other, so
that the ensuing rapidity gap in a given event (that is, for a
given final state) may be actually larger than $Y_0$. (See Figs.
\ref{GluonFS} and \ref{DipFS} for an example.) In fact, the
largest allowed gap is equal to $Y$ and corresponds to elastic
scattering\footnote{This is consistent with the fact that, for
$Y_0=Y$, the diffractive probability (\ref{Pdiff}) reduces to the
corresponding elastic one, cf. Eq.~(\ref{Pel1}) : that is, $P_{\rm
diff}^{\rm\, inel}(Y=Y_0)=0$.}, i.e., to the situation where all
the gluons (or dipoles) in the final state recombine back before
their detection into the original dipole $(\x,\y)$. This is
depicted in Fig. \ref{DipEl}.

\begin{figure}[t]
    \centerline{\epsfxsize=14cm\epsfbox{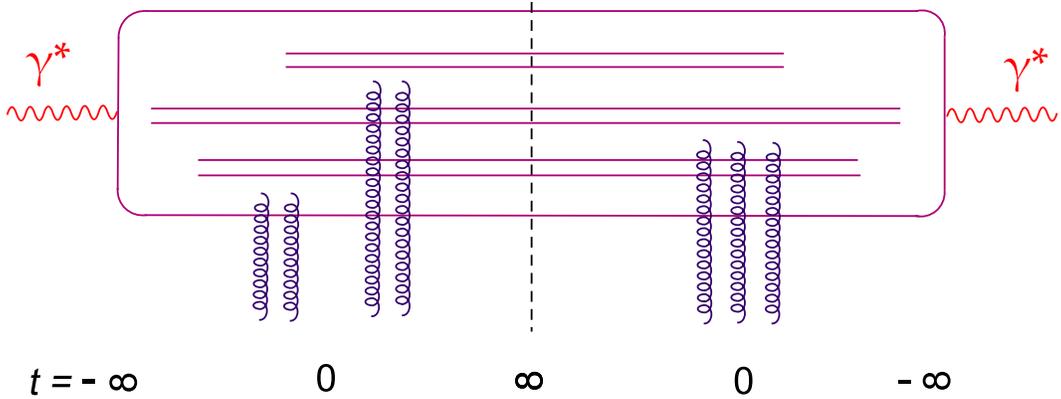}}
    \caption{\sl  A final example for the diffractive process $\gamma^* h \to X
h$. Here there are no final state interactions at all and the
rapidity gap in this case is equal to its minimum value $Y_0$.
    \label{DipInel}\bigskip}
    \end{figure}

On the other hand, within the present assumptions, the final gap
can clearly not be {\em smaller} than $Y_0$: This is so because of
our restriction to processes in which the target undergoes {\em
elastic} scattering, so that none of the virtual quanta initially
contained within the target wavefunction (and which are
distributed in rapidity from 0 to $Y_0$) can be released in the
final state. Since, on the other hand, there is no activity in the
projectile wavefunction at rapidities smaller than $Y_0$, it is
clear that all the hadrons emerging in the final states and coming
from the dissociation of the virtual photon must have a rapidity
equal, or superior, to $Y_0$. Therefore, the minimal rapidity gap
is indeed $Y_0$. (See Fig. \ref{DipInel}.)

In principle, it should be possible to compute the probability
$P_{\rm diff}(\x,\y; Y, Y_{\rm gap})$ for any given value for the
{\em minimal} rapidity gap $Y_{\rm gap}$  within the range $Y\ge
Y_{\rm gap}\ge Y_0$. To that aim, one should explicitly follow the
evolution of the outgoing state after the collision, as indicated
in Eq.~(\ref{Psifin}), and then project $|\Psi_{\rm
diff}({\infty})\rangle$ onto final states in which the hadrons
released through the fragmentation of the projectile have
rapidities $Y\ge Y_{\rm gap}$. This procedure would select those
states within $|\Psi_{\rm diff}({0^+})\rangle$ in which all the
projectile gluons with rapidities between $Y_0$ and $Y_{\rm gap}$
have recombined back in the evolution from $t=0^+$ up to
$t=\infty$. Very recently, a more general formalism has been
proposed \cite{W05}, which in principle allows one to achieve this
goal, and thus to compute the diffractive scattering in a generic
frame (and for generic values of $N_c$). It would be interesting
to see how our present factorization emerges from that formalism.

However, our main point here is precisely that the calculation of
$P_{\rm diff}(\x,\y; Y, Y_{\rm gap})$ can be drastically
simplified by conveniently choosing the Lorentz frame
--- namely, by choosing $Y_0=Y_{\rm gap}$ ---, since with this
particular choice one does not need to worry about the final state
interactions. Indeed, the condition that the gap be bigger than,
or equal to, $Y_0$ is automatically satisfied within this
kinematics, and it does not introduce any constraint on the final
state of the projectile. Hence, the probability for such a
diffractive event is measured by the norm $\langle \Psi_{\rm
diff}({\infty})|\Psi_{\rm diff}({\infty})\rangle$ of the final
state, which is the same as the norm of the state $|\Psi_{\rm
diff}({0^+})\rangle$ immediately after the collision, because of
the unitarity of the operator $U(\infty,0)$. One sometimes says
that `the final state interactions cancel between the direct
amplitude and the complex conjugate one', but in the present case
this cancelation is an almost trivial consequence of our choice
for the frame (see, e.g., Refs. \cite{ChenMueller,KM98,KL00} for
less trivial cancelations of this type).

More precisely, in computing $P_{\rm diff}$ one needs to consider
the deviation between the outgoing diffractive state and the
incoming state $|\Psi({0^-})\rangle$ (indeed, the final state must
be different from the initial one in order to have a real
scattering). Thus, if one defines:
 \be |\delta\Psi_{\rm diff}({0^+})\rangle\,\equiv\,
  |\Psi_{\rm diff}({0^+})\rangle \,-\,|\Psi({0^-})\rangle\,,\ee
then in this particular frame one can write
 \be\label{Pdiff0}
 P_{\rm diff}(\x,\y; Y, Y_{0})&\,=\,&\langle \delta\Psi_{\rm
 diff}({0^+})|\delta\Psi_{\rm diff}({0^+})\rangle \nn
 &\,=\,& \sum_N \,|c_N(Y-Y_0)|^2\ \Big|\big\langle \prod_{i=1}^N\,S_i[A]
 - 1
 \big\rangle_{Y_0}\Big|^2
 \,,\ee
where in writing the second line we have used the fact that the
partonic states $|N\rangle$ form an orthonormal basis. At large
$N_c$, the norm $|c_N(Y-Y_0)|^2$ of the state $|N\rangle$ (summed
over polarizations and color indices) can be identified with the
probability $P_N(Y-Y_0)$ for a $N$--dipole configuration in the
dipole picture. With this identification, Eq.~(\ref{Pdiff0}) is
finally recognized as our originally proposed expression,
Eq.~(\ref{Pdiff}).

The expression (\ref{Aforward}) for the forward scattering
amplitude can be similarly obtained : By definition, ${\cal
A}\equiv 1 -{\cal S}$ with ${\cal S}$ the forward $S$--matrix
element, computed as
 \comment{ as measured by the
overlap between the outgoing state (\ref{Psiout}) and the incoming
one (\ref{Psiin}). (Strictly speaking, one should use the
corresponding states at $t=\infty$, namely
$U(\infty,0)|\Psi({0^+})\rangle$ and respectively
$U(\infty,0)|\Psi({0^-})\rangle$, but the evolution operator
$U(\infty,0)$ cancels again in the scalar product.) Thus,}
 \be\label{Aforward0}
 {\cal S}(\x,\y;Y)&\,=\,&\langle \Psi({0^-}) |\Psi({0^+})\rangle
 \nn
 &\,=\,& \sum_N \,|c_N(Y-Y_0)|^2\,\int D[A]\, \big|\Phi[A](Y_0)\big|^2\,
  \prod_{i=1}^N\,S_i[A]
  \,,\ee
which immediately leads to Eq.~(\ref{Aforward}), as anticipated.

The previous discussion also explains our limitation to
diffractive processes in which the hadronic target emerges intact
from the collision. Of course, it would be very interesting
(especially in view of applications to the phenomenology) to be
able to describe the more general processes $\gamma^* h\,\to\,XY$
in which the hadron dissociates into some low--mass hadronic
system $Y$ separated by a rapidity gap $Y_{\rm gap}$ from the
diffractive state $X$. But to that aim, one cannot avoid a
detailed study of the final state interactions, which in the case
of the target requires moreover a model for the hadronic
structure.

It is finally interesting to compare our formula (\ref{Pdiff}) for
the diffractive probability to related results in the literature.
In the next subsection we shall show that, under an additional,
mean field, approximation, our expression for $P_{\rm diff}$ obeys
a non--linear equation originally proposed by Kovchegov and Levin
\cite{KL00}. But before that we would like to point out that
Eq.~(\ref{Pdiff}) also encompasses other results from the previous
literature. When the target is not too dense so that multiple
scattering is negligible, Eq.~(\ref{Pdiff}) reduces to the
expression employed by Bialas and Peschanski \cite{BP96}. Namely,
in the single scattering approximation, we have ${\cal A}_N\approx
\sum_{i=1}^{N} T(i)$ (c.f., Eq.~(\ref{STN})), and then
Eq.~(\ref{Pdiff}) becomes
 \be\label{PBFKL}
 P_{\rm diff}(\x,\y; Y,Y_0)&\,\approx&
  \int\limits_{\bm{u},\bm{v}}
 n(\bm{u},\bm{v};Y-Y_0)\
  |\langle T(\bm{u},\bm{v})\rangle_{Y_0}| ^2\nn &{}& +
 \int\limits_{\bm{u}_i,\bm{v}_i}
 n^{(2)}(\bm{u}_1,\bm{v}_1;\bm{u}_2,\bm{v}_2;Y-Y_0)
  \langle T(\bm{u}_1,\bm{v}_1)\rangle_{Y_0} \langle
 T(\bm{u}_2,\bm{v}_2)
 \rangle_{Y_0}^*\,, \ee
where $n$ and $n^{(2)}$ denote, respectively, the average dipole
number density and the average density of (distinct) dipole pairs
in the projectile (see, e.g., Sect. 5 Ref. \cite{IT04} for the
corresponding definitions) and the amplitude $\langle
T\rangle_{Y_0}$ obeys the BFKL equation, as appropriate for a
dilute target. Hence, Eq.~(\ref{PBFKL}) represents the BFKL
approximation to $P_{\rm diff}$. Physically, the first term in the
r.h.s. of Eq.~(\ref{PBFKL}) represents the probability that the
same dipole scatters in both the direct and the complex conjugate
amplitude, whereas the second term describes the scattering of
different dipoles.

On the other hand, for a generic, dense, target but a relatively
small rapidity for the projectile, such that $\bar\alpha_s(Y-Y_0)
\ll 1$, one can restrict oneself to onium configurations which
contain only two dipoles at the time of scattering. In that case,
Eq.~(\ref{Pdiff}) reduces to results previously obtained by Kovner
and Wiedemann \cite{KW01} and by Munier and Shoshi \cite{MS03}.
The corresponding expressions will be presented in Sect.
\ref{SECT_QQ}, where their high--energy limit will be also
investigated.

Let us emphasize here, however, some important differences in our
treatment of the target expectation values as compared to Refs.
\cite{KW01,MS03}. \texttt{(i)} In Ref. \cite{MS03}, the target
averages have been estimated in a mean field approximation (cf.
Sect. \ref{SECT_MFA}) which neglects the correlations induced by
gluon--number fluctuations; such an approximation is reasonable at
intermediate energies, but it eventually fails at sufficiently
high energies (see the discussion in Sect. \ref{SECT_PL}).
\texttt{(ii)} In an attempt to relax the restriction to elastic
scattering on the target side, the authors of Ref. \cite{KW01}
have suggested to replace the target averaging at the level of the
{\em amplitude} with an averaging at the level of the
(diffractive) {\em probability}. With our present notations, their
suggestion amounts to the following replacement
 \be
\big|\langle \,{\cal A}\, \rangle_{Y_0}\big|^2\ \longrightarrow\
 \langle \,{\cal A}\,P_{\rm sing}\,{\cal A}^* \rangle_{Y_0}\,\ee
within the r.h.s. of Eq.~(\ref{Pdiff}). Here,  $P_{\rm sing}$ is
the projector onto color singlet states for the projectile
wavefunction, and it has been introduced to ensure that ${\cal A}$
involves only gauge--invariant operators; e.g., $P_{\rm sing}
{\cal A}={\cal A}_N$, cf. Eq.~(\ref{STN}), for a $N$--dipole
state. (In our previous developments, such a projection was
automatically ensured by the target averaging at the level of the
amplitude, cf. Eq.~(\ref{Psidiff}).) However, with this new
prescription, the target averaging is tantamount to summing over
all the possible gauge--invariant final states for the target,
without any restriction on their distribution in rapidity. But
prior to scattering, the target wavefunction had developed virtual
excitations at all the rapidities $y$ ranging from $y=0$ up to
$y=Y_0$, and in the absence of any rapidity veto on the final
state, there is no reason why these quanta should not materialize
into hadrons occupying this whole rapidity interval. In other
terms, with the target expectations values evaluated as in Ref.
\cite{KW01}, one is actually including processes which have no
rapidity gap whatsoever. Such processes do not qualify as
``diffractive'' according to the usual terminology, and will be
not considered in what follows.

\subsection{Evolution equations in the mean field approximation}
\label{SECT_MFA}

Although our main focus in this paper will be on the effects of
{\em fluctuations} (i.e., of the {\em deviations} from the mean
field behaviour) in so far as the target expectation values are
concerned, it is nevertheless interesting at this level to
slightly deviate from the main stream of the presentation and
discuss the mean--field version of our precedent results. This is
useful for, at least, two reasons: First, it will allow us to make
contact with a non--linear equation for the diffractive
probability previously proposed by Kovchegov and Levin
\cite{KL00}, and thus clarify the conditions for the validity of
the latter. Second, the {\em mean field approximation} (MFA) that
we shall introduce here will later serve us as a term of
comparison, to better emphasize the consequences of the
fluctuations at high energy.

Specifically, the MFA consists in the following factorization
assumption
 \be\label{FACT}
 \langle S(1)S(2)\cdots S(N)\rangle_{Y_0} \,\approx\, \langle
S(1)\rangle_{Y_0}\, \langle S(2)\rangle_{Y_0}\cdots \langle
S(N)\rangle_{Y_0}\,,
 \ee
which neglects the correlations among the dipoles induced by their
scattering off the target. Ultimately, this is an assumption about
the absence of correlations in the target gluon distribution. (For
instance, such a factorization holds indeed, at large $N_c$,
within the framework of the McLerran--Venugopalan model \cite{MV}
for the gluon distribution of a large nucleus. More generally, it
amounts to use a Gaussian approximation for the CGC weight
function in Eq.~(\ref{CGCaverage}) \cite{GAUSS}.) In what follows
we shall demonstrate that, with this additional assumption,  our
formula (\ref{Pdiff}) for the diffractive probability obeys indeed
to the Kovchegov--Levin (KL) equation of Ref. \cite{KL00}. During
this procedure, and under similar assumptions, we shall also
provide a rapid derivation of the Balitsky--Kovchegov (BK)
equation for the forward amplitude \cite{B,K}.

For more clarity, we shall denote with a bar quantities computed
in the mean field approximation. For instance:
 \be\label{Pdiffbar}
 \bar P_{\rm diff}(\x,\y; Y,Y_0)\,=\,
\sum_{N=1}^{\infty} \int \rmd \Gamma_N \,
 P_N(\{\bm{z}_i\}; Y-Y_0)
 \ \big| 1- s_1 s_2\cdots
 s_n
 \big|^2, \ee
and similarly
 \be\label{Sbar}
 \bar S(\x,\y; Y)\,=\,
\sum_{N=1}^{\infty} \, \int \rmd \Gamma_N \,
 P_N(\{\bm{z}_i\}; Y-Y_0)
 \ s_1 s_2\cdots
 s_n
  \, .\ee
In these formul\ae , $s_i$ is a simplified notation for $\langle
S(\bm{z}_{i-1},\bm{z}_i)\rangle_{Y_0}$, which in the present
context is of course the same as $\bar S(\bm{z}_{i-1},\bm{z}_i;
{Y_0})$.

%dipole, as computed within the mean field
%approximation\footnote{Therefore $\langle
%S(\bm{x},\bm{y})\rangle_{Y_0}$ itself evolves with $Y_0$ according
%to the BK equation (\ref{bk}).}.
%We have, of course, $\bar S(\x,\y;
%Y)= \langle S(\bm{x},\bm{y})\rangle_{Y}$ for any $Y$.

In what follows, we shall exploit the evolution of the
probabilities $P_N(Y-Y_0)$ within the dipole picture in order to
deduce a set of non--linear evolution equations for the quantities
defined in Eqs.~(\ref{Pdiffbar})--(\ref{Sbar}). To that aim, it is
more convenient to use the original version of the equations for
$P_N$, due to Mueller \cite{AM94}. Specifically, Mueller has
derived an equation for the following generating functional
 \begin{align}\label{Zdip}
  Z_{\x\y}[Y,u]=\sum_{N=1}^\infty \int \textrm{d}\Gamma_{N}\,
 P_N(\z_1...\z_{N-1};Y)\,u_{1}u_{2}\cdots u_{N},
 \end{align}
where $u_i\equiv u(\z_{i-1},\z_{i})$ is an arbitrary `source'
field, and $(\x,\y)=(\z_0,\z_N)$, as usual. The dipole
probabilities $P_N(Y)$ can be deduced from $Z_{\x\y}[Y,u]$ by
functionally differentiating with respect to $u$ and then letting
$u\to 0$. The evolution equation satisfied by $Z_{\x\y}$ reads
 \begin{align}\label{Zevol}
\frac{\partial Z_{\x\y}}{\partial Y}= \frac{\bar{\alpha}_s}{2\pi}
\int\limits_{\bm{z}}
\mathcal{M}(\x,\y,\z)\left(-Z_{\x\y}+Z_{\x\z}Z_{\z\y}\right),
 \end{align}
to be solved with the following initial condition, which follows
from Eq.~(\ref{P0}) :
\begin{align}
  Z_{\x\y}[0,u]=u_{\x\y}.
  \end{align}
As demonstrated in Ref. \cite{LL03}, this non--linear equation is
equivalent with the infinite hierarchy of linear, master,
equations for the probability densities $P_N$ \cite{IM031},
although the respective pictures of the evolution are quite
different: According to Eq.~(\ref{Zevol}), the one--step evolution
consists in the splitting of the original, high--rapidity, dipole
into two child dipoles, which then separately evolve and produce
their own distribution of dipoles. By contrast, the master
equations of Ref. \cite{IM031} focus on the splitting of the
low--rapidity dipoles produced in the previous steps of the
evolution.

Eq.~(\ref{Zevol}) looks formally similar to the BK equation for
the $S$--matrix  \cite{K}, and indeed the latter can be easily
derived from the former, as we show now: Eqs.~(\ref{Sbar}) and
(\ref{Zdip}) imply $\bar S(\x,\y; Y) = Z_{\x\y}[Y,s]$, which
together with Eq.~(\ref{Zevol}) immediately leads to (with the
simplified notation $\bar S_{\x\y}\equiv \bar S(\x,\y; Y)$)
 \begin{align}
\frac{\partial \bar S_{\x\y}}{\partial Y}=
\frac{\bar{\alpha}_s}{2\pi} \int\limits_{\bm{z}}
\mathcal{M}(\x,\y,\z)\left(-\bar S_{\x\y}\,+ \,\bar S_{\x\z} \bar
S_{\z\y}\right), \label{bk}
 \end{align}
which is the BK equation \cite{B,K}, as anticipated. It is more
customary to write this equation in terms of the scattering
amplitude $\bar T_{\x\y}\equiv 1- \bar S_{\x\y}$, in which case it
reads:
 \be\label{BK}
 \frac{\partial \bar T_{\x\y}}{\partial Y}= \frac{\bar{\alpha}_s}{2\pi}
\int\limits_{\bm{z}} \mathcal{M}(\x,\y,\z)\,\big(
 \bar T_{\x\z}+ \bar T_{\z\y}- \bar T_{\x\y}-\bar T_{\x\z}
 \bar T_{\z\y}\big).
 \ee

The evolution of the diffractive probability (\ref{Pdiffbar}) can
be similarly addressed. Eqs.~(\ref{Pdiffbar}) and (\ref{Zdip})
imply (the dependence upon the rapidity gap $Y_0$ is kept
implicit, since this variable is fixed in the following
manipulations)
 \be  \bar P^{\textrm{diff}}_{\x\y}(Y)
 &\,=\,&1-2 \bar S_{\x\y}(Y)+Z_{\x\y}[Y,s^2] \nonumber \\ &=&
 -1+2 \bar T_{\x\y}(Y)+Z_{\x\y}[Y,s^2], \ee
where we have assumed that $s$ is real, as appropriate at high
energy. This rewriting of $ \bar P^{\textrm{diff}}_{\x\y}$
together with the previous equations (\ref{Zevol}) and (\ref{BK})
then imply
 \be  \frac{\partial
 \bar P^{\textrm{diff}}_{\x\y}}{\partial
Y}&\,=\,&2\frac{\bar{\alpha}_s}{2\pi} \int\limits_{\bm{z}}
\mathcal{M}\,(\x,\y,\z)\left(
  \bar T_{\x\z}+  \bar T_{\z\y}-  \bar T_{\x\y}-
   \bar T_{\x\z} \bar T_{\z\y}\right) \nonumber \\
&{}& \quad  -\frac{\bar{\alpha}_s}{2\pi} \int\limits_{\bm{z}}
\mathcal{M}(\x,\y,\z)\left(Z_{\x\y}[s^2]-Z_{\x\z}[s^2]Z_{\z\y}[s^2]\right)
\nonumber \\ &\,=\,&\frac{\bar{\alpha}_s}{2\pi}
\int\limits_{\bm{z}}
 \mathcal{M}(\x,\y,\z)\,\Bigl(2 \bar
 T_{\x\z}+2 \bar T_{\z\y}-2 \bar T_{\x\y}-2 \bar T_{\x\z} \bar T_{\z\y}
 \nonumber \\ &{}& \qquad  -( \bar P^{\textrm{diff}}_{\x\y}+1-2 \bar T_{\x\y})
 +( \bar P^{\textrm{diff}}_{\x\z}+1-2 \bar T_{\x\z})( \bar
 P^{\textrm{diff}}_{\z\y}
 +1-2 \bar T_{\z\y}) \Bigr) \ee
After some simple manipulations, the expression in the r.h.s. can
be recast into the form
 \be\label{KL}
 \frac{\partial  \bar P^{\textrm{diff}}_{\x\y}}{\partial Y}&\,=\,&
\frac{\bar{\alpha}_s}{2\pi} \int\limits_{\bm{z}}
\mathcal{M}(\x,\y,\z)\,\Bigl( \bar
 P^{\textrm{diff}}_{\x\z}+  \bar P^{\textrm{diff}}_{\z\y}-
 \bar P^{\textrm{diff}}_{\x\y}+ \bar P^{\textrm{diff}}_{\x\z}
  \bar P^{\textrm{diff}}_{\z\y}\nonumber
\\ &{}& \qquad \qquad \qquad  -2 \bar T_{\x\z}
 \bar P^{\textrm{diff}}_{\z\y}-2 \bar P^{\textrm{diff}}_{\x\z}
  \bar T_{\z\y}+2 \bar T_{\x\z} \bar T_{\z\y}
 \Bigr), \ee
which is recognized, as anticipated, as the equation proposed by
Kovchegov and Levin \cite{KL00}. It is interesting to notice that,
in Ref. \cite{KL00}, this equation has been obtained by working in
the target rest frame, which required a more intricate analysis of
the final state interactions.

The non---linear terms in Eqs.~(\ref{BK}) and (\ref{KL}) describe
(incoherent) multiple scattering between the dipoles in the
projectile and the color fields in the target and are responsible
for unitarization: In the high energy limit, the solutions $\bar T
(Y)$ and $\bar P_{\rm diff}(Y,Y_0)$ approach the `black--disk'
fixed points $\bar T = \bar P_{\rm diff} =1$, in agreement with
the corresponding properties of the more general formul\ae\
(\ref{Pdiff}) and (\ref{Aforward}).

But although they do respect the unitarity bounds, Eqs.~(\ref{BK})
and (\ref{KL}) cannot be used in a study of the high--energy
limit, because of the mean field approximation (\ref{FACT})
inherent in their derivation and which fails at high energy
\cite{IM032,MS04}. To properly include the relevant fluctuations,
one must replace the BK equation with the hierarchy of Pomeron
loop equations \cite{IT04,IT05} for the $N$--dipole amplitudes
$\langle S(1)S(2)\cdots S(N)\rangle_Y$. Once these amplitudes are
thus computed, they can be used to evaluate the diffractive
probability according to Eq.~(\ref{Pdiff}), which replaces the
solution to the KL equation for sufficiently high energy. Although
the general solution to the Pomeron loop equations is not known,
the correspondence between high--energy QCD and problems in
statistical physics \cite{MP03,IMM04} has allowed one to deduce
valuable information about the behaviour of the dipole amplitudes
in the high--energy limit \cite{MS04,IMM04,IT04}. This will be
explained in the next section.

\section{Dipole amplitudes at high energy: Fluctuations \& Diffusive scaling}
\setcounter{equation}{0}\label{SECT_PL}

The essential ingredient required by a calculation of the various
scattering probabilities introduced in the previous section are
the $N$--dipole amplitudes $\langle T^{(N)} \rangle_{Y}$ which
describe the scattering between the target and a system of
dipoles. In this section, we shall describe the calculation of
these amplitudes in the high--energy regime where the MFA breaks
down, because of the strong influence of {\em gluon--number
fluctuations} \cite{MS04,IMM04,IT04}.

Most of the results to be presented below have already appeared in
Refs. \cite{MS04,IMM04,IT04}, but since these are recent
developments and, moreover, are of utmost importance for the
present analysis, our respective discussion here will be quite
complete. In doing so, we shall also clarify some points which
have not been addressed in the previous studies, like the
borderline between the intermediate--energy regime, where the BK
equation is a reasonable approximation and {\em geometric} scaling
applies, and the high--energy regime, where the evolution is
dominated by fluctuations leading to {\em diffusive} scaling.

The subsequent picture will be given in a frame in which all of
the energy is carried by the hadronic target, so the projectile is
a bare dipole, or a set of few such dipoles. To compute a
target--averaged amplitude like $\langle T^{(N)} \rangle_{Y}$, one
needs to \texttt{(i)} evaluate the scattering amplitude
$T_Y(\x,\y)$ for a single dipole and in a {\em single event}
(meaning, for a given configuration of the gluon fields in the
target, as generated through a single evolution from $y=0$ up to
$y=Y$) and \texttt{(ii)} average the result over the {\em ensemble
of events} (that is, over all the possible target evolutions from
$y=0$ to $y=Y$). It turns out that, for sufficiently large values
of $Y$, both these operations become relatively simple and their
results are {\em universal} (i.e., independent of the initial
conditions at low energy), as they are fully determined by the
quantum evolution with $Y$ \cite{MS04,IMM04,IT04}. In fact, this
universality is even stronger, in the sense that the details of
the evolution matter for a couple of parameters which enter the
final results (namely,  the coefficients $\lambda$ and $D_{\rm
fr}$ which appeared in the Introduction), but not for the
functional form of the average amplitudes \cite{IT04}.

\subsection{The event--by--event dipole amplitude}

From now on, we shall neglect any non--trivial dynamics in the
impact parameter space, that is, we shall assume that the
evolution with increasing energy is quasi--local in $b$ and we
shall often omit the $b$--argument from the amplitudes. This is a
reasonable approximation so long as one is interested (as we are
here) in the high--energy limit of the DIS cross--sections at
fixed impact parameter. Then, the dipole amplitude in a single
event can be written as
 \be\label{Tev1}
 T_Y(\x,\y)  \,\equiv  \, T_Y(r) \,\equiv  \, T_Y(\rho)\,\ee
 %  \,\simeq T(\rho-\rho_s(Y)) \,\ee
where $r=|\x-\y|$ is the dipole size and $\rho$ represents $r$ in
logarithmic units, like in Fig. \ref{Diag} : $\rho\equiv \ln
(1/r^2 Q_0^2)$, with $Q_0$ a scale of reference introduced by the
initial conditions at low energy. Note that large $\rho$
corresponds to small dipole sizes, or to large transverse momenta
($\rho\sim \ln k_\perp^2$) after a Fourier transform.

For sufficiently high energy, the amplitude takes the form of a
{\em traveling wave} \cite{MP03} :
 \be\label{Tev0}
 T_Y(\rho) \,\simeq T(\rho-\rho_s(Y)) \,.\ee
This is a front which interpolates between a strong scattering
regime at $\rho <\rho_s(Y)$, where the unitarity bound $T=1$ has
been saturated, and a weak scattering regime at $\rho
>\rho_s(Y)$, where $T_Y\ll 1$, and which
propagates towards larger values of $\rho$ when increasing $Y$.
Here, $\rho_s(Y)$ is the position of the front --- conventionally
defined as the dipole size $\rho\,$ for which $T(\rho)=1/2$ ---
and it increases linearly with $Y$:
$\rho_s(Y)\simeq\lambda\bar\alpha_s Y$. The position of the front
also serves as a definition for the {\em target saturation
momentum} :
 \be
 \rho_s(Y)\,\equiv \,\ln \big(Q_s^2(Y)/Q_0^2\big)\ \Longrightarrow \
 Q_s^2(Y)\,\simeq\,Q_0^2\,{\rm e}^{\lambda \bar\alpha_s Y}\,. \ee
Accordingly,  $\lambda$ will be alternatively referred to as the
{\em saturation exponent}, or the {\em front velocity}. The BK
equation (\ref{BK}) predicts $\lambda= \lambda_0 \approx 4.88$
\cite{SCALING,MT02}, but this result is significantly lowered by
gluon number fluctuations \cite{MS04,IMM04} (see below) and also
by the next--to--leading order corrections in perturbative QCD
\cite{DT02}.

We shall also need later some analytic control over the {\em
shape} of an individual front. As indicated in Eq.~(\ref{Tev0}),
this depends only upon the difference $\rho-\rho_s(Y)$, meaning
that the front propagates without distortion: in a co--moving
frame, the shape of the front is independent of $Y$. This property
has been first noticed within the framework of the mean field
approximation, that is, for the solution to the BK equation
\cite{AB01,Motyka,LL01,SCALING,MT02,MP03,RW03,MS05}, but for the
individual fronts this remains true also in the presence of
fluctuations \cite{BD,IMM04}. Since $\rho-\rho_s= \ln (1/r^2
Q_s^2)$, we deduce that the dipole amplitude $T_Y(r)$ depends upon
its two kinematical variables $r$ and $Y$ only trough the
dimensionless product $r^2 Q_s^2(Y)$
--- the property usually referred to as {\em geometric scaling}
\cite{geometric,SCALING,MT02,DT02,MP03}. This property has far
reaching physical consequences, as it provides a natural
explanation \cite{SCALING} for a new scaling behaviour observed in
the small--$x$ data at HERA \cite{geometric} and plays also an
essential role in understanding the particle production in
deuteron--gold collisions at RHIC \cite{KLM02,IIT04}. Note,
however, that the quantity which matters for the calculation of
physical observables is the {\em target--averaged} dipole
amplitude $ \lan T(\rho) \ran_Y$ and, as we shall later explain,
the geometric scaling property characteristic of a single event
does not necessarily translate to the average quantities, because
of {\em fluctuations} \cite{MS04,IMM04,IT04}.

The fluctuations are inherent in the evolution with increasing
energy, because the instantaneous configuration of the gluons in
the target (as probed by the external dipole) can abruptly change
from one rapidity step to another via the emission of new gluons.
At high energy, the fluctuations are relatively important only in
the dilute regime at large transverse momenta ($k_\perp \gg Q_s$),
where the gluon occupation numbers are small. Since the
high--$k_\perp$ gluons scatter preponderantly with small dipoles,
of size $r\sim 1/k_\perp$, we deduce that the fluctuations should
significantly affect only the tail of the dipole amplitude
$T_Y(\rho)$ at large values of $\rho$. Simple physical
considerations \cite{BD,IMM04} (see also below) show that the
fluctuations have a relatively small effect on the {\em overall
shape} of an individual front --- they modify only the foremost
part of the front, where $T$ is very small: $T\sim \alpha_s^2$
---, but they strongly influence the {\em dynamics} of the front,
thus considerably reducing its velocity $\lambda$ with respect to
the mean field prediction $\lambda_0$. This behaviour, which is
confirmed by numerical simulations within both QCD and statistical
physics (see, e.g., \cite{BD,Moro042,GS05,EGBM05}), reflects the
{\em pulled} nature of the saturation front: the propagation is
entirely driven by the dynamics in the tail of the front at
$\rho\gg \rho_s(Y)$, and thus is very sensitive to even small
changes in this tail due to fluctuations.

Returning to a discussion of the shape of the front, it is
important to stress here (in anticipation of Sect.
\ref{SECT_DISP}) that, in order to compute the {\em average}
amplitudes at high energy, one needs only a very limited
information about the shape of the individual fronts: it is in
fact enough to know that $T_Y(\rho)=1$ for $\rho <\rho_s(Y)$
\cite{IT04}. But precisely in order to develop the corresponding
argument in Sect. \ref{SECT_DISP}, and also in view of numerical
tests in the forthcoming sections, we shall need to know a little
bit more about the function $T_Y(\rho)$. Specifically, the
following piecewise approximation, which can be inferred from the
relevant literature \cite{BD,IMM04,BDMM}, will be sufficient for
our purposes:
 \begin{equation}\label{Tevent}
    T(z)=
    \begin{cases}
        \displaystyle{1} &
        \text{ for\,  $z < 0$}
        \\*[0.2cm]
        \displaystyle{A\,
        {\rm e}^{-\gamma_0 z}} &
        \text{ for\,  $1 < z < L$}
        \\*[0.2cm]
        \displaystyle{B\,
        {\rm e}^{- z}} &
        \text{ for\,  $L \ll z $}.
    \end{cases}
\end{equation}
Here, $z\equiv \rho - \rho_s$ is the scaling variable, $A$ and $B$
are undetermined normalization factors, $\gamma_0\approx 0.63$,
and $L$ is the `width of the front', namely, the distance $\rho -
\rho_s$ over which the amplitude falls off from its saturation
value $T=1$ to a value of order $\alpha_s^2$, where the
fluctuations become important. This condition immediately yields:
 \be\label{L}
  L\,\simeq\,\frac{1}{\gamma_0}\,\ln\,\frac{1}{\alpha_s^2}
  \,+ \,O(1)\,.
 \ee
Note that the physical regions in which the solution is
approximately given have no overlap with each other, so, not
surprisingly, the expressions shown in Eq.~(\ref{Tevent}) do not
provide a continuous interpolation for $T(z)$. The first line in
Eq.~(\ref{Tevent}) is, clearly, the {\em saturation region\,}; the
second line describes the {\em front} of the traveling wave, with
the characteristic `anomalous dimension' $1-\gamma_0\approx 0.37$
which is a hallmark of the BFKL evolution in the presence of
saturation \cite{SCALING,MT02}; finally, the third line represents
the perturbative QCD tail, where the amplitude exhibits {\em color
transparency} ($T(r)\sim r^2$), and which is generated via the
large--$\rho$ behaviour of the BFKL kernel. More precisely, within
the BFKL evolution, the color transparency is amended by quantum
corrections in the ``double--logarithmic approximation'' (DLA).
These are not explicitly shown here, since the contribution of
this large--$\rho$ tail to physical quantities is anyway
negligible in the high--energy regime of interest. It is however
important to notice that such DLA effects violate geometric
scaling, which is therefore restricted to the region at $z\simle
L$.

The single--event amplitude in Eq.~(\ref{Tevent}) looks quite
similar to the solution to the BK equation (see, e.g., Ref.
\cite{IIT04}), but it differs from the latter in two essential
points:

\texttt{a)} The front is {\em compact} in Eq.~(\ref{Tevent}),
i.e., it has a finite width $L$ which is independent of $Y$. By
contrast, the front of the BK solution extends up to distances
$\rho - \rho_s\sim \sqrt{Y}$, and thus its width increases with
$Y$. In the context of the BK equation, this front is often
referred to as the {\em geometric scaling window}
\cite{SCALING,MT02}. We see that, as a consequence of
fluctuations, the width of the geometric scaling window for a {\em
single} event (and for sufficiently high energy) is considerably
reduced with respect to the mean field approximation.

\texttt{b)} The front velocity $\lambda$ which enters
Eq.~(\ref{Tevent}) via $\rho_s$ is considerably smaller than the
corresponding BK velocity $\lambda_0$. This is seen numerically
for generic values of $\alpha_s$, and is confirmed by an analytic
calculation valid in the weak coupling limit $\alpha_s^2\to 0$,
which shows that in this limit $\lambda$ converges towards
$\lambda_0$, but {\em only very slowly} \cite{BD,MS04,IMM04} :
 \begin{equation}
\lambda\,\simeq\,\lambda_0\,-\, \frac{\cal
C}{\ln^2(1/\alpha_s^2)}\qquad {\rm when}\quad \alpha_s^2\ll 1\,.
\label{satscal}
\end{equation}
Here, ${\cal C}$ is a known number, which is determined by the
linear, BFKL, dynamics and turns out to be quite huge : ${\cal
C}=\pi^2 \gamma_0 \chi''(\gamma_0)\approx 150$. (Here,
$\chi(\gamma)$ is the Mellin transform of the BFKL kernel
\cite{MS04,IMM04}. Note also that $\lambda_0=
\chi(\gamma_0)/\gamma_0 \approx 4.88$.) The tendency of $\lambda$
to decrease when increasing $\alpha_s^2$ is also confirmed by a
study of the `strong noise limit', which shows that in that limit
$\lambda$ vanishes as a power of $1/\alpha_s^2$ \cite{MPS05}.

More precisely, the above picture applies only for high enough
rapidities, larger than the {\em formation time} of the front,
estimated as \cite{IMM04,Saar} :
  \be\label{Dtau}
 \bar\alpha_s \, Y_{\rm form}
 \,\sim\,
 \frac{\ln^2 (1/\alpha_s^2)}{2 \chi''(\gamma_0)\,\gamma_0^2}\,.\ee
This is the rapidity evolution necessary for the front to reach
its final form in Eq.~(\ref{Tevent}), and for the velocity to
reach its asymptotic value in Eq.~(\ref{satscal}), starting with
some generic initial condition at $Y=0$. For $Y < Y_{\rm form}$,
the width of the front increases with $Y$ as $\rho - \rho_s\sim
\sqrt{Y}$, via BFKL diffusion, while for $ Y > Y_{\rm form}$, this
width gets stuck at its maximal value, equal to $L$. As indicated
by the  `geometric scaling' region in the diagram in Fig.
\ref{Diag}, a similar behaviour is shown also by the {\em average}
amplitude $ \lan T(\rho) \ran_Y$, but only up to some maximal
rapidity $Y \sim Y_{\rm DS}$, beyond which the effects of
fluctuations become overwhelming (see the discussion in the next
subsection).

Let us conclude this presentation with a few physical
considerations on the role of fluctuations \cite{IMM04}. The
previous arguments exhibit the special role played by the
parameter $\alpha_s^2$ in the study of fluctuations (in
particular, the mean field limit is obtained as $\alpha_s^2\to
0$), and this deserves a comment. In the dilute regime where the
fluctuations are important, and for large $N_c$, the target itself
can be described as a collection of dipoles, and then the
amplitude $T(r)$ describes the scattering between the external
dipole $r$ and the target dipoles. The dipole--dipole amplitude
$T(r,r')$ is of order $\alpha_s^2$ and is peaked at sizes $r'\sim
r$. Hence, to a good approximation, $T(r)\sim \alpha_s^2 f(r)$,
where $f(r)$ is the {\em dipole occupation number} in the target,
that is, the number of dipoles with size $r$ (per unit of $\ln
1/r^2$) within an area $r^2$ around the impact parameter of the
external dipole. In a given event, $f$ is discrete
($f=0,1,2,\dots$), so the scattering amplitude takes on only
discrete values, which are multiples of $\alpha_s^2$. In
particular, the minimal non--zero value for $T$ is $\alpha_s^2$,
showing that the tail of the front must abruptly end when $T$
becomes of $O(\alpha_s^2)$. This explains the compact nature of
the front.

Furthermore, the particle number fluctuations follow a normal
distribution: $\delta f\sim \sqrt{f}$, showing that $\delta T \sim
\alpha_s^2\sqrt{f} \sim \sqrt{\alpha_s^2 T}$. Thus, the mean field
approximation becomes reliable, in the sense that $\delta T \ll
T$, as soon as $T\gg \alpha_s^2$. In particular, this becomes a
good approximation everywhere (i.e., for any $r$) in the limit
$\alpha_s^2\to 0$.

\subsection{Front dispersion through fluctuations}
\label{SECT_DISP}

Let us consider now the average over the target wavefunction,
which in the present context amounts to an average over the {\em
statistical ensemble of fronts} : the fronts associated to all the
possible evolutions of the target over a rapidity interval $Y$. It
turns out that at high energy --- sufficiently high for the fronts
to reach their canonical form, cf. Eq.~(\ref{Dtau}) --- this
averaging is quite simple: Since all the fronts in the ensemble
have the same shape, they can differ from each other only by a
translation. That is, the only random variable in the problem is
the position $\rho_s$ of the front, which can be argued to be
distributed according to the following, Gaussian, probability
density \cite{BD,IMM04,BDMM}
\begin{equation}\label{probdens}
    P_Y(\rho_s) =
    \frac{1}{\sqrt{\pi}\sigma}\,
    \exp \left[
    -\frac{\left( \rho_s - \langle \rho_s \rangle \right)^2}{\sigma^2}
    \right].
\end{equation}
In this equation, $\langle \rho_s \rangle\simeq
\lambda\bar\alpha_s Y$ is the average position of the front and
increases with $Y$ with the velocity shown in Eq.~(\ref{satscal}),
since this is the common average velocity for all the fronts in
the ensemble. Furthermore, $\sigma^2/2\equiv \lan\rho_s^2\ran -
\langle\rho_s\rangle^2$ is the {\em front dispersion}, which rises
linearly with $Y$, $\sigma^2(Y) \simeq D_{\rm fr}\bar\alpha_s Y$,
since the stochastic process is a random walk around the average
front\footnote{Note that our present normalizations for the
coefficients $\lambda$ and $ D_{\rm fr}$ differ by a factor
$\bar\alpha_s$ from those used in the Introduction.}. It has been
first suggested by numerical simulations \cite{BD}, and very
recently confirmed through analytic arguments \cite{BDMM} that, in
the weak coupling limit $\alpha_s^2\to 0$, the {\it front
diffusion coefficient} $D_{\rm fr}$ scales as
 \be\label{Dfr} D_{\rm
fr}\,\simeq\,\frac{\cal
 D}{\ln^3(1/\alpha_s^2)} \qquad {\rm when}\quad \alpha_s^2\,\ll\, 1\,.\ee
This vanishes, as expected, in the mean field limit $\alpha_s^2\to
0$, but only very slowly. The coefficient ${\cal D}$ has been
explicitly computed in Ref. \cite{BDMM}. (Once again, this is
determined by the linear dynamics in the tail.)

Then the average amplitude $\lan T \ran_Y$ is determined by
\begin{equation}\label{Tavedef}
    \lan T(\rho) \ran_Y =
    \int \limits_{-\infty}^{\infty}
    \rmd\rho_s\, P_Y(\rho_s)\, T(\rho -\rho_s),
\end{equation}
with $T(\rho -\rho_s)$ the single--event front in
Eq.~(\ref{Tevent}).  Higher--point correlations can be computed
similarly. Note that the ensuing average amplitudes will naturally
depend upon the difference $z\equiv \rho - \lan \rho_s \ran $, but
they will also show additional dependencies upon $Y$, via the
front dispersion $\sigma$. That is, unlike the individual fronts,
the average amplitudes will generally {\em not} show geometric
scaling \cite{MS04,IMM04}. In fact, as we shall shortly discover,
at sufficiently high energies the geometric scaling is replaced by
a new, {\em diffusive}, scaling.

Given the Gaussian nature of the probability distribution
(\ref{probdens}), it is straightforward to compute the average in
Eq.~(\ref{Tavedef}) for the piecewise--defined front in
Eq.~(\ref{Tevent}), and we shall do so indeed in the more detailed
studies in the next subsections. Here, however, we would like to
point out (following an original analysis in Ref. \cite{IT04})
that, for sufficiently high energy, the dominant contributions to
the average amplitudes $\langle T^{(N)} \rangle_{Y}$ with any $N$
are given by the saturation region in Eq.~(\ref{Tevent}), and thus
they are all expressed through a universal function.

Let us start by better specifying what we mean by ``sufficiently
high energy'' : As anticipated in the Introduction (cf. Fig.
\ref{Diag}), this is the regime in which the target expectation
values are dominated by dense fluctuations, which in turn requires
relatively high values of $Y$ and not too large values of $\rho$
(but $\rho$ can still be much larger than the average saturation
scale $\lan \rho_s \ran$; see below). More precisely, we shall
demonstrate here that within the {\em high--energy regime} defined
as
 \be\label{HER}
 \mbox{High--energy regime}\ \  \ \ \ \Longleftrightarrow \ \ \ \ \
 \sigma \gg 1/\gamma_0\ \ \   \mbox{and} \
 \ \ \rho - \lan \rho_s \ran \ll
 \gamma_0\sigma^2\,,\ee
the dominant contributions to the average dipole amplitudes can be
simply computed with a $\Theta$--function saturation front:
  \be\label{Ttheta} T_\theta(\rho)\,\equiv\,\Theta(\rho_s-\rho)\,,\ee
which in turn implies \cite{IT04}
 \be\label{Thighsigma}
    \lan T(\rho) \ran_Y \,=\,
    \frac{1}{2}\, {\rm Erfc}\left(\frac{z}{\sigma} \right)\,,
 \ee
and, more generally,
 \be\label{ThighsigmaN}
    \lan T(\rho_1) T(\rho_2)\cdots T(\rho_N)\ran_Y \,=\,
    \frac{1}{2}\, {\rm Erfc}\left(\frac{z_>}{\sigma} \right)\,,
 \ee
where $z= \rho - \lan \rho_s \ran $ and $z_>= \rho_> - \lan \rho_s
\ran $, with $\rho_> =$ the largest among the $\rho_i$'s. In these
equations, ${\rm Erfc(x)}$ is the complementary error function,
 \be \label{erfcdef}
    {\rm Erfc}(x)\,\equiv\,\frac{2}{\sqrt{\pi}}
    \int\limits_x^\infty {\rm
    d}t\,{\rm e}^{-t^2}\,,\ee
which arises here as the integral of the probability density
(\ref{probdens}) over all values of $\rho_s$ larger than $\rho$
(respectively, $\rho_>$): The r.h.s. of Eq.~(\ref{Thighsigma}) is
simply the probability to find a front which is at saturation on
the resolution scale $\rho$.

Eqs.~(\ref{Thighsigma})--(\ref{ThighsigmaN}) show that the average
scattering amplitudes at high energy are dominated by those fronts
within the statistical ensemble at $Y$ which are at saturation at
the highest resolution scale $\rho_>$ set by the incoming
projectile. This is true, in particular, for relatively large values
of $\rho$, well above $\lan \rho_s \ran $, where {\em on the
average} the scattering is weak, $\lan T(\rho) \ran_Y\ll 1$, yet
this average value is dominated by the relatively rare
configurations (`black spots') which are at saturation for that
(large) value of $\rho$. The subdominant contributions not shown in
Eqs.~(\ref{Thighsigma})--(\ref{ThighsigmaN}) are suppressed by, at
least, one power of\footnote{Such estimates should more properly
read, e.g., $z/\gamma_0\sigma^2$, but here and in what follows we
shall often use the fact that $\gamma_0\approx 0.63$ is a number of
$O(1)$ to simplify the parametric estimates.} $1/\sigma$ and/or
$z/\sigma^2$ (respectively, $z_>/\sigma^2$), and are sensitive to
the detailed shape of the single--event front. A similar sensitivity
holds for the average amplitudes outside the high--energy regime
defined in Eq.~(\ref{HER}).

Note that the universal behaviour exhibited in
Eqs.~(\ref{Thighsigma})--(\ref{ThighsigmaN}) emerges also in the
limit of {\em strong fluctuations}, as recently shown in Ref.
\cite{MPS05}. This observation is consistent with the analysis in
Ref. \cite{IT04} and the present considerations, since the limit
of strong fluctuations is formally obtained by letting
$\sigma\to\infty$ within our formul\ae , in which case the
`high--energy regime' of Eq.~(\ref{HER}) extends everywhere.

To demonstrate the above results, we need to study the influence
of the shape of the front on the average amplitudes with
increasing energy. To that aim, it is sufficient to consider the
following, simplified, profile for an individual event :
 \be\label{Tevent1} T_\gamma(\rho)=\Theta(\rho_s-\rho) + \Theta(\rho-\rho_s)
 \,{\rm e}^{- \gamma(\rho-\rho_s)}\,.\ee
For $\gamma=\gamma_0\approx 0.63$, this simulates the front region
in the second line of  Eq.~(\ref{Tevent}),  while for $\gamma=1$
it rather corresponds to the `color transparency' region in the
third line there. By using Eqs.~(\ref{Tevent1}) and
(\ref{probdens}) it is straightforward to deduce
 \begin{equation}\label{Tave}
    \lan T (z)\ran_Y=
    \frac{1}{2}\, {\rm Erfc}\left(\frac{z}{\sigma} \right)
    +\frac{1}{2}
    \exp\left( \frac{\gamma^2 \sigma^2}{4} - \gamma z\right)
    \left[ 2 - {\rm Erfc}\left(\frac{z}{\sigma} -
    \frac{\gamma \sigma}{2} \right)\right],
\end{equation}
with $z= \rho - \lan \rho_s \ran $. The first term is the same as
in Eq.~(\ref{Thighsigma}) and comes from the saturation piece of
$T_\gamma$, while the second term comes from the exponential piece
of Eq.~(\ref{Tevent1}).

The following limiting behaviours of the function ${\rm Erfc}(x)$
will be useful in what follows:
 \begin{equation}\label{erfc}
    {\rm Erfc}(x)=
    \begin{cases}
        \displaystyle{2-\frac{\exp(-x^2)}{\sqrt{\pi}\,|x|}} &
        \text{ for\,  $x \ll -1$}
        \\*[0.1cm]
        \displaystyle{1} &
        \text{ for\,  $x=0$}
        \\*[0.1cm]
        \displaystyle{\frac{\exp(-x^2)}{\sqrt{\pi}x}} &
        \text{ for\,  $x \gg 1$}.
    \end{cases}
 \end{equation}
\texttt{i)} Consider first the (relatively) low energy situation
at $\sigma \ll 1/\gamma$. By using Eqs.~(\ref{Tave})--(\ref{erfc})
one finds that the average amplitude retains its single event
profile, except in the short interval $|z|\simle \sigma$ where it
gets smoothed:
 \begin{equation}\label{Tavelow}
    \lan T (z)\ran_Y=
    \begin{cases}
        \displaystyle{1} &
        \text{ for\,  $z \ll - \sigma$,}
        \\*[0.2cm]
        \displaystyle{
        {\rm e}^{-\gamma z}} &
        \text{ for\,  $\sigma \ll z $},
    \end{cases}
     \qquad\quad \mbox{for}\quad \sigma \ll 1/\gamma\,.
\end{equation}
In this regime, geometric scaling is manifest at the level of the
average amplitudes. The situation is similar to the mean field
approximation, except for the facts that the real front is compact
and has a smaller velocity, cf. Eq.~(\ref{satscal}). (The
compactness becomes manifest, of course, only if one starts with
the more realistic profile in Eq.~(\ref{Tevent}).) We thus
conclude
 \be\label{LER}
 \mbox{Negligible dispersion}\ \  \  \ \Longleftrightarrow \ \ \ \
 \sigma\, \ll \,1/\gamma_0\ \  \  \ \Longleftrightarrow \ \ \ \
 \bar\alpha_s Y \, \ll \,\frac{\ln^3(1/\alpha_s^2)}{\gamma_0^2
 {\cal  D}}\,.\ee
The upper limit above, namely $\bar\alpha_s Y_{\rm DS} \sim
\ln^3(1/\alpha_s^2)$, is parametrically larger than the `critical'
rapidity $\bar\alpha_s Y_c \sim \ln(1/\alpha_s^2)$ for the onset
of the unitarity corrections, cf. Eq.~(\ref{Ymaxproj}), and also
than the formation time $\bar\alpha_s Y_{\rm form} \sim
\ln^2(1/\alpha_s^2)$,  cf. Eq.~(\ref{Dtau}), for the individual
fronts (as it should for the present discussion to make sense).

\texttt{ii)} For $\sigma \gg 1/\gamma$, the first term in
Eq.~(\ref{Tave}) dominates everywhere except at extremely large
distances ahead of the front, such that $z\simge \gamma\sigma^2$.
For instance, in the interesting range $\sigma \ll z \ll \gamma
\sigma^2$, where the {\em average} amplitude is small, we obtain
 \begin{equation}\label{TaveHER}
    \lan T (z)\ran_Y\,\simeq\,
    \frac{1}{2\sqrt{\pi}}\, \frac{\sigma}{z}\,
    \left\{1 \,+\, \frac{2z}{\gamma\sigma^2}
    \right\}\,{\rm e}^{-z^2/\sigma^2}%{\exp (-z^2/\sigma^2)}
    \qquad \mbox{for}\quad \sigma \ll z \ll \gamma\sigma^2,
\end{equation}
where the first term inside the curly brackets comes from the
`saturation' piece in the r.h.s. of Eq.~(\ref{Tave}), while the
second term corresponds to the `exponential' piece there, and is
suppressed with respect to the first term by a factor
$z/{\gamma\sigma^2}\ll 1$, as anticipated. One can similarly check
that Eq.~(\ref{ThighsigmaN}) yields indeed the dominant behaviour
for the $N$--dipole amplitude $\langle T^{(N)} \rangle_{Y}$ within
the range specified by Eq.~(\ref{HER}).

On the other hand, it is easy to check that, for much larger
values of $\rho$, such that $z\gg \sigma^2$, the dominant
contribution to the average amplitude comes from the second term
in Eq.~(\ref{Tave}), that is, from the exponential tail in
Eq.~(\ref{Tevent1}). One then finds
 \begin{equation}\label{TaveLRHO}
    \lan T (z)\ran_Y\,\simeq\,%{\rm e}^{-\gamma z}\,
     \exp( - \gamma z)
%   {\rm e}^{\gamma^2 \sigma^2/4}
 \, \exp\left( \frac{\gamma^2 \sigma^2}{4}\right)
    \qquad \mbox{for}\quad  z \gg \sigma^2,
 \end{equation}
where we recognize the same exponential decay with $\rho$ as for a
single front, Eq.~(\ref{Tevent1}).

The piecewise expression of the average amplitude at high energy
reads therefore
 \begin{equation}\label{Tavehigh}
    \lan T (z)\ran_Y=
    \begin{cases}
        \displaystyle{(1/2){\rm Erfc}({z}/{\sigma})} &
        \text{ for\,  $-\infty < z \ll \gamma_0\sigma^2$,}
        \\*[0.4cm]
        \displaystyle{\,\sim\,
        {\rm e}^{- z}} &
        \text{ for\,  $\gamma_0\sigma^2 \ll z $},
    \end{cases}
    \qquad\quad \mbox{for}\quad \sigma \gg 1/\gamma_0,
 \end{equation}
where we have directly considered the more realistic profile in
Eq.~(\ref{Tevent}), so the exponential tail at very large $z$ is
determined by the `color transparency' piece of
Eq.~(\ref{Tevent}).

\begin{figure}[t]
    \centerline{\epsfxsize=14cm\epsfbox{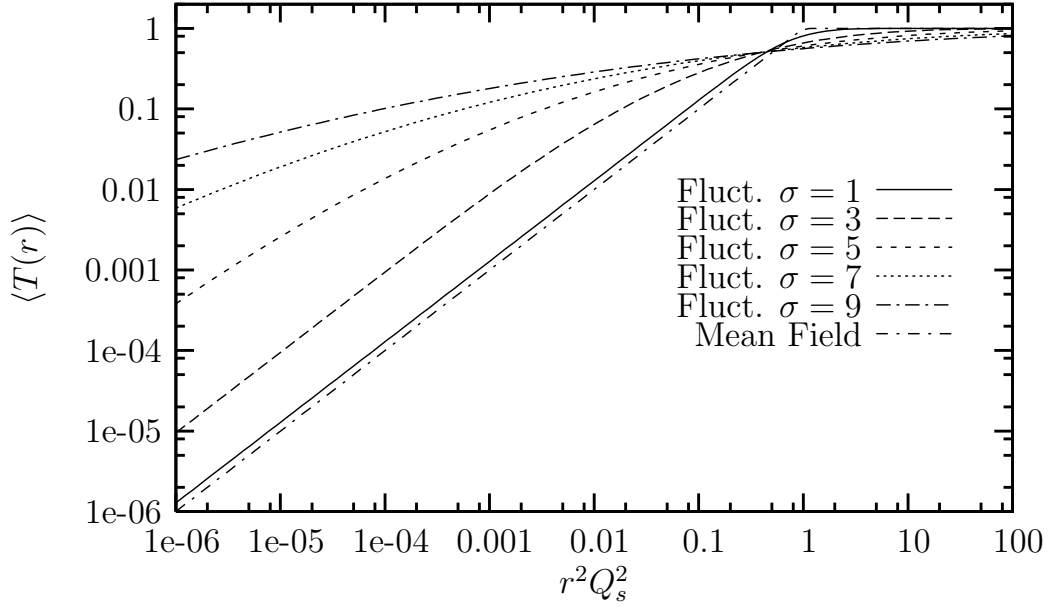}}
    \caption{\sl  The average dipole--hadron forward amplitude of
    Eq.~(\ref{Tave}) (with $\gamma=1$) as a function of
the ``standard'' scaling variable $r^2 \langle Q_s^2 \rangle$ and
for various values of the front dispersion $\sigma$.
    \label{front}}
    \end{figure}

\begin{figure}[t]
    \centerline{\epsfxsize=14cm\epsfbox{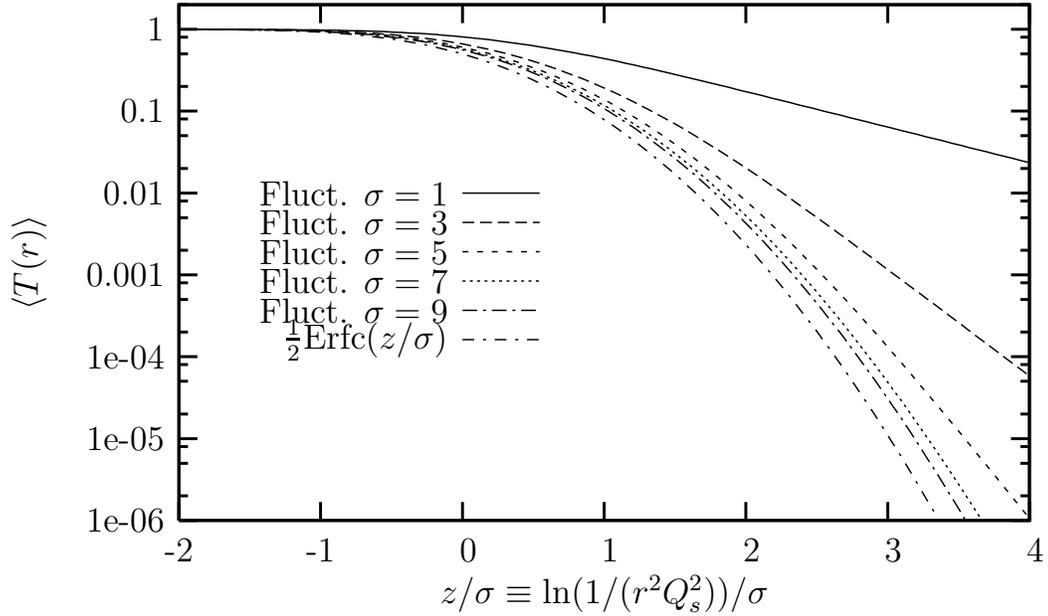}}
    \caption{\sl  The same as in Fig.~10 as a function of the ``diffusive''
scaling variable $z=-\ln(r^2 \langle Q_s^2 \rangle)/\sigma$. As
$\sigma$ increases the amplitude approaches the limiting behavior
in Eq.~(\ref{Thighsigma}).
    \label{DS_front}}
    \end{figure}

For all but very large values of $z$, the average front profile in
Eq.~(\ref{Tavehigh}) has no resemblance at all with the
single--event profile in Eq.~(\ref{Tevent}): There is no trace of
the BFKL  `anomalous dimension' $\gamma_0$, nor of\, `geometric
scaling'. Rather, so long as $z  \ll \gamma_0\sigma^2$, the
average amplitude obeys a new type of scaling --- it scales as a
function of $z /\sigma$
--- which was previously observed in Refs. \cite{IMM04,IT04},
and for which we propose the name of {\em diffusive scaling\,}.
Note that the kinematical region for diffusive scaling coincides
with the high--energy regime of Eq.~(\ref{HER}), and is
represented as the region at $Y >   Y_{\rm DS}$ and $\rho$ at the
left of the curve ``$Q_d^2$'' in Fig. \ref{Diag}. In particular,
within the weak scattering regime at $\sigma \ll z \ll
\gamma_0\sigma^2$, the average amplitude decreases as a Gaussian
in $z=\rho-\lan \rho_s \ran$, and not as an exponential (as it
happens for the single--event front within the geometric scaling
window).

This behaviour is illustrated in Figs. \ref{front} and
\ref{DS_front}, which show the average amplitude in
Eq.~(\ref{Tave}) for $\gamma=1$ and various values of $\sigma$, as
a function of either the `geometric scaling' variable $r^2 \langle
Q_s^2 \rangle$ (in Fig. \ref{front}), or the `diffusive scaling'
variable $z=\ln(1/r^2 \langle Q_s^2 \rangle)/\sigma$ (in Fig.
\ref{DS_front}). The ``mean field'' curve in Fig. \ref{front}
represents the limit $\sigma\to 0$, that is, the single--event
profile in Eq.~(\ref{Tevent1}). This shows geometric scaling, and
would be the actual behaviour in the absence of fluctuations. As
manifest on Figs. \ref{front} and \ref{DS_front}, when increasing
$\sigma$, the average amplitude deviates more and more from
geometric scaling and eventually approaches the limiting profile
in Eq.~(\ref{Thighsigma}) (displayed as the ``$\erfc$'' curve in
Fig. \ref{DS_front}), which shows diffusive scaling.

These considerations show that the mean field approximations (like
the BK equation) completely fail to describe the dipole scattering
at high energy. This is particularly striking when one considers
multi--dipole correlations : in the high--energy regime
(\ref{HER}) these correlations are given by
Eq.~(\ref{ThighsigmaN}), which shows that the mean--field
factorization (\ref{FACT}) is {\em maximally violated}. For
instance, within the range $\sigma \ll z \ll  \sigma^2$, the
average amplitudes are weak, yet strongly correlated :
 \begin{equation}\label{Tave1N}
   \lan T(\rho)\ran \,\simeq\,\lan T^2(\rho)\ran \,\simeq\,
       \lan T^N (\rho)\ran\,\approx\,
    \frac{1}{2\sqrt{\pi}}\, \frac{\sigma}{z}\,
    \,{\rm e}^{-z^2/\sigma^2}
     \qquad \mbox{for} \quad \sigma \ll z \ll \sigma^2\,.
 \end{equation}
Accordingly, the $N$--dipole amplitudes with enter the evolution
equations with Pomeron loops become as important as the
one--dipole amplitude $\lan T(\rho) \ran_Y$ already in the weak
scattering regime, and not only in the region where the unitarity
corrections are important. This is so because the individual
fronts which contribute to these {\em average} amplitudes are
themselves at saturation, and thus they are sensitive to unitarity
corrections {\em in the event--by--event description}.

This dominance of saturated gluon configuration within the
diffusive scaling region has another interesting consequence, to
which we shall refer as the {\em rigidity} of the average
amplitude: This is the property that, within the intermediate weak
scattering regime at $\sigma \ll z \ll \sigma^2$, the average
amplitude increases unusually slow when increasing $Y$ at fixed
dipole size (i.e., at fixed $\rho$). To better appreciate this
property, recall first the rapid, BFKL, increase of the average
amplitude in the weak scattering regime at low and intermediate
energies, say, in the geometric scaling window (cf.
Eqs.~(\ref{TBK})):
 \be\label{DALE}
 \frac{\del \lan T(r)\ran_Y}{\del Y}\ \sim \
 \gamma_0 \lambda \abar\lan T(r)\ran_Y\,\qquad\mbox{when}\qquad
 r^2 \langle Q_s^2\rangle\,\ll\,1\,.\ee
On the other hand, Eq.~(\ref{Thighsigma}) implies
 \be\label{DAHE}
  \frac {\del \lan T(r)\ran_Y}
    {\del Y}&\,=\,&\frac{1}{2}\, \erfc^\prime
    \left(\frac{z}{\sigma}\right)\,
        \left(\frac{1}{\sigma}\frac {\rmd z}{\rmd Y}
        -\frac{z}{\sigma^2}\frac {\rmd \sigma}{\rmd Y}
        \right)\,\nn
       &\,=\,&\abar
       \left(\lambda + \frac{D_{\rm fr}}{2}\frac{z}{\sigma^2}\right)\,
       \frac{1}{\sqrt{\pi}\sigma}\,
       \,
    \exp\left(-\frac{z^2}{\sigma^2}\right)\,\sim\,\lambda \abar \,
     \,\frac{z}{\sigma^2}\lan T(r)\ran_Y\,,
    \ee
where we have used  $z=\rho-\langle \rho_s \rangle$,  $\langle
\rho_s \rangle\simeq \lambda\bar\alpha_s Y$ and $\sigma^2 \simeq
D_{\rm fr}\bar\alpha_s Y$, and the last estimate holds for $\sigma
\ll z \ll \sigma^2$ (compare to Eq.~(\ref{Tave1N})). Clearly, as
compared to the intermediate energy (or mean--field) regime in
Eq.~(\ref{DALE}), the relative variation of the high--energy
amplitude in Eq.~(\ref{DAHE}) is smaller by a factor
${z}/{\sigma^2}\ll 1$. This is so since the individual fronts
which contribute to $\lan T(r)\ran_Y$ in this regime are all at
saturation, so by themselves they cannot increase when further
increasing $Y$; the comparatively slow growth visible  in
Eq.~(\ref{DAHE}) is rather due to the progression of the average
front towards larger values of $\rho$.

\section{Fluctuation effects on deep inelastic scattering
at high energy} \setcounter{equation}{0}\label{SECT_FLUCT}

In this section, we shall consider the high--energy limit of the
inclusive and diffractive cross--sections introduced in Sect.
\ref{SECT_DIPOLE}, with the purpose of demonstrating the
fundamental change of behaviour introduced in this regime by the
gluon--number fluctuations. On this occasion, we shall present the
first calculation of the DIS cross--sections including the effects
of the Pomeron loops, and we shall compare our results with the
corresponding predictions of the mean field approximation (the BK
equation).

\subsection{The general set--up}
\label{SECT_GEN}

To perform this analysis, we need to complete our previous
factorization formul\ae\ with the electromagnetic vertex
describing the dissociation of the virtual photon into a $q \bar
q$ pair. Within the present, high--energy, approximations, this
vertex factorizes out and the DIS cross--sections for the
(inclusive and diffractive) $\gamma^* h$ scattering can be
computed as \cite{AM90,NZ91}
 \be\label{sigmatot}
 \frac{\rmd\sigma^{\gamma}_{\rm tot}}
 {\rmd^2 {b}}\,(Y,Q^2)
 =\int_0^1 \rmd v\int \rmd^2 {\bm r}\,\sum_{\alpha=L,T}
 \vert \psi^{\gamma}_{\alpha}(v, r; Q)\vert^2 \,
 P_{\rm tot}({\bm b}, {\bm r}; Y), \ee
and, respectively (recall that $Y_{\rm gap}$ denotes the {\em
minimal} rapidity gap),
 \be\label{sigmadiff}
 \frac{\rmd\sigma^{\gamma}_{\rm diff}}
 {\rmd^2 {b}}\,(Y,Y_{\rm gap},Q^2)
 =\int_0^1 \rmd v \int \rmd^2 {\bm r}\,\sum_{\alpha=L,T}
 \vert \psi^{\gamma}_{\alpha}(v, r; Q)\vert^2 \,
 P_{\rm diff}({\bm b}, {\bm r}; Y, Y_{\rm gap}). \ee
In these equations, $\vert \psi^{\gamma}_{T/L}\vert^2$ are
probability densities for the $q \bar q$ dissociation of a virtual
photon with either transverse ($T$) or longitudinal ($L$)
polarization (these can be found in the literature and are
displayed, for convenience, in Appendix A), ${\bm r}= \x-\y$ and
${\bm b}= (\x+\y)/2$ are the transverse size and the impact
parameter of the $q \bar q$ pair, $r\equiv |{\bm r}|$, $v$ is the
fraction of the photon longitudinal momentum taken away by the
quark, and $P_{\rm tot}$ and $P_{\rm diff}$ are the total and,
respectively, diffractive probabilities for onium--hadron
scattering, as introduced in Eqs.~(\ref{Pdiff}) and (\ref{Ptot}).

The above formul\ae\ are {\em a priori} frame--independent, but in
order to be able to use our expression (\ref{Pdiff}) for $P_{\rm
diff}$, we need to evaluate the diffractive cross--section
(\ref{sigmadiff}) in the frame where $Y_0=Y_{\rm gap}$. On the
other hand, the inclusive cross--section is most simply evaluated
in the frame where $Y_0=Y$ and the projectile is an  elementary
$q\bar q$ pair:
  \be \label{Ptotfin}
 \frac{\rmd\sigma^{\gamma}_{\rm tot}}
 {\rmd^2 {b}}\,
 =\int_0^1 \rmd v\int \rmd^2 {\bm r}\,\sum_{\alpha=L,T}
 \vert \psi^{\gamma}_{\alpha}(v, r; Q)\vert^2 \ 2\,{\rm Re}\,
 \lan T({\bm b}, {\bm r}) \ran_Y\,.
 \ee
The factorization formula (\ref{sigmadiff}) holds within the same
approximations as previously used in the calculation of $P_{\rm
diff}$, namely, the leading logarithmic approximations with
respect to both $\ln (1/x_\mathbb P)\equiv Y_{\rm gap}$  and $\ln
(1/\beta)\equiv Y- Y_{\rm gap}$. Under these assumptions, the
diffractive cross--section depends upon $x_\mathbb P$ and $\beta$
only via the high--energy evolutions of the target and the
projectile, respectively. In the high--energy regime of interest,
these approximations are certainly correct in so far as the
$x_\mathbb P$--dependence is concerned (since $x_\mathbb P$ is
truly small), but they fail to describe the actual
$\beta$--dependence in the vicinity of $\beta=1$ (i.e., in the
case of a relatively small mass $M_X^2\simle Q^2$ for the
diffractive system). In fact, within the present approximations,
the diffractive cross--section for $\beta\simeq 1$ reduces to its
`elastic' piece in Eqs.~(\ref{sigmadiffint}) and (\ref{sigmael}),
which is independent of $\beta$. However, for $\beta\simeq 1$, the
projectile is simply a $q\bar q$ pair, and for that pair the
generalization of Eq.~(\ref{sigmadiff}) which provides the correct
$\beta$--dependence is known \cite{BP96,W97} and will be exhibited
in Appendix B (see Eq.~\eqref{sigmaqq1}). It is therefore quite
straightforward to improve the subsequent analysis in the vicinity
of $\beta=1$, whenever this is needed. However, since our main
interest here is rather in the high--energy limit $x_\mathbb P\to
0$ at fixed $\beta$, we find it more convenient to focus on
generic values of $\beta\ll 1$, for which the relatively simple
factorization formula (\ref{sigmadiff}) applies as written.

By using $P_{\rm diff}= P_{\rm el}+P_{\rm diff}^{\rm\, inel}$, cf.
Eq.~(\ref{Pinel}), it is further possible to separate the
`elastic' and `inelastic' components of the diffractive
cross--section (\ref{sigmadiff}), as shown in
Eq.~(\ref{sigmadiffint}). Although these components cannot be
separately measured in DIS, as explained at the end of Sect. 2.1,
this decomposition is still useful because, as we shall argue
below, {\em the elastic component dominates at high energy}, up to
relatively large values of $Q^2$ (within the region for diffusive
scaling). Namely, we shall find that within the high--energy
region defined by Eq.~(\ref{HER}), the diffractive probability is
predominantly elastic:
 \be\label{Pdiffel}
 P_{\rm diff}(r; Y,Y_{\rm gap})\,\simeq\,P_{\rm el}(r; Y)
 \qquad\mbox{when}\quad
    \sigma \gamma_0 \,\gg\, 1\quad\mbox{and}\quad
 \ln \frac{1}{r^2\langle Q_s^2 \rangle}\,\ll \,\sigma^2.\ee
Since, moreover, the convolutions with the virtual photon
wavefunction in Eqs.~(\ref{sigmatot}) and (\ref{sigmadiff}) are
dominated by dipole sizes $r\sim 1/Q$ within the whole kinematical
region of interest (this will be explicitly checked below), the
dominance of the elastic scattering extends to the diffractive
cross--section, as anticipated :
 \be\label{sdiffel}
 \frac{\rmd\sigma^{\gamma}_{\rm diff}}
 {\rmd^2 {b}}\,(Y,Y_{\rm gap},Q^2)\,\simeq\,
 \frac{\rmd\sigma^{\gamma}_{\rm el}}
 {\rmd^2 {b}}\,(Y,Q^2)
 \qquad\mbox{when}\quad
    \sigma \gamma_0\, \gg\, 1\quad\mbox{and}\quad
 \ln \frac{Q^2}{\langle Q_s^2 \rangle}\,\ll \,\sigma^2.\ee
This in turn implies that, within the present approximations at
high energy, $\sigma_{\rm diff}$ is quasi--independent of the
minimal rapidity gap $Y_{\rm gap}$ and can be simply computed as
the elastic scattering of the elementary $q\bar q$ pair in the
frame where $Y_0=Y$ (cf. Eq.~(\ref{Pel1})):
  \be\label{sigmadiffHE}
 \frac{\rmd\sigma^{\gamma}_{\rm el}}
 {\rmd^2 {b}}\,=\,
 \int_0^1 \rmd v \int \rmd^2 {\bm r}\,\sum_{\alpha=L,T}
 \vert \psi^{\gamma}_{\alpha}(v, r; Q)\vert^2 \ \, %\times
 |\lan T({\bm b}, {\bm r}) \ran_{Y}|^2\,. \ee
In view of this and of Eq.~(\ref{Ptotfin}), we conclude that, for
sufficiently high energy and up to relatively large $Q^2\gg
\langle Q_s^2 \rangle$ (cf. Eq.~(\ref{sdiffel})), the calculation
of both the inclusive and the diffractive cross--sections requires
merely the knowledge of the forward amplitude $\langle T({\bm b},
{\bm r})\rangle_{Y}$ for the $q\bar q$ pair in the frame where the
target carries most of the total energy.

The dominance of the elastic over the inelastic scattering in
diffractive DIS at large $Q^2$ and  relatively small $\beta$ is an
essential consequence of fluctuations
--- in the mean field approximation, the opposite situation holds
! \cite{GBW99,MS03}
---, and it might even seem counterintuitive at a first sight:
At small $\beta$, the projectile contains many gluons, yet
Eq.~(\ref{Pdiffel}) implies that the relative normalization of
these many Fock states remains unchanged after the collision. But
in view of the discussion in Sect. \ref{SECT_PL}, this property is
in fact natural, as we explain now:

Note first that Eq.~(\ref{Pdiffel}) becomes natural in the
saturation regime at $r^2 \langle Q_s^2 \rangle\gg 1$, where $P_{\rm
diff}$ and $P_{\rm el}$ attain their respective unitarity bounds:
$P_{\rm diff}=P_{\rm el}=1$. In that regime, all the Fock states in
the projectile are equally absorbed, since they are {\em completely}
absorbed. Thus, for low $Q^2\simle \langle Q_s^2 \rangle $ at least,
the inelastic diffraction is naturally suppressed. Now, in the
discussion in Sect. \ref{SECT_PL} we have noticed the tendency of
the fluctuations to `push--up the saturation physics', that is, to
extend a behaviour which looks natural in the saturation regime up
to relatively large values of $Q^2$, within a distance  $\ln
(Q^2/\langle Q_s^2 \rangle)\sim \sigma^2$ above the average
saturation momentum which is fixed by the front dispersion.
Physically, this is so since, within the relevant region at large
$Q^2$ (the `diffusive scaling' region), the cross--section is
controlled by `black spots' in the target wavefunction, i.e., by
rare gluon configurations which are already at saturation. Hence, in
any scattering event which contributes significantly to the
diffractive probability (\ref{Pdiff}), the projectile with its all
Fock space components is {\em completely} absorbed --- the
diffraction is purely elastic ! ---, and the reason where $P_{\rm
diff}$ is nevertheless smaller than one is just because the
probability to find such a `dense spot' is relatively small when
$Q^2\gg \langle Q_s^2 \rangle $.

Still from the discussion in  Sect. \ref{SECT_PL}, we expect the
inelastic probability $P_{\rm diff}^{\rm\, inel}$ to be suppressed
with respect to the elastic one $P_{\rm el}$ by powers of
$1/\sigma$ and/or $z/\sigma^2$, and this will be indeed confirmed
by the subsequent calculations. But although suppressed in so far
as the contribution to the cross--section (\ref{sigmadiff}) ---
which, we recall, is `inclusive' with respect to the rapidity gap
(in the sense of including all the gap values from $Y_{\rm gap}$
to $Y$) ---  is concerned, the inelastic diffraction represents
the leading contribution to the {\em differential} cross--section
per unit rapidity gap at sufficiently small $\beta\ll 1$, and
hence cannot be ignored when computing this particular quantity
(even at very high energy). In fact, within the present
approximations, the contribution of the {\em elastic} scattering
to the differential cross--section $\rmd \sigma_{\rm diff}/\rmd
\ln (1/\beta)$ is a $\delta$--function peaked at $\beta=1$. This
contribution gets smeared when going beyond the leading--log
approximation with respect to $\ln (1/\beta)$ (see
Eq.~\eqref{sigmaqq1}), but is in any case localized near
$\beta=1$; hence, for $\beta \ll 1$ we expect most of the
respective contribution to come from the {\em inelastic}
diffraction:
 \be\label{sigmadiff2}
 \frac{\rmd\sigma^{\gamma}_{\rm diff}}
 {\rmd^2 {b} \ \rmd \ln (1/\beta)}\simeq
 \int_0^1 \rmd z \int \rmd^2 {\bm r}\sum_{\alpha}
 \vert \psi^{\gamma}_{\alpha}(z, r; Q)\vert^2 \left(- \frac{\del
 P_{\rm diff}^{\rm\, inel}}{\del Y_{\rm gap}}({\bm b}, {\bm r}; Y,
 Y_{\rm gap})\right)\quad{\rm for}\quad \beta \ll 1,\nn \ee
where in the r.h.s. $Y_{\rm gap}\equiv Y-\ln (1/\beta)$.

To summarize, within the high--energy regime specified in
Eq.~(\ref{sdiffel}), the diffractive cross--section
(\ref{sigmadiff}) is dominated by its elastic component and can be
computed with the $q\bar q$ pair alone (cf.
Eqs.~(\ref{sdiffel})--(\ref{sigmadiffHE})), so like the inclusive
cross--section (cf. Eq.~(\ref{Ptotfin})).  But the calculation of
the inelastic diffraction is still necessary, for two reasons:
\texttt{(i)} to explicitly check the dominance of the elastic over
the inelastic diffraction, and \texttt{(ii)} to compute the
differential diffractive cross--section (\ref{sigmadiff2}) at
small $\beta\ll 1$. Within the present approximations, $P_{\rm
diff}^{\rm\, inel}$ is given by Eqs.~(\ref{Pinel}) and
(\ref{Pdiff}) which, we recall, are valid so long as the
saturation effects in the projectile wavefunction remain
negligible (this requires $\abar \ln (1/\beta) \ll
\ln(1/\alpha_s^2)$, cf. Eq.~(\ref{Ymaxproj})). For any $\beta$
satisfying this condition,  Eq.~(\ref{Pdiff}) can be in principle
computed by using the dipole probabilities $P_N$ obtained from
Monte--Carlo simulations of the dipole picture \cite{Salam95}
together with the dipole scattering amplitudes constructed in
Sect. \ref{SECT_PL} (or, more generally, obtained by numerically
solving the Pomeron loop equations \cite{IT05}).

But in order to illustrate the main points of physics, it is
preferable to use analytic calculations, and this is what we shall
do in what follows. For these calculations to be tractable, we
shall restrict our computation of $P_{\rm diff}^{\rm\, inel}$ to
the simplest non--trivial case --- that where the onium
wavefunction at the time of scattering contains only two
components: a one--dipole state representing the original $q\bar
q$ pair, and a two--dipole state corresponding to a $q\bar q g$
configuration. This is the physical situation in a frame in which
the rapidity $Y-Y_0=\ln (1/\beta)$ of the projectile is relatively
large, so its high--energy evolution cannot be neglected, but not
{\em too} large, so that one can restrict oneself to a single step
in this evolution: the emission of one gluon with longitudinal
momentum fraction $\beta$. This restriction on $\beta$ does not
entail any serious loss of generality in so far as the proof of
Eq.~(\ref{sdiffel}) is concerned --- as previously explained, the
dominance of elastic over inelastic diffraction at high energy
follows from general arguments, and our subsequent calculation is
merely intended to illustrate such arguments and to render them
more quantitative. On the other hand, this restriction would
become of problem for phenomenological studies of diffraction at
very small values of $\beta$; but in that case, it should be
straightforward
--- at least, at conceptual level --- to extend our analysis by
using numerical simulations of the dipole picture within
Eq.~(\ref{Pdiff}). Note that the  $q\bar q$ and the $q\bar q g$
components have been already included in previous studies of DIS
in the presence of saturation
\cite{BP96,GBW99,BGBK,IIM03,W97,BJW99,KM99,Gotsman00,KST00,K01,KW01,MS03,CM04,GBM05,MariaDiff05},
which were however based on the mean field approximation for the
target correlations. The comparison between such previous studies
in the literature and our subsequent results will make it easier
to emphasize the role of the gluon--number fluctuations at high
energy.

To conclude this subsection, let us summarize here, for more
clarity, the strategy that we shall adopt in our analysis: In what
follows, we shall evaluate the cross--sections in
Eqs.~(\ref{Ptotfin}), (\ref{sigmadiffHE}) and (\ref{sigmadiff2})
(the latter, at the level of the $q\bar q g$ Fock state) by using
the high--energy estimates (\ref{Thighsigma})--(\ref{ThighsigmaN})
for the dipole amplitudes. Note that, whereas the functional form
of these amplitudes is {\em universal} (in the sense explained in
the Introduction), the coefficients $\lambda$ and $D_{\rm fr}$
which implicitly enter their expressions (via the
$Y$--dependencies of $\langle\rho_s\rangle$ and $\sigma^2$) are
{\em not}\,: they depend upon the details of the QCD evolution,
and presently they are not well under control, except in the limit
$\alpha_s\to 0$ (cf. Sect. \ref{SECT_DISP}). To avoid specifying
these parameters, we shall perform our subsequent analysis in
terms of the variables $\rho-\langle\rho_s\rangle$ and $\sigma$.
That is, we shall increase the energy by increasing the value of
$\sigma$, and we shall measure all dimensionful quantities, like
$r^2$ or $Q^2$, in units of $\langle Q_s^2 \rangle$. This strategy
is physically meaningful, since the proper way to approach the
``high--energy limit'' is to simultaneously increase $Y$ and
$Q^2$, in such a way that the ratio  $Q^2/\langle Q_s^2 \rangle$
remains within the interesting scaling region in Fig. \ref{Diag}.

To better emphasize the effects of the fluctuations, we shall also
estimate the relevant convolutions with the following
`mean--field' dipole amplitude
 \begin{equation}\label{TGBW}
    \bar T (r,Y)=
    \begin{cases}
        \displaystyle{(r^2 Q^2_s(Y))^\gamma} &
        \text{ for\,  $r \le 1/Q_s(Y)$}
        \\*[0.2cm]
        \displaystyle{\quad
        1} &
        \text{ for\,  $r > 1/Q_s(Y)$},
    \end{cases}
    % \qquad\quad \mbox{for}\quad \sigma \ll 1/\gamma\,.
 \end{equation}
which is representative for either the  Golec-Biernat and
W\"usthoff (GBW) `saturation model' \cite{GBW99} (in that case,
$\gamma=1$), or for the solution to the BK equation in the regime
of geometric scaling (when $\gamma=\gamma_0\approx 0.63$). More
generally, for our present purposes, the amplitude (\ref{TGBW}) is
also representative for the actual situation in QCD at {\em
intermediate} energies, that is, so long as the dispersion of the
fronts remains negligible (cf. Eq.~(\ref{LER})). To study the
transition between this regime and the {\em high--energy} regime
defined by Eq.~(\ref{HER}), where the amplitude takes its
``error--function'' form in Eq.~(\ref{Thighsigma}), we shall
perform numerical simulations with the dipole amplitude in
Eq.~(\ref{Tave}). Indeed, the latter provides a smooth
interpolation between the mean--field amplitude (\ref{TGBW}) at
low energy ($\sigma\to 0$) and the error--function amplitude
(\ref{Thighsigma}) at high energy ($\sigma\to \infty$). In most of
the explicit calculations below, we shall choose $\gamma=1$, for
simplicity.

To obtain order--of--magnitude estimates in what follows, we shall
rely on simplified versions of the convolutions appearing in
Eqs.~(\ref{sigmatot})--(\ref{sigmadiff}), which in addition to
being simpler, have also the merit to clearly separate the
physical origin of the various contributions. The relevant
simplifications exploit the properties of the ``virtual photon
wavefunctions'' $\vert \psi^{\gamma}_{T/L}\vert^2$, as explained
in the Appendix A. The ensuing, simplified, formul\ae\ read  then
 \be \label{sigmaTfin} \int \rmd v \,\rmd^2 {\bm r}\,
 \vert \psi^{\gamma}_{T}(v, r; Q)\vert^2 \, f(r)\,\sim\,
 {\alpha_{\rm em}N_c}\sum_f e_f^2 \left[\,\int\limits_0^{2/Q}
 \frac{\rmd
 r}{r}\ f(r)\, + \,\frac{1}{Q^2}\int\limits_{2/Q}^{\infty}
 \frac{\rmd r}{r^3}\,
 f(r)\,\right],\ee
and, respectively,
 \be\hspace*{-.1cm}
  \label{sigmaLfin} \int \rmd v\, \rmd^2 {\bm r}\,
 \vert \psi^{\gamma}_{L}(v, r; Q)\vert^2 \, f(r)\,\sim\,
{\alpha_{\rm em}N_c}\sum_f e_f^2 \left[{Q^2}\int\limits_0^{2/Q}
\rmd
 r \,r f(r) +\f1{Q^4}\int\limits_{2/Q}^{\infty}\f{\rmd r}{r^5}
 \,f(r)\right],\ee
where in writing the r.h.s.'s we have kept the various parametric
dependencies, but ignored all numerical factors. For each of these
expressions, the first term within the square brackets corresponds
to the {\em symmetric configurations}, for which $v\sim 1/2$ (in
such configurations, the longitudinal momentum of the incoming
photon is ``democratically'' divided among the quark and the
antiquark), and the second one, to the  {\em aligned jet
configurations}, for which $v$ is either close to zero, or close
to one (either the quark, or the antiquark, carries most of the
total longitudinal momentum).

\subsection{The $q \bar q$ component}
 \label{SECT_QQ}

In this subsection, we shall evaluate the DIS cross--sections
which can be computed from the scattering of the  $q\bar q$ alone,
namely, the inclusive cross--section (\ref{Ptotfin}) and the
elastic component of the diffractive cross--section,
Eq.~(\ref{sigmadiffHE}). As anticipated in the previous
discussion, Eq.~(\ref{sigmadiffHE}) is in fact the dominant
contribution to the diffractive cross--section (\ref{sigmadiff})
at high energy. Accordingly, the cross--sections that we shall
compute in this subsection represent our main results in this
paper.

\begin{figure}[t]
    \centerline{\epsfxsize=8cm\epsfbox{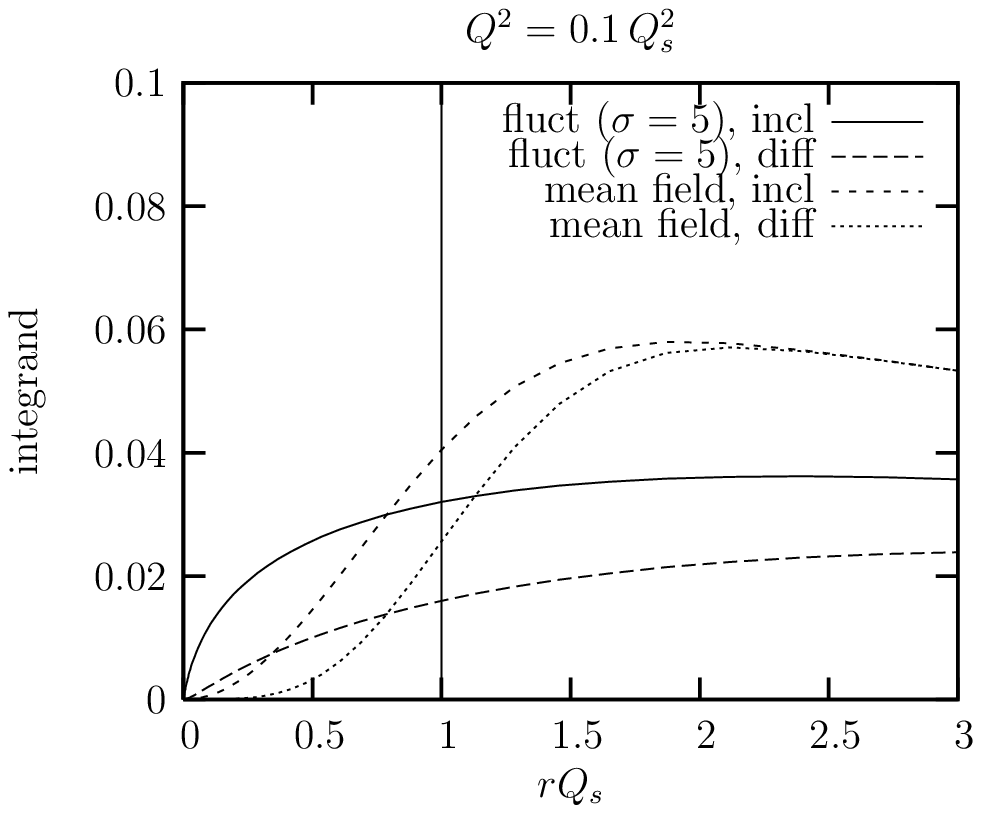}
    \hspace*{.2cm}
    \epsfxsize=8cm\epsfbox{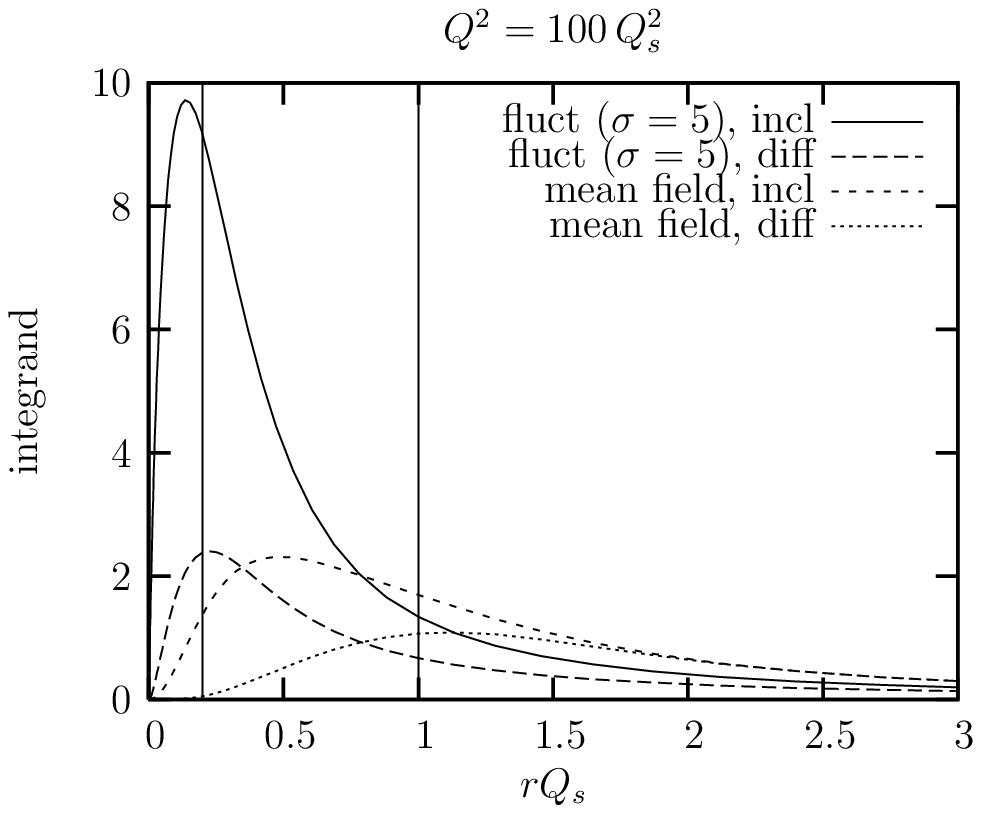}}
    \caption{\sl The $q\bar q$ contributions to the integrands in
    Eqs.~(\ref{Ptotfin}) and (\ref{sigmadiffHE}) as computed with
    two different expressions
    for the dipole amplitude --- the mean--field
    amplitude (\ref{TGBW}) with $\gamma=1$
    and the `fluctuation' amplitude
    (\ref{Thighsigma}) --- are plotted as functions of $rQ_s$ for two
    values of $Q^2$ : $Q^2=0.1 Q_s^2$ (left) and
    $Q^2=100 Q_s^2$ (right).
 \label{integrand}\bigskip}
\end{figure}

To render the subsequent discussion more intuitive, it is
instructive to see the plots of the integrands in
Eqs.~(\ref{Ptotfin}) and (\ref{sigmadiffHE}), as numerically
computed with the two limiting expressions for the dipole
amplitude: the high--energy approximation (\ref{Thighsigma}) (the
`fluctuation' piece in the plots in Fig. \ref{integrand}) and the
mean--field model of Eq.~(\ref{TGBW}) (with $\gamma=1$ for
definiteness).  As manifest on Fig. \ref{integrand}, and it will
be analytically demonstrated in what follows, the strength of the
integration is shifted towards smaller dipoles sizes after
including the effects of fluctuations. This phenomenon becomes
truly spectacular at high $Q^2$, where the integrand corresponding
to `fluctuations' appears to be strongly peaked at $r\sim 2/Q$
(for both inclusive and diffractive cross--sections), while that
for the mean field approximation is rather smoothly distributed at
all sizes $r\simge 2/Q$ (with only mild maxima at $r\sim 2/Q$ for
$\sigma_{\rm tot}$ and, respectively, at $r\sim 1/Q_s$ for
$\sigma_{\rm diff}$).

\subsubsection{The mean--field approximation}
\label{SECT_MFDIS}

Let us first present the corresponding mean--field estimates,
which are well--known (see, e.g., Refs.
\cite{KM99,GBW99,MariaDiff05}), but are interesting here as a term
of comparison with the forthcoming results which will include the
effects of fluctuations. These estimates will be obtained here by
using the dipole amplitude (\ref{TGBW}) together with the
approximations in Eqs.~(\ref{sigmaTfin})--(\ref{sigmaLfin}). To
avoid a proliferation of physical situations, we shall focus on
the case $Q^2 \gg Q_s^2$, where the contrast with the high energy
behaviour turns out to be most important. For that case, we shall
distinguish between the total and the diffractive cross--sections,
and also between transverse and longitudinal polarizations for
$\gamma^*$. To simplify writing, we shall omit the overall factor
${\alpha_{\rm em}N_c}\sum_f e_f^2$, which is common to all the
cross--sections.

\texttt{i)} {\em Total cross--sections}

Choosing $\gamma=1$ in Eq.~(\ref{TGBW}), the {\em transverse}
contribution to Eq.~(\ref{Ptotfin}) is found as
 \be\label{sigmaIAT}
 \frac{\rmd\sigma^{T}_{\rm tot}}
 {\rmd^2 {b}}\ \sim\
 \int\limits_0^{2/Q}
 \frac{\rmd
 r}{r}\, r^2 Q^2_s\, + \,\frac{1}{Q^2}\int\limits_{2/Q}^{1/Q_s}
 \frac{\rmd r}{r^3}\,r^2 Q^2_s\,
 + \,\frac{1}{Q^2}\int\limits_{1/Q_s}^\infty
 \frac{\rmd r}{r^3}
 %\,\nn &\,\sim\,&
 \ \, \sim\ \, \frac{Q_s^2}{Q^2}\,\ln
 \left(\frac{Q^2}{Q_s^2}\right),\ee
where the dominant term, as isolated in the r.h.s., comes from the
aligned--jet configurations with dipole sizes $r$ within the range
$2/Q<r<1/Q_s$. For the {\em longitudinal} sector, one similarly
obtains
 \be\label{sigmaIAL}
 \frac{\rmd\sigma^{L}_{\rm tot}}
 {\rmd^2 {b}}\ \sim\
 {Q^2}\int\limits_0^{2/Q} \rmd
 r \,r^3 Q_s^2\, +\,\f1{Q^4}\int\limits_{2/Q}^{1/Q_s}
 \f{\rmd r}{r^3}\, Q_s^2\, + \,\f1{Q^4}
 \int\limits_{1/Q_s}^\infty \f{\rmd r}{r^5}
 \,\,\sim\ \, \frac{Q_s^2}{Q^2}\,,\ee
where the leading--order term comes from dipole sizes $r\sim 1/Q$,
and it receives equally important contributions from both the
symmetric and the aligned--jet configurations.

The above results show that, at sufficiently large $Q^2$, the
transverse sector dominates over the longitudinal one, by a
logarithm  $\ln ({Q^2}/{Q_s^2})$. However, this dominance
disappears at lower values of $Q^2$, within the geometric scaling
window in Fig. \ref{Diag}. Indeed, in that regime one must rather
use $\gamma=\gamma_0\approx 0.63$, and then one finds that both
$\sigma_{\rm tot}^T$ and $\sigma_{\rm tot}^L$ are of the same
order, namely $\sim ({Q_s^2}/{Q^2})^{\gamma_0}$, and they are
dominated by dipole sizes $r\sim 1/Q$.

After similarly treating the case $Q^2 \ll Q_s^2$, one finds the
following limiting behaviours for the total ${\gamma^*}h$
cross--section in this mean--field scenario  :
  \beq\label{sigmatotGBW}
    \frac{\rmd\sigma^{\gamma}_{\rm tot}}
 {\rmd^2 {b}} \approx
    \begin{cases}
        \displaystyle{\ln(Q_s^2/Q^2)} &
        \text{ for\,  $Q^2 \ll Q_s^2$}
        \\*[0.2cm]
        \displaystyle{Q_s^2/Q^2 \ln (Q^2/Q_s^2)} &
        \text{ for\,  $Q^2 \gg Q_s^2$}\,.
    \end{cases}
    \eeq

\texttt{ii)} {\em Diffractive cross--sections}

The diffractive cross--section (\ref{sigmadiffHE}) involves the
dipole amplitude {\em squared}, and this leads to important
differences with respect to the inclusive cross--section, at least
in this mean--field scenario. The contribution of the small
dipoles is now strongly suppressed, because $\bar T^2\simeq (r^2
Q^2_s)^{2\gamma}$ for $r \ll 1/Q_s$. Accordingly, the dominant
contribution to $\sigma_{\rm diff}$ comes from relatively large
dipole sizes, of the order of the saturation length: $r\sim 1/Q_s$
\cite{GLR,LW94,BW93,GBW99}.

To see this, let us focus on the contribution of the {\em
aligned--jet} configurations within the {\em transverse} sector
(this turns out to be the dominant piece). The relevant integral
reads
 \be\label{sigmaIIAT}
 \frac{\rmd\sigma^{T}_{\rm diff}}
 {\rmd^2 {b}}&\ \sim\ &\frac{1}{Q^2}\int\limits_{2/Q}^{\infty}
 \frac{\rmd r}{r^3}\ \bar T^{\, 2} \ \sim\
 \frac{1}{Q^2}\int\limits_{2/Q}^{1/Q_s}
 \frac{\rmd r}{r^3}\,(r^2Q^2_s)^{2\gamma}\,
 + \,\frac{1}{Q^2}\int\limits_{1/Q_s}^\infty
 \frac{\rmd r}{r^3}
 \ \sim\ \frac{Q_s^2}{Q^2}\,,\ee
where for more clarity we have split the integration domain into
two: $2/Q<r<1/Q_s$ and $r >1/Q_s$. It can be easily checked that,
so long as $\gamma > 1/2$, both domains contribute on equal
footing to the leading order result $\sim {Q_s^2}/{Q^2}$, which is
generated by dipole sizes $r\sim 1/Q_s$, as anticipated. Indeed,
the integral over the first domain is dominated by large values of
$r$, of the order of the upper cutoff $1/Q_s$, while that over the
second domain is saturated by small values of $r$, close to the
respective lower cutoff $1/Q_s$. By comparison, the other
contributions to diffraction --- that of the {\em symmetric}
transverse  configurations, and the whole {\em longitudinal}
contribution
--- are of `higher--twist order', that is, they are suppressed by
powers of ${Q_s^2}/{Q^2} \ll 1$ with respect to the leading
contribution (\ref{sigmaIIAT}).

We thus arrive at the (by now) standard picture of DIS diffraction
in the presence of saturation \cite{LW94,BW93,GBW99}, in which
$\sigma_{\rm diff}$ is a leading--twist quantity at high--$Q^2$,
so like the inclusive cross--section (\ref{sigmaIAT}).
Accordingly, the ratio between the diffractive and the inclusive
cross--sections is only slowly varying with $Q^2$ and with the
total energy \cite{GBW99,BGBK}
 \be\label{ratioMF}
 R\,\equiv\, \frac{(\rmd\sigma_{\rm diff} /\rmd^2 {b})}
 {(\rmd\sigma_{\rm tot} /\rmd^2 {b})}
 \ \sim\ \f1{\ln ({Q^2}/{Q_s^2(Y)})}\,,\ee
(this behaviour is illustrated by the `mean--field' curve in Fig.
\ref{ratio_diff_incl}), in rough agreement with the pattern
observed in the corresponding HERA data at small $x$. This
agreement has represented one of the main successes of the
saturation models applied to the phenomenology at HERA
\cite{GBW99,BGBK}.

Notice the {\em qualitative} difference between the ways that the
inclusive and diffractive cross--sections get constructed
--- $\sigma_{\rm tot}$ receives leading--twist contributions from
all dipole sizes, with a slight preference though for $r\sim 1/Q$,
whereas $\sigma_{\rm diff}$ is dominated by dipoles with $r\sim
1/Q_s$
---, which nevertheless leads to {\em quantitatively} similar
results (cf. Eq.~(\ref{ratioMF})). For later reference, let us
display here the $q\bar q$ contribution to the diffractive
cross--section in the mean--field approximation for both small and
large $Q^2$:
  \beq\label{sigmaqqGBW}
    \frac{\rmd\sigma^{q\bar q}_{\rm diff}}
 {\rmd^2 {b}} \approx
    \begin{cases}
        \displaystyle{\ln(Q_s^2/Q^2)} &
        \text{ for\,  $Q^2 \ll Q_s^2$}
        \\*[0.2cm]
        \displaystyle{Q_s^2/Q^2} &
        \text{ for\,  $Q^2 \gg Q_s^2$}\,.
    \end{cases}
    \eeq
The above results depend crucially upon the property of the
mean--field amplitude $\bar T(r)$ to decrease very fast when
decreasing $r$ below $1/Q_s$ : $\bar T\sim (r^2Q_s^2)^{\gamma}$
with $\gamma > 1/2$. From the discussion in Sect. \ref{SECT_PL},
we know that this behavior is eventually washed out by
fluctuations when increasing energy. As we shall shortly see, this
change of behaviour has dramatic consequences for the high--energy
limit of both inclusive and diffractive DIS cross--sections.

\subsubsection{The high--energy
behaviour: `black spots' and diffusive scaling} \label{SECT_HEDIS}

We now turn to the most interesting case for us here, namely the
high--energy regime of Eq.~(\ref{HER}). The kinematical region
where the effects of the fluctuations are most visible, and that
we shall concentrate on in what follows, is the high--$Q^2$ range
defined by
 \be\label{HEDIS}
 \sigma(Y) \gg 1/\gamma_0\qquad   \mbox{and} \qquad
  \sigma(Y) \ \ll \ Z\equiv \ln \frac{Q^2}{\langle Q_s^2 \rangle}\
  \ll \ \sigma^2(Y)\,.\ee
For much smaller $Q^2$, such that $Q^2\ll Q_s^2(Y)$, the
cross--sections are dominated by `saturated' dipoles with sizes
$r\simge  1/Q_s$, for which the mean field approximation works
quite well. For much larger $Q^2$, such that $Z\gg \sigma^2(Y)$,
one enters the standard perturbative regime where the amplitudes
show color transparency and evolve according to the DGLAP
equation.

We shall soon verify that, within the interesting range
(\ref{HEDIS}), the convolutions yielding the DIS cross--sections
are dominated by dipole sizes $r\sim 1/Q$ (at variance with the
mean field case, especially in the case of diffraction). It is
therefore appropriate to evaluate these convolutions with the
expression (\ref{Thighsigma}) for the dipole amplitude, which is
valid for $z\ll \sigma^2(Y)$. By using that, we shall be able to
compute {\em exactly} the dominant behaviour of the
cross--sections at high energy. But before describing the exact
calculations, we shall present some parametric estimates which
have the merit to simply demonstrate the physical points that we
would like to emphasize here.

\bigskip
\texttt{i)} {\em Total cross--sections}

We start with the {\em transverse} sector and, for more clarity,
we separate the respective contributions of the symmetric and the
aligned--jet configurations:
 \be\label{sigmaSA}
 \frac{\rmd\sigma^{T}_{\rm tot}}
 {\rmd^2 {b}}\ = \ S^{T}_{\rm tot} \ + \ A^{T}_{\rm tot}\,,
 \ee
where (with $z\equiv \ln (1/r^2\langle Q_s^2 \rangle)$\,) :
 \be\label{symIIT}
 S^{T}_{\rm tot}&\ \sim \ &
\int\limits_0^{2/Q}
 \frac{\rmd
 r}{r}\,\lan T(r)\ran_Y\ \sim \
 \int_{Z}^{\infty}\rmd z \ {\rm Erfc}\left(\frac{z}{\sigma} \right)
 \,,
 \ee
 \be\label{alignIIT}
 A^{T}_{\rm tot}&\ \sim \ &\,\frac{1}{Q^2}\int\limits_{2/Q}^{\infty}
 \frac{\rmd r}{r^3}\,\lan T(r)\ran_Y \ \sim \
 \rme^{-Z}\int_{-\infty}^{Z}\rmd z \ \rme^z\
 {\rm Erfc}\left(\frac{z}{\sigma} \right).
 \ee
Consider the large--$Q^2$ case, such that $\sigma \ll Z \ll
\sigma^2$. It is then easy to verify that: \texttt{(a)} the
integrals giving $S^{T}_{\rm tot}$ and $A^{T}_{\rm tot}$ are
dominated by their respective endpoints at $z=Z$ (i.e., by $r\sim
1/Q$), \texttt{(b)} the dominant behaviour at large $Z$ is given
by the Gaussian ${\exp}(-z^2/\sigma^2)$, and \texttt{(c)} after
also computing the prefactors in front of this Gaussian, the
dominant contribution is found to come from the {\em symmetric}
configurations.

To demonstrate this, we shall first consider the symmetric
integral:
  \be\label{symZ}
 S^{T}_{\rm tot} &\  \sim \ &
 \sigma \int_{Z/\sigma}^{\infty}{\rmd x}\ {\rm Erfc}(x)
 \  \sim \ {\sigma}%{\sqrt{\pi}}
  \int_{Z/\sigma}^{\infty} \frac{\rmd x}{x} \ \rme^{-x^2}
 \ \sim \ {\sigma}
 \int_{Z^2/\sigma^2}^{\infty} \frac{\rmd t}{t} \ \rme^{-t}
 \nn &\  \sim \ & \frac{\sigma^3}{Z^2}\
 \exp\left(-\frac{Z^2}{\sigma^2} \right),\ee
where $x=z/\sigma$, $t=x^2$, and we have used the asymptotic
behaviour of $ {\rm Erfc}(x)$, cf. Eq.~(\ref{erfc}) (recall that
$Z/\sigma \gg 1$). It is now clear that the integral over $t$ is
dominated by its lower end at $t= Z^2/\sigma^2$, i.e., by $z=Z$,
as anticipated.

The aligned--jet contribution can be similarly treated: because of
the exponential factor $\rme^z$ within the integrand, the integral
in Eq.~(\ref{alignIIT}) is dominated by its upper cutoff at $z=Z$,
which yields
  \be\label{alignZ}
 A^{T}_{\rm tot} \  \sim \
 \frac{\sigma}{Z}\
 \exp\left(-\frac{Z^2}{\sigma^2} \right).\ee
As anticipated, this contribution is suppressed (by a factor
$Z/\sigma^2 \ll 1$) with respect to the contribution (\ref{symZ})
of the symmetric configurations.

Similar conclusions hold for the {\em longitudinal} sector : the
dominant contribution is again generated by dipole sizes $r\sim
1/Q$ and it is of the same order as $A^{T}_{\rm tot}$,
Eq.~(\ref{alignZ}),  hence it is parametrically suppressed. We
thus conclude that the inclusive cross--section in the
high--energy limit and for relatively large $Q^2$ is dominated by
symmetric dipole configurations with typical sizes $r\sim 1/Q$
within the transverse sector.

It is in fact quite straightforward to go beyond the parametric
estimate in Eq.~(\ref{symZ}) and compute the dominant behaviour
{\em exactly}. We present here only the final results, since a
similar calculation will be explained in more detail below, for
the case of diffraction. Specifically, after also reintroducing
the overall electromagnetic factor $F\equiv (N_c \alpha_{\rm em}/2
\pi^2) \sum_ f e_f^2$, one finds that the longitudinal part of the
total cross--section is given by
    \beq\label{sigmaLtotfinal}
    \frac{\rmd\sigma^{L}_{\rm tot}}
 {\rmd^2 {b}}\,\simeq\,
    \frac{\pi F}{3}\,
    \erfc\left(\frac{Z}{\sigma}\,\right),
    \eeq
while the respective transverse part reads
    \beq\label{sigmaTtotfinal}
     \frac{\rmd\sigma^{T}_{\rm tot}}
 {\rmd^2 {b}}\, \simeq\,
    \frac{\pi F}{3}\,\sigma\,
    \Phi_1\left(\frac{Z}{\sigma}\,\right).
    \eeq
Here we have found convenient to define the function
    \beq\label{phi1def}
    \Phi_1(x) \equiv \int\limits_{x}^{\infty}
    \dif v\,\erfc(v) =
    \frac{1}{\sqrt{\pi}}\, \exp(-x^2)
    -x \, \erfc(x)
    \eeq
which has the following behavior in the various limits
    \beq\label{phi1lim}
    \Phi_1(x)=
    \begin{cases}
        \displaystyle{2|x| + \frac{\exp(-x^2)}{2 \sqrt{\pi} x^2}} &
        \text{ for\,  $x \ll -1$}
        \\*[0.33cm]
        \displaystyle{\frac{1}{\sqrt{\pi}}} &
        \text{ for\,  $x=0$}
        \\*[0.33cm]
        \displaystyle{\frac{\exp(-x^2)}{2 \sqrt{\pi} x^2}} &
        \text{ for\,  $x \gg 1$}.
    \end{cases}
    \eeq
Eqs.~(\ref{sigmaLtotfinal}) and (\ref{sigmaTtotfinal}) hold for
$Z\ll \sigma^2$ and are correct up to corrections of relative
order $Z/\sigma^2$ and/or $1/\sigma$. They confirm the previous
estimates like (\ref{symZ}) and imply the following, final, result
for the total DIS cross--section in the high--energy regime:
    \beq\label{sigmatotfin}
     \frac{\rmd\sigma^{\gamma}_{\rm tot}}
 {\rmd^2 {b}}
 \, \simeq\,
    \frac{\rmd\sigma^{T}_{\rm tot}}{\rmd^2 {b}}\,
   \simeq\,
    \frac{\pi F}{3}\ \sigma\,
    \Phi_1\left(\frac{\ln (Q^2/\langle Q_s^2 \rangle)}{\sigma}\right)
    \quad\mbox{for\
$\sigma \gamma_0 \gg 1$ \ and\ } \ln \frac{Q^2}{\langle Q_s^2
\rangle} \ll \sigma^2.
    \eeq
Note that the quantity $({\rmd\sigma^{\gamma}_{\rm tot}}/ \rmd^2
{b})/\sigma$ shows {\em diffusive scaling}, i.e., it depends upon
the kinematical variables $Q^2$ and $Y$ only via the dimensionless
variable $Z/\sigma\equiv {\ln (Q^2/\langle Q_s^2
\rangle)}/{\sigma}$.

\begin{figure}[t]
    \centerline{\epsfxsize=14cm\epsfbox{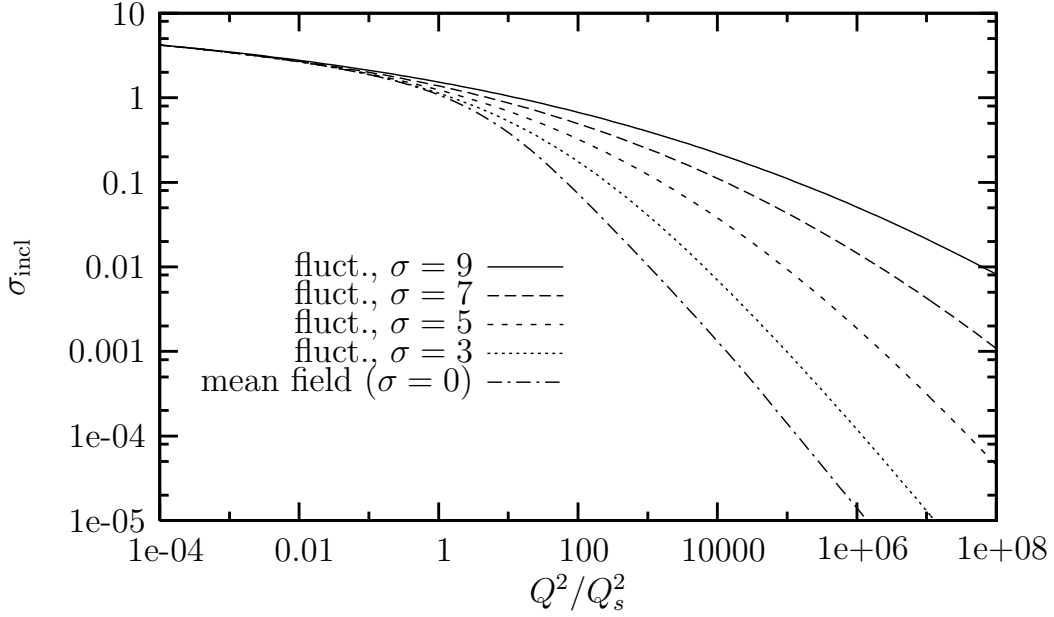}}
    \caption{\sl  The $\gamma^* h$ total cross section as a function of
$Q^2/\langle Q_s^2 \rangle$ and for various values of the front
dispersion $\sigma$.
    \label{sigma_incl}\vspace*{0.5cm}}
    \end{figure}

\begin{figure}[t]
    \centerline{\epsfxsize=14cm\epsfbox{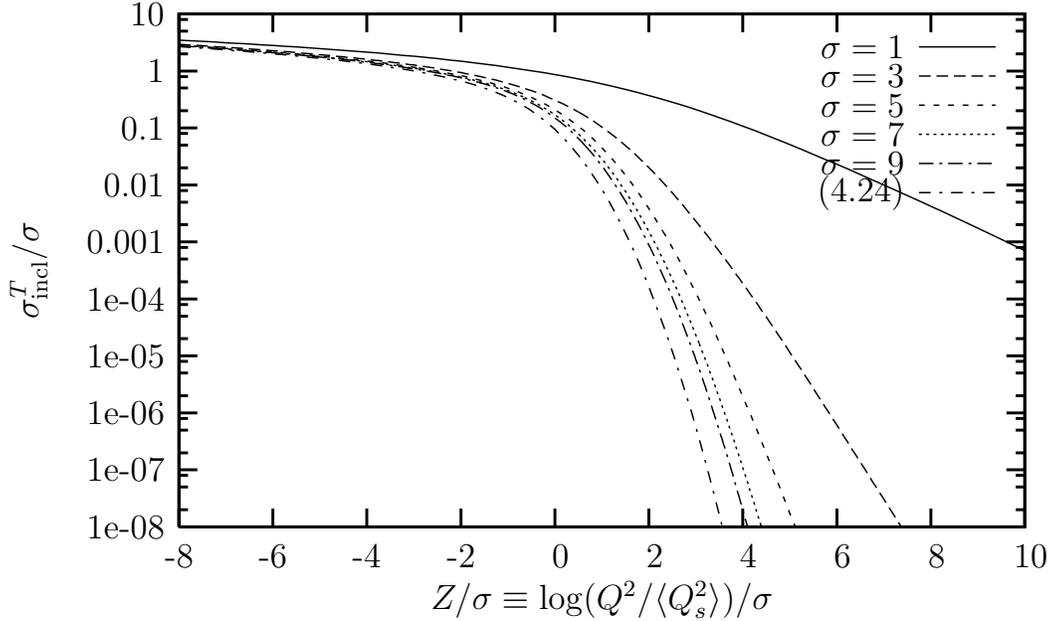}}
    \caption{\sl  The $\gamma^* h$ total cross section (divided by the front
dispersion) as a function of the ``diffusive'' scaling variable
$Z=\ln(Q^2/\langle Q_s^2 \rangle)/\sigma$.
    \label{DS_incl_mod1}\vspace*{0.5cm}}
    \end{figure}

Eq.~(\ref{sigmatotfin}) also allows us to study the behaviour of
the cross--section at relatively low $Q^2$ (below and around
$\langle Q_s^2 \rangle$). Deeply inside the saturation region, one
finds
   \beq\label{sigmatotlowQ}
     \frac{\rmd\sigma^{\gamma}_{\rm tot}}
 {\rmd^2 {b}}\,\approx\,\frac{2\pi F}{3}\, \ln
 \frac{\langle Q_s^2 \rangle}{Q^2}\qquad \mbox{for}\qquad \ln
 (\langle Q_s^2\rangle /{Q^2}) \,\gg\, \sigma\,,\eeq
which is in agreement, as expected, with the mean--field behaviour
in Eq.~(\ref{sigmatotGBW}). However, the difference with respect
to the mean field situation occurs already in the vicinity of the
(average) saturation line $Z=0$ : Within a wide interval
$|\ln(Q^2/\langle Q_s^2 \rangle)| \ll \sigma$ around this line,
the cross--section (\ref{sigmatotfin}) is rather large --- since
proportional to $\sigma$ --- and $Q^2$--independent:
 \beq\label{sigmatotsat}
     \frac{\rmd\sigma^{\gamma}_{\rm tot}}
 {\rmd^2 {b}}\,\approx\,\frac{\sqrt{\pi} F}{3}\, {\sigma}
 \qquad \mbox{for}\qquad |\ln(Q^2/\langle Q_s^2 \rangle)|
  \,\ll\, \sigma\,.\eeq
The cross--section starts to fall only when $Q^2$ becomes so large
that $\ln(Q^2/\langle Q_s^2 \rangle) \gg \sigma$, and then it has
a rapid fall, of the Gaussian type (cf. Eq.~(\ref{symZ})).

This behaviour is illustrated in Figs. \ref{sigma_incl} and
\ref{DS_incl_mod1}, which are the analog at the level of the DIS
inclusive cross--section of Figs. \ref{front} and \ref{DS_front}
for the dipole amplitude. Namely, these figures represent results
obtained via the numerical evaluation of the total cross--section
(\ref{Ptotfin}) with the dipole amplitude in Eq.~(\ref{Tave}) with
$\gamma=1$. Fig. \ref{sigma_incl} shows the increase in the
deviation from the mean--field behaviour (and thus from geometric
scaling) when increasing $\sigma$, whereas Fig. \ref{DS_incl_mod1}
demonstrates the emergence of the diffusive scaling and the
convergence of the cross--section towards its high--energy
asymptotic in Eq.~(\ref{sigmatotfin}).

 \bigskip
 \texttt{ii)} {\em Diffractive cross--sections}

Moving to the diffractive sector, one can immediately notice an
important difference with respect to the mean--field scenario of
Sect. \ref{SECT_MFDIS}: Whereas in that case, the replacement of
the dipole amplitude by its square has strongly suppressed the
small--dipole configurations and thus shifted the strength of the
integration from $r\sim 1/Q$ to $r\sim 1/Q_s$, in the present,
high--energy, case, there is no similar suppression, because both
the Gaussian ${\exp}(-z^2/\sigma^2)$ and its square
${\exp}(-2z^2/\sigma^2)$ decay rather slowly (as compared to the
exponential $\rme^{-z}$) within the interesting range at
$\sigma\ll z\ll \sigma^2$. Accordingly, the aligned--jet
contribution to high--energy $q\bar q$ diffraction, namely
(compare to Eq.~(\ref{sigmaIIAT}))
  \be\label{aligndif}
 A^{T}_{\rm diff}&\ \sim \ &\,\frac{1}{Q^2}\int\limits_{2/Q}^{\infty}
 \frac{\rmd r}{r^3}\,\lan T(r)\ran^2_Y \ \sim \
 \rme^{-Z}\int_{-\infty}^{Z}\rmd z \ \rme^z\
 {\rm Erfc}^2\left(\frac{z}{\sigma} \right).
 \ee
is dominated by its lower end at $r\sim 1/Q$ (that is, by $z=Z$),
so like the corresponding contribution (\ref{alignIIT}) to the
inclusive cross--section.

This points towards an important physical difference between the
mean--field (or intermediate energy) behaviour and the behaviour at
high energy, which deserves a more qualitative explanation. To that
aim, notice that the measure ${\rmd r}/{r^3}$ in the aligned--jet
integral favors the small dipoles, so the integral would be
naturally dominated by its lower end at $r\sim 1/Q$ if there was not
for the strong suppression of the dipole sizes $r\ll 1/Q_s$
introduced by the dipole amplitude squared $T^2$. Such a strong
suppression occurs, as we have seen, for the mean--field amplitude
$\bar T$, and also for the {\em typical} events in the statistical
ensemble in QCD at high energy and large $Q^2\gg \lan Q_s^2 \ran$ :
indeed, a typical event has $Q_s\sim \lan Q_s \ran$ and thus yields
a small contribution, of order $\lan Q_s^2 \ran/Q^2=\rme^{-Z}$, to
the integral. However, due to the front dispersion in the presence
of fluctuations, the statistical ensemble contains also fronts which
are at saturation at the minimal size $1/Q$, and each such a front
yields a relatively large contribution, of $O(1)$. Although such
fluctuations are relatively {\em rare}, their contribution weighted
by the respective probability $\sim {\exp}(-Z^2/\sigma^2)$ is still
larger than that of the typical fronts, which behaves like
$\rme^{-Z}$. In other terms, the convolution peaks up those rare
gluon configurations, or `black spots', in the target wavefunction
which are at saturation at the scale $Q^2$ set by the virtual
photon.

Therefore, at high energy --- and in contrast to what happens in
the mean--field approximation, or at intermediate energies --- the
inclusive and diffractive cross--sections receive their dominant
contributions from the {\em same} physical configurations, namely
the small dipoles with size $r\sim 1/Q$. In fact, a qualitative
analysis similar to that in Eqs.~(\ref{sigmaSA})--(\ref{alignZ})
reveals that the analogy between inclusive and diffractive
processes at high energy is even stronger: In the large--$Q^2$
regime defined by Eq.~(\ref{HEDIS}), both types of cross--sections
are dominated by symmetric dipole configurations within the
transverse sector. In particular, the dominant behaviour of
$\sigma^{q\bar q}_{\rm diff}$ can be estimated as in
Eq.~(\ref{symZ}), and reads
  \be\label{symZdiff}
       \frac{\rmd\sigma^{q\bar q}_{\rm diff}}
 {\rmd^2 {b}}  \ \simeq\
 S^{T}_{\rm diff} \  \sim \  \frac{\sigma^4}{Z^3}\
 \exp\left(-\frac{2 Z^2}{\sigma^2} \right)
 \qquad\mbox{for}\qquad \sigma \ll \, Z \, \ll \sigma^2\,,\ee
which implies the following estimate for the
(diffractive/inclusive) ratio $R$ :
 \be\label{ratioHE}
 %\frac{(\rmd\sigma_{\rm diff} /\rmd^2 {b})}
 %{(\rmd\sigma_{\rm tot} /\rmd^2 {b})}
 R\ \sim\ \f{\sigma}{Z}\,\exp\left(-\frac{Z^2}{\sigma^2}\right)
 \qquad\mbox{for}\qquad \sigma \ll \, Z \, \ll \sigma^2\,.\ee
This is a {\em scaling} function, which decreases quite rapidly
with $Q^2$ at fixed $Y$ (unlike the respective mean field estimate
in Eq.~(\ref{ratioMF})), but one should notice that this decrease
becomes significant only at very large $Q^2$, such that
$Z\gg\sigma$.

\begin{figure}[t]
    \centerline{\epsfxsize=14cm\epsfbox{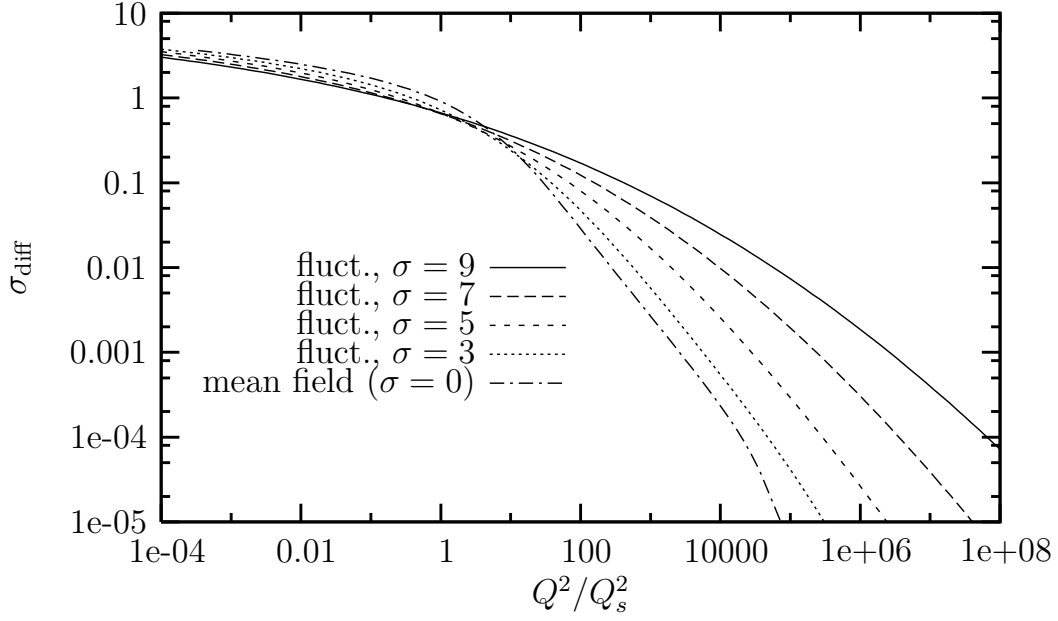}}
    \caption{\sl  The contribution of the $q\bar{q}$ component of the onium
wavefunction to the diffractive cross section as a function of
$Q^2/\langle Q_s^2 \rangle$ and for various values of the front
dispersion.
    \label{sigma_diff}\vspace*{0.5cm}}
    \end{figure}

\begin{figure}[t]
    \centerline{\epsfxsize=14cm\epsfbox{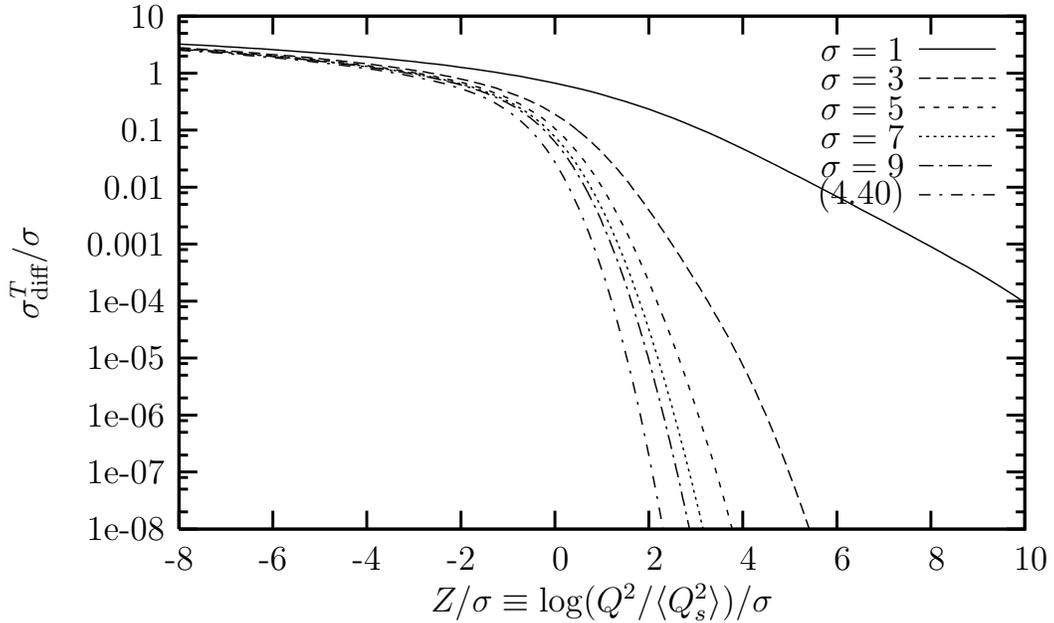}}
    \caption{\sl  The same as in Fig.~15 (divided by the front dispersion
$\sigma$) as a function of $Z/\sigma$.
    \label{DS_diff_mod1}\vspace*{0.5cm}}
    \end{figure}

To make these estimates more precise, let us compute the dominant
behaviour of $\sigma^{q\bar q}_{\rm diff}$ in the regime $\sigma
\gamma_0 \gg 1$ and $Z \ll \sigma^2$. This is obtained by
evaluating the convolution in Eq.~(\ref{sigmadiffHE}) with the
dipole amplitude $\avg{T(r)}$ in Eq.~(\ref{Thighsigma}). Consider
first the {\em longitudinal} cross section. Using Eq.~(\ref{PsiL})
and defining $u=r\,Q$ and $\tau=Q/\avg{Q_s}$ we have
    \beq\label{sigmaLqq}
    \frac{\rmd\sigma^{L}_{\rm diff}}{\rmd^2 {b}}\,  =\,
    \frac{\pi F}{2}
    \int\limits_{0}^{1} \dif v
    \int\limits_{0}^{\infty} \dif u\,
    4 v^2 (1-v)^2\, u\,
    \rmK_0^2\left(u \sqrt{v (1-v)}\right)\,
    \erfc^2\left(\frac{\ln (\tau^2/u^2)}{\sigma}\right).
    \eeq
When $\sigma$ is large we can set $u=1$ in the argument of the
error function; indeed, as previously explained, the convolution
is dominated by $r\sim 1/Q$. Then the integrals over $u$ and $v$
are easily performed and we arrive at (note that $\ln \tau^2 =
\ln(Q^2/\langle Q_s^2 \rangle) = Z$)
    \beq\label{sigmaLqqfinal}
    \frac{\rmd\sigma^{L}_{\rm diff}}{\rmd^2 {b}}\, =\,
    \frac{\pi F}{6}\,
    \erfc^2\left(\frac{Z}{\sigma}\,\right).
    \eeq
Now let us look at the {\em transverse} cross section. It reads
    \beq\label{sigmaTqq}
    \hspace*{-0.35cm}
     \frac{\rmd\sigma^{T}_{\rm diff}}{\rmd^2 {b}}\,  =\,
    \frac{\pi F}{2}
    \int\limits_{0}^{1} \dif v
    \int\limits_{0}^{\infty} \dif u\,
    v (1-v) [v^2 + (1-v)^2 ]\,
    u\, \rmK_1^2\left(u \sqrt{v (1-v)}\right)\,
    \erfc^2\left(\frac{\ln (\tau^2/u^2)}{\sigma}\right).\nn
    \eeq
The behavior of the last factor in the small--$u$ region is
crucial since it cancels the logarithmic singularity of the
remaining part of the integrand at $u=0$. Hence, we cannot set
$u=1$ as we did for the longitudinal case. Here it is convenient
to define the function
    \beq\label{phi2def}
    \Phi_2(x) \equiv \int\limits_{x}^{\infty}
    \dif v\,\erfc^2(v) =
    \frac{2}{\sqrt{\pi}}\, \exp(-x^2)\, \erfc(x)
    -x \, \erfc^2(x)
    -\sqrt{\frac{2}{\pi}}\, \erfc(\sqrt{2}\,x),
    \eeq
which has the following behavior in the various limits
    \beq\label{phi2lim}
    \Phi_2(x)=
    \begin{cases}
        \displaystyle{4|x| - \frac{2\sqrt{2}}{\sqrt{\pi}}
        -\frac{\exp(-2 x^2)}{4 \pi |x|^3}} &
        \text{ for\,  $x \ll -1$}
        \\*[0.33cm]
        \displaystyle{\frac{2-\sqrt{2}}{\sqrt{\pi}}} &
        \text{ for\,  $x=0$}
        \\*[0.33cm]
        \displaystyle{\frac{\exp(-2 x^2)}{4 \pi x^3}} &
        \text{ for\,  $x \gg 1$}.
    \end{cases}
    \eeq
Using Eq.~(\ref{phi2def}) we can write the last factor in
Eq.~(\ref{sigmaTqq}) as
    \beq\label{erfcphi2}
    \erfc^2\left(\frac{\ln (\tau^2/u^2)}{\sigma}\right)=
    \frac{\sigma}{2}\,u\,\frac{\dif}{\dif u}\,
    \Phi_2\left(\frac{\ln (\tau^2/u^2)}{\sigma}\right),
    \eeq
and then the $u$--integration in Eq.~(\ref{sigmaTqq}) can be
performed by parts. The boundary term vanishes and therefore the
virtual photon transverse cross section becomes
    \beq\label{sigmaTqqtemp}
     \frac{\rmd\sigma^{T}_{\rm diff}}{\rmd^2 {b}}\,  =\,
    \frac{\pi F}{2}\,\sigma
    \int\limits_{0}^{1} \dif v
    \int\limits_{0}^{\infty} \dif u\,&&
    [v (1-v)]^{3/2}\, [v^2 + (1-v)^2 ]\,
    \Phi_2\left(\frac{\ln (\tau^2/u^2)}{\sigma}\right)
    \nn \times &&
    u^2\, \rmK_0\left(u \sqrt{v (1-v)}\right)
    \rmK_1\left(u \sqrt{v (1-v)}\right).
    \eeq
Now we can safely set $u=1$ in the argument of the $\Phi_2$
function since the remaining part of the integrand is
well--defined for any value of $u$. The integration over $u$ and
$v$ becomes straightforward  and we obtain
    \beq\label{sigmaTqqfinal}
     \frac{\rmd\sigma^{T}_{q\bar{q}}}{\rmd^2 {b}}\,  =\,
    \frac{\pi F}{6}\,\sigma\,
    \Phi_2\left(\frac{Z}{\sigma}\,\right).
    \eeq
Comparing Eqs.~(\ref{sigmaLqqfinal}) and (\ref{sigmaTqqfinal}) we
see that the transverse cross section dominates over the
longitudinal one within the whole region of interest (namely,
$-\infty < Z \ll \sigma^2$ and $\sigma \gamma_0 \gg 1$), so we
finally arrive at
    \beq\label{sigmaqq}
     \frac{\rmd\sigma^{\gamma}_{\rm diff}}{\rmd^2 {b}}
    \, \simeq\,
    \frac{\rmd\sigma^{T}_{\rm diff}}{\rmd^2 {b}}\,
   \simeq\,
    \frac{\pi F}{6}\ \sigma\
    \Phi_2\left(\frac{\ln (Q^2/\langle Q_s^2 \rangle)}{\sigma}\right),
    \eeq
up to corrections of relative order $Z/\sigma^2$ and/or
$1/\sigma$. As anticipated by the notation in Eq.~(\ref{sigmaqq}),
this is in fact the {\em complete} dominant contribution to the
diffractive cross--section in the high--energy regime of interest
(see the discussion in Sect. \ref{SECT_QQG}). This quantity has
the same qualitative behaviour as the inclusive cross--section in
Eq.~(\ref{sigmatotfin}), and this behaviour is illustrated in
Figs. \ref{sigma_diff} and \ref{DS_diff_mod1} (to be read by
analogy with Figs. \ref{sigma_incl} and \ref{DS_incl_mod1}). In
particular, the rescaled quantity $({\rmd\sigma^{\gamma}_{\rm
diff}}/ \rmd^2 {b})/\sigma$ exhibits diffusive scaling.

\begin{figure}[t]
    \centerline{\epsfxsize=14cm\epsfbox{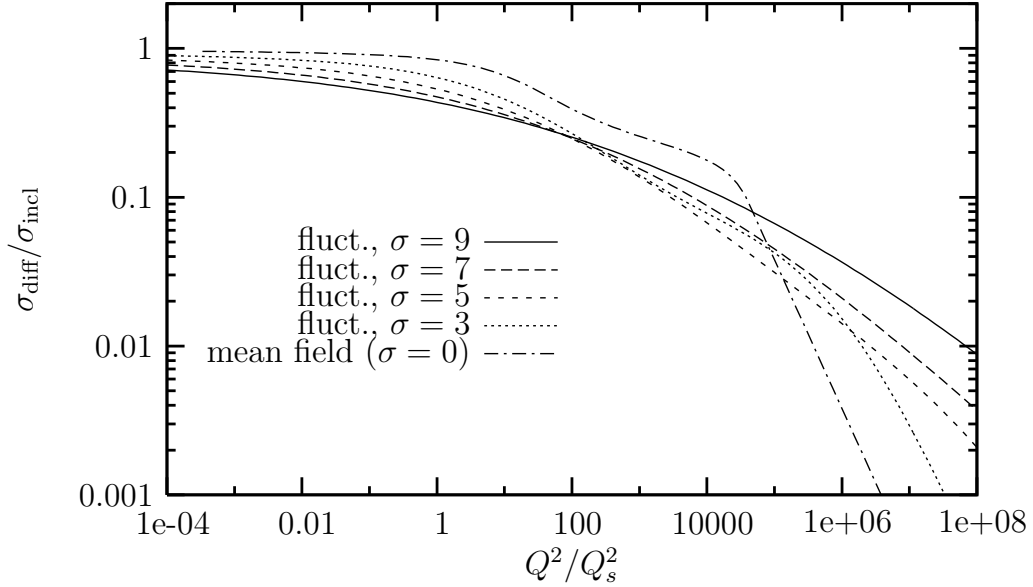}}
    \caption{\sl  The ratio of the diffractive (contribution of the $q\bar{q}$
component) to the total cross section as a function of
$Q^2/\langle Q_s^2 \rangle$.
    \label{ratio_diff_incl}}
    \end{figure}

Finally, the ratio $R$ of the diffractive to the total cross
section is easily obtained from Eqs.~(\ref{sigmaqq}) and
(\ref{sigmatotfin}).  It reads
    \beq\label{ratio}
    R = \frac{1}{2} \,
    \frac{\Phi_2\left(\ln (Q^2/\avg{Q_s}^2)/\sigma\right)}
    {\Phi_1\left(\ln (Q^2/\avg{Q_s}^2)/\sigma\right)}.
    \eeq
For small values of $Q$ such that $\ln(\langle Q_s^2 \rangle/Q^2)
\gg \sigma$ the ratio approaches 1, it is constant and equal to
$1-\sqrt{2}/2=0.293$ in the wide interval $|\ln(Q^2/\langle Q_s^2
\rangle)| \ll \sigma$ around the saturation line, while it falls
very fast with increasing momentum when $\ln(Q^2/\langle Q_s^2
\rangle) \gg \sigma$ with the precise form easily inferred from
Eqs.~(\ref{phi2lim}) and (\ref{phi1lim}). This behaviour is
graphically illustrated in Fig. \ref{ratio_diff_incl}, and
compared to the corresponding mean--field behaviour.

\subsection{The $q \bar q g$ component}
\label{SECT_QQG}

In this subsection,  we shall analyze the first non--trivial
contribution to the inelastic diffraction, which arises when the
wavefunction of the projectile at the time of scattering is a
superposition of two Fock space components: the original $q \bar
q$ pair (one dipole) and its $q \bar q g$ excitation (i.e., two
dipoles at large $N_c$). This configuration is physically relevant
for relatively large diffractive mass $M_X^2\gg Q^2$, or small
values of $\beta\ll 1$, but such that $\abar\ln (1/\beta)\ll 1$,
in order for the effects of the higher Fock states in the
projectile to be negligible. As we shall see, the two--component
configuration generates contributions of order $\abar\ln
(1/\beta)$ to both the elastic and the inelastic pieces of the
diffractive cross--section, but at high energy, the inelastic
piece is parametrically suppressed with respect to the elastic one
in the whole kinematical range of interest (including the
large--$Q^2$ domain defined by Eq.~(\ref{HEDIS}), which extends
with $Y$). This means that, if one starts at $Y=Y_{\rm gap}$,
where the diffraction is purely elastic (cf. Eq.~(\ref{Pel1})),
and one increases $Y$ for a fixed value of $Y_{\rm gap}$ (i.e.,
one increases the rapidity of the projectile), then the elastic
piece of diffraction will rise faster than the inelastic one, and
thus will remain the dominant contribution to $\sigma_{\rm diff}$
at any $Y\ge Y_{\rm gap}$ (i.e., for arbitrarily
small\footnote{Strictly speaking, our argument applies so long as
$\abar \ln (1/\beta) \ll \ln(1/\alpha_s^2)$, i.e., for a
projectile which is non--saturated, since this is the validity
limit of Eq.~(\ref{Pdiff}).} $\beta$). This situation occurs for
sufficiently high values of $Y_{\rm gap}$, such that
$\sigma(Y_{\rm gap})\gg 1$, and is in sharp contrast with the
corresponding mean--field scenario \cite{Gotsman00,MS03}, where
for $Q^2 \gg Q_s^2$ the inelastic diffraction rises faster than
the elastic one with $1/\beta$ and thus rapidly dominates over the
latter.

One should also remind here that the diffractive cross--section
that we have in mind, cf. Eqs.~(\ref{sigmadiffint}) and
(\ref{sigmadiff}), is integrated over all the values of the
rapidity gap between $Y_{\rm gap}$ and $Y$. But, of course, the
{\em differential} cross--section $\rmd \sigma_{\rm diff}/\rmd \ln
(1/\beta)$ at sufficiently small $\beta\ll 1$ is always dominated
by inelastic scattering (cf. Eq.~(\ref{sigmadiff})), because the
corresponding elastic contribution is peaked near $\beta=1$. In
what follows, we shall compute the lowest order contribution to
this differential cross--section, as given by the $q \bar q g$
state.

\subsubsection{Diffractive DIS with one or two dipoles}
\label{SECT_2DIP}

Within the present approximations, the scattering probabilities
for an onium made with two dipoles are obtained by taking $N=2$ in
the general formul\ae\ in Sect. \ref{SECT_DIPOLE} and using the
dipole probabilities $P_1$ and $P_2$ given by Eq.~(\ref{PdY}) with
$\rmd Y=Y-Y_0=\ln (1/\beta)$.

We are primarily interested in diffraction, but in order to
separate its elastic and inelastic components, we also need the
forward amplitude ${\cal A}(\x,\y; Y)$, so let us start with this
latter. According to Eq.~(\ref{Aforward}), the forward amplitude
can be expressed either as the scattering of a single $q\bar q$
dipole in the frame where the target has rapidity $Y$ :
 \be\label{AQQ}
 {\cal A}(\x,\y; Y)&\,=\,&\langle T({\bm{x},\bm{y}})
   \rangle_{Y}\,,\ee
or as the scattering of an onium composed of two dipoles, in the
frame where the target has rapidity $Y_0$ (with the simplified
notation $T_{\bm{x}\bm{y}}\equiv  T({\bm{x},\bm{y}})$) :
 \be\label{DAQQG}
 {\cal A}(\x,\y; Y)&\,=\,&\langle T({\bm{x},\bm{y}})
   \rangle_{Y_0} \,+\,\ln (1/\beta)\,\Delta {\cal A}(\x,\y;
   Y_0)\,,\nn
   \Delta {\cal A}(\x,\y; Y_0)&\,\equiv\,&
 \atpi \int\limits_{\bm{z}}
    \mcal{M}(\bm{x},\bm{y},\bm{z})
    \lan {T_{\bm{x}\bm{z}}} +{T_{\bm{z}\bm{y}}}
    - {T_{\bm{x}\bm{z}} T_{\bm{z}\bm{y}}}
     - {T_{\bm{x}\bm{y}}} \ran_{Y_0}\,.
    \ee
The quantity $\Delta {\cal A}$ describes the changes in ${\cal A}$
brought in by the one--step evolution, namely, the scattering of
the two--dipole state (the first three terms within the brackets)
together with the `virtual' correction to the scattering of the
one--dipole state (the fourth term there). By comparing
Eqs.~(\ref{AQQ}) and (\ref{DAQQG}) and taking the limit $\rmd
Y\equiv \ln (1/\beta)\to 0$, one can immediately deduce an
evolution equation for $\langle T_{\bm{x}\bm{y}} \rangle_Y$ :
 \be\label{Tevol}
 \frac{\del \lan T_{\bm{x}\bm{z}}\ran_Y }
    {\del Y}\,=
 \atpi \int\limits_{\bm{z}}
    \mcal{M}(\bm{x},\bm{y},\bm{z})
    \lan {T_{\bm{x}\bm{z}}} +{T_{\bm{z}\bm{y}}} -
    {T_{\bm{x}\bm{y}}} - {T_{\bm{x}\bm{z}} T_{\bm{z}\bm{y}}}
    \ran_{Y}\,,
    \ee
which is recognized as the first equation in both the Balitsky
hierarchy \cite{B} and in the hierarchy of equations with Pomeron
loops \cite{IT05}. (The differences between these two hierarchies,
which express the effects of gluon number fluctuations, start with
the second equation, as obeyed by the two--dipole amplitude $\lan
T(\bm{x}_1,\bm{y}_1)T(\bm{x}_2,\bm{y}_2)\ran_Y$.) Within the mean
field approximation (\ref{FACT}), Eq.~(\ref{Tevol}) reduces to the
BK equation (\ref{BK}).

Consider similarly the elastic probability, Eq.~(\ref{Pel}). This
too can be evaluated in either frame (the `$Y$--frame' and the
`$Y_0$--frame'), with the following results :
  \be\label{PelQQG}
   P_{\rm el}(\x,\y; Y)&\,=\,&
    | \langle T({\bm{x},\bm{y}})
   \rangle_{Y}|^2
   \,\simeq\, |\langle T({\bm{x},\bm{y}})
   \rangle_{Y_0} |^2 \,+\,\ln (1/\beta)\,\Delta P_{\rm el}(\x,\y; Y_0)
  \nn  \Delta P_{\rm el}(\x,\y; Y_0)
 &\,\equiv\,&2\,
   {\rm Re}\{
    \langle T({\bm{x},\bm{y}})
   \rangle_{Y_0}\,
   \Delta {\cal A}(\x,\y; Y_0)\}\,,\ee
where the expression in the $Y_0$--frame  holds to the order of
interest in $\abar \ln (1/\beta)$. Clearly, $\Delta P_{\rm el}$
represents the $q \bar q g$ contribution to the elastic
probability.

For the diffractive probability (\ref{Pdiff}), on the other hand,
the choice of the $Y_0$--frame is mandatory, since $Y_0$ plays
also the role of the physical rapidity gap. We thus find:
 \be\label{PdiffQQG}
 P_{\rm diff}(\x,\y; Y,Y_0)\,=\,
 P_{\rm diff}^{q\bar q}(\x,\y; Y_0)\,+\,
 \ln (1/\beta)\,\Delta P_{\rm diff}(\x,\y; Y_0),\ee
where
 \be\label{Pdiffqq}
 P_{\rm diff}^{q\bar q}(\x,\y; Y_0)\,\equiv \,|\lan T({\bm x},
 {\bm y}) \ran_{Y_0}|^2\,=\,P_{\rm el}(\x,\y; Y_0)\ee
is the contribution of the elementary $q\bar q$ pair, and is the
same as the elastic probability corresponding to a total rapidity
$Y_0$ (cf. Eq.~(\ref{PelQQG})), while
 \be\label{PQQG}
 \Delta P_{\rm diff}(\x,\y; Y_0)&\,\equiv\,&
 \atpi \int\limits_{\bm{z}}
    \mcal{M}(\bm{x},\bm{y},\bm{z})\left\{
 |\langle 1\,-\, S_{\bm{x}\bm{z}}S_{\bm{z}\bm{y}}
 \rangle_{Y_0}|^2 \,-\, |\langle 1\,-\, S_{\bm{x}\bm{y}}
   \rangle_{Y_0}|^2\right\}\nn
 &\,=\,&\atpi \int\limits_{\bm{z}}
    \mcal{M}(\bm{x},\bm{y},\bm{z})\left\{
 |\langle T_{\bm{x}\bm{z}}+ T_{\bm{z}\bm{y}}
 - T_{\bm{x}\bm{z}}T_{\bm{z}\bm{y}}
 \rangle_{Y_0}|^2 - |\langle T_{\bm{x}\bm{y}}
   \rangle_{Y_0}|^2\right\}
    \ee
represents the additional contribution due to the $q \bar q g$
state. This contains both elastic and inelastic components, and in
order to separate the latter we use $P_{\rm diff}^{\rm\,
inel}=P_{\rm diff}- P_{\rm el}$, together with Eqs.~(\ref{PelQQG})
and (\ref{Pdiffqq}). This yields:
 \be\label{PQQGfin}
  P_{\rm diff}^{\rm\, inel}(\x,\y; Y,Y_0)&\, =\,&
  \ln (1/\beta)\, \Delta
  P_{\rm inel}(\x,\y; Y_0)\nn
  \Delta P_{\rm inel}(\x,\y;Y_0)&\,\equiv\,&
 \Delta P_{\rm diff}(\x,\y; Y_0)\,-\,
 \Delta P_{\rm el}(\x,\y;Y_0)\,.\ee
From the expressions (\ref{DAQQG}), (\ref{PelQQG}) and
(\ref{PQQG}), one immediately finds
 \be\label{PinelQQG}
    \Delta P_{\rm inel}(\x,\y;Y_0)&\,=\,&
  \atpi \int\limits_{\bm{z}}
    \mcal{M}(\bm{x},\bm{y},\bm{z})\,
    \big|\lan {T_{\bm{x}\bm{z}}} +{T_{\bm{z}\bm{y}}} -
    {T_{\bm{x}\bm{y}}} - {T_{\bm{x}\bm{z}} T_{\bm{z}\bm{y}}}
    \ran_{Y_0}\big|^2\,.
    \ee
Note that the inelastic diffraction starts at order  $\abar \ln
(1/\beta)$, i.e., it requires at least one soft gluon in the
wavefunction of the projectile.

From  Eqs.~(\ref{sigmadiff2}) and  (\ref{PQQGfin}), one can deduce
our lowest--order approximation for the differential diffractive
cross--section, valid for $\beta \ll 1$ but such that $\abar\ln
(1/\beta)\ll 1$ :
 \be\label{dsigmaGdb}
 \frac{\rmd\sigma^{\gamma}_{\rm  diff}}
 {\rmd^2 {b} \ \rmd \ln (1/\beta)}\,
 \simeq\int_0^1 \rmd z \int \rmd^2 {\bm r}\,\sum_{\alpha=L,T}
 \vert \psi^{\gamma}_{\alpha}(z, r; Q)\vert^2 \
 \Delta P_{\rm inel}({\bm b}, {\bm r}; Y_0)\,. \ee
Note that, for the simple onium configuration at hand, the
inelastic diffraction is, by construction, the probability to have
two dipoles in the final state. Hence,  the quantity
(\ref{dsigmaGdb}) represents also the {\em cross--section for
gluon production} in diffractive DIS \cite{K01,CM04}, at the
present level of accuracy.

\subsubsection{The inelastic component of diffraction}
\label{SECT_PINEL}

The inelastic diffraction probability associated with the $q \bar
q g$ state is given by Eq.~(\ref{PinelQQG}), which after the
trivial change of variable $\bm{z}\equiv\bm{x}-\bm{s}$ is
rewritten as (with $\bm{r}=\x-\y$) :
 \be\label{Pinels}
  \Delta P_{\rm inel}(r)\,=\,
  \atpi \int\limits_{\bm{s}}\,\frac{\bm{r}^2}{\bm{s}^2(\bm{r}
  -\bm{s})^2}\,
    \lan T(\bm{s}) + T(\bm{r} -\bm{s}) - T(\bm{r}) -
    T(\bm{s})T(\bm{r} -\bm{s})
    \ran^2\,.
    \ee
In line with the previous approximations, we have assumed the
high--energy amplitudes to be real and ignored their impact
parameter dependence. It is also implicit in our notations that
the target expectation values are evaluated at a rapidity $Y_0$,
and therefore they involve the (average) saturation momentum $\lan
Q_s^2\ran(Y_0)$ and the front dispersion $\sigma^2(Y_0) =D_{\rm
fr}\bar\alpha_s Y_0$. We shall evaluate this expression in both
the mean--field approximation and the general case with
fluctuations, to emphasize the difference between these two
situations.

\bigskip
\texttt{i)} {\em The mean--field approximation}

In this approximation, the two--dipole amplitude is assumed to
factorize : $\lan T(\bm{s})T(\bm{r} -\bm{s})\ran \approx  \bar
T(\bm{s})\,\bar T(\bm{r} -\bm{s})$, with $\bar T(\bm{r})$ the
mean--field amplitude of Eq.~(\ref{TGBW}). The most interesting
situation is when the external dipole is small, $r \ll 1/Q_s$, and
for that case one can distinguish between three different physical
regions for the internal dipoles $\bm{s}$ and $\bm{r} -\bm{s}\,$:
\texttt{(i)} one of the two dipoles, say $\bm{s}$, is much smaller
than the external one: $s\ll r$; \texttt{(ii)} both internal
dipoles are much larger than the external one, but they are still
small as compared to the saturation length: $r\ll s\simeq |\bm{r}
-\bm{s}|\ll 1/Q_s\,$; \texttt{(iii)} both internal dipoles are at
saturation: $s\simeq |\bm{r} -\bm{s}| \simge 1/Q_s$. Corresponding
to these three cases, the integral in Eq.~(\ref{Pinels}) is
decomposed into three pieces, which are estimated as (up to
irrelevant numerical factors)
   \beq\label{sigmaDGBW}
    \Delta P_{\rm inel}(r) \, \sim\, \abar\left[
    \int\limits_{0}^{r}  \frac{\dif s}{s}\,(s^2
    Q_s^2)^{2 \gamma}\,+\, r^2
    \int\limits_{r}^{1/Q_s} \frac{\dif s}{s^3}\,(s^2
    Q_s^2)^{2 \gamma}
    \,+\, r^2 \int\limits_{1/Q_s}^{\infty} \frac{\dif
    s}{s^3}\right]
     \sim\,  \abar\,r^2 Q_s^2\,
    \eeq
The dominant contribution, as isolated in the r.h.s., comes from
relatively large internal dipoles with sizes $s\sim 1/Q_s\gg r$
(i.e., from domains \texttt{(ii)} and \texttt{(iii)} alluded to
above). Note the similarity between this calculation and that of
the $q\bar q$ contribution to the diffractive cross--section in
the MFA (cf. Eq.~(\ref{sigmaIIAT})) : the external dipole size $r$
plays here the same role as the resolution scale $1/Q$ in
Eq.~(\ref{sigmaIIAT}), and the dipole kernel has limiting
behaviours similar to those of the transverse virtual photon
wavefunction (cf. Eq.~(\ref{sigmaTfin})).
%This explains the parallel between the present discussion and the
%corresponding one in Sect. \ref{SECT_QQ}, which will be manifest
%at several levels in the subsequent analysis.

The case of a relatively large dipole $r\gg 1/Q_s$ is technically
more involved, but the analysis in Ref. \cite{MS03} shows that, in
that case, $\Delta P_{\rm inel}(r)$ vanishes very fast --- at
least as fast as $\bar S^2(r) \equiv (1 -\bar T({r}))^2$ --- when
increasing $r$.

\begin{figure}[t]
    \centerline{\hspace*{-.5cm}
    \epsfxsize=14cm\epsfbox{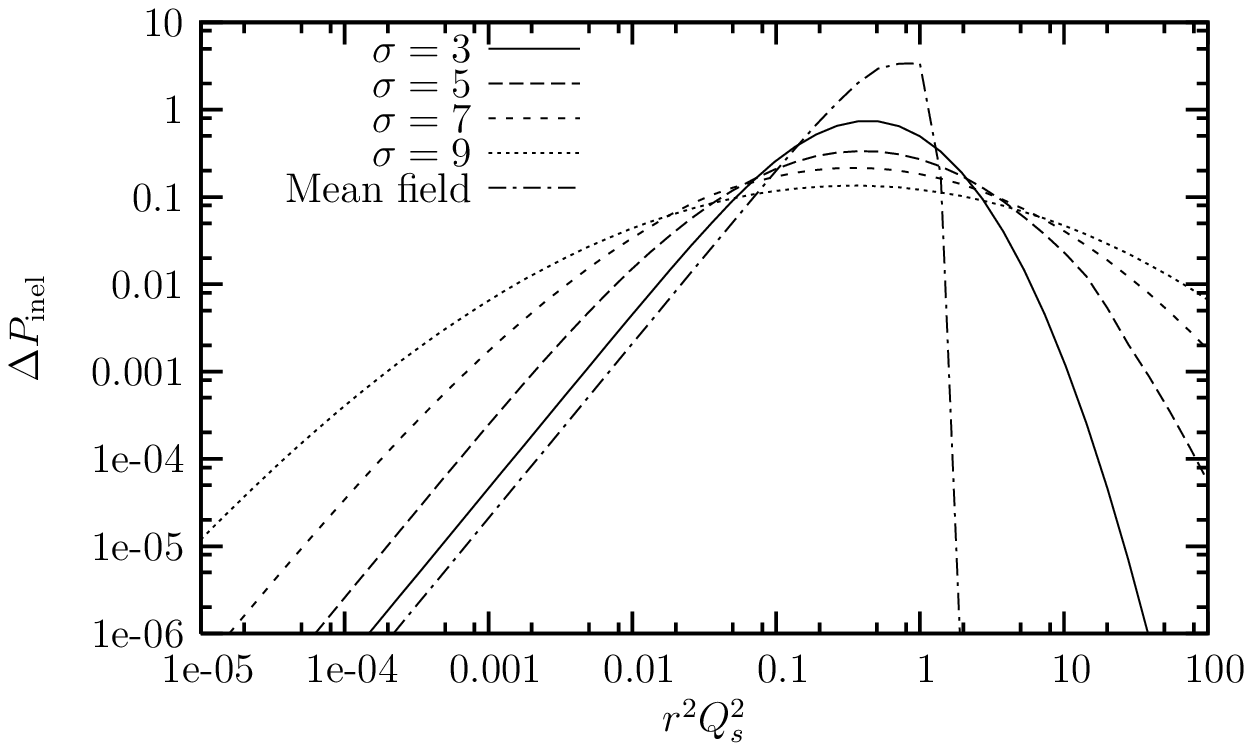}}
    \caption{\sl  The contribution of the $q\bar{q}g$ component of the onium
wavefunction to the inelastic diffraction probability as a
function of $r^2\langle Q_s^2 \rangle$ and for various values of
the front dispersion.
    \label{deltaP_inel}\vspace*{0.5cm}}
    \end{figure}

To summarize, in the mean field approximation, $\Delta P_{\rm
inel}(r)$ increases with $r$ like $r^2$ at $r^2 Q_s^2\ll 1$, it
decreases very fast at $r^2 Q_s^2\gg 1$, and it develops a maximum
at $r\sim 1/Q_s$ \cite{MS03}. This behaviour is manifest on the
``mean field'' curve in Fig. \ref{deltaP_inel}.

\bigskip
\texttt{ii)} {\em The high--energy behaviour with fluctuations}

We shall now extract the dominant behaviour of  the integral in
Eq.~(\ref{Pinels}) in the high--energy regime at $\sigma \gg
1/\gamma_0$ and $z \equiv \ln(1/r^2 \langle Q_s^2 \rangle) \ll
\sigma^2$.  We anticipate that, in this regime, the dominant
contribution comes from internal dipoles with sizes comparable to
that of the external one, $s\sim |\bm{r} -\bm{s}|\sim r$, hence
one can use the high--energy estimates
(\ref{Thighsigma})--(\ref{ThighsigmaN}) for all the amplitudes
which appear in Eq.~(\ref{Pinels}).

Eq.~(\ref{ThighsigmaN}) implies  $\lan T(\bm{s})T(\bm{r}
-\bm{s})\ran = \lan T(r_<)\ran$, where $r_<= \min(s,|\bm{r}
-\bm{s}|)$; thus, the quadratic and one of the linear terms in
Eq.~(\ref{Pinels}) cancel each other, and the integral reduces to
    \beq\label{sigmaDtempa}
    \Delta P_{\rm inel} = 2\,\atpi \int\limits_{\bm{s}}
    \frac{\bm{r}^2}{\bm{s}^2 (\bm{r}-\bm{s})^2}\,
    \big[\avg{T(\bm{s})} -
    \avg{T(\bm{r})}\big]^2,
    \eeq
where the integration is now restricted to the half plane
$|\bm{s}| \geq |\bm{r}-\bm{s}|$. It is convenient to change the
integration variable to $u=s/r$. Then the above equation becomes
    \beq\label{sigmaDtempb}
    \Delta P_{\rm inel} = 4\,\atpi \int\limits_{1/2}^{\infty}
    \frac{\dif u}{u}
    \int\limits_{0}^{\phi_0}
    \frac{\dif \phi}{1 + u^2 - 2 u \cos \phi}\,
    \big[\avg{T(ur)} -
    \avg{T(r)}\big]^2,
    \eeq
where the upper limit is $\phi_0 = {\rm arccos}(1/2 u)$. Since the
scattering amplitudes in Eq.~(\ref{sigmaDtempb}) do not depend on
the angle $\phi$, one can perform the angular integration to
obtain
    \beq\label{sigmaDtempc}
    \Delta P_{\rm inel} = 8\,\atpi \int\limits_{1/2}^{\infty}
    \dif u\,
    \frac{{\rm arctan}\left(\frac{(1+u)}{|1-u|}
    \sqrt{\frac{2u-1}{2u+1}}\right)}
    {u (1+u) |1-u|}\,
    \big[\avg{T(ur)} -
    \avg{T(r)}\big]^2.
    \eeq
One can check that the above integrand vanishes as $1/u^3$ for
large $u$. Thus, the dominant contribution comes from $u={O}(1)$,
that is, from internal dipoles sizes $s$ such that $s \sim r$, as
anticipated. By using this property together with the specific
form of the amplitude given in Eq.~(\ref{Thighsigma}), we can
expand
    \beq\label{Texpand}
    \avg{T(ur)} - \avg{T(r)} \simeq
    -\frac{\ln u^2}{\sigma}\,
    \frac{\del \avg{T}}{-\del(\ln(r^2 \langle Q_s^2 \rangle)/\sigma)}=
    \frac{\ln u^2}{\sqrt{\pi}\sigma}\,
    \exp\left(-\frac{\ln^2 (r^2 \langle Q_s^2 \rangle)}{\sigma^2}\right),
    \eeq
where we have kept only the first term in the Taylor expansion,
since the other terms are suppressed by higher powers of $\ln (
u^2)/\sigma$ and thus are truly negligible at high energy. After
the expansion (\ref{Texpand}), the result of the integration is a
pure number, so we arrive at
    \beq\label{sigmaDfinal}
    \Delta P_{\rm inel}(r) \,\simeq\, \atpi\, \frac{J}{\pi \sigma^2}\,
    \exp\left(-\frac{2z^2}{\sigma^2}\right)
    \qquad\mbox{for}\qquad  z \, \ll \, \sigma^2\,,
    \eeq
with $J$ numerically computed as $J=15.1$.

\begin{figure}[t]
    \centerline{\hspace*{-1.cm}
    \epsfxsize=14cm\epsfbox{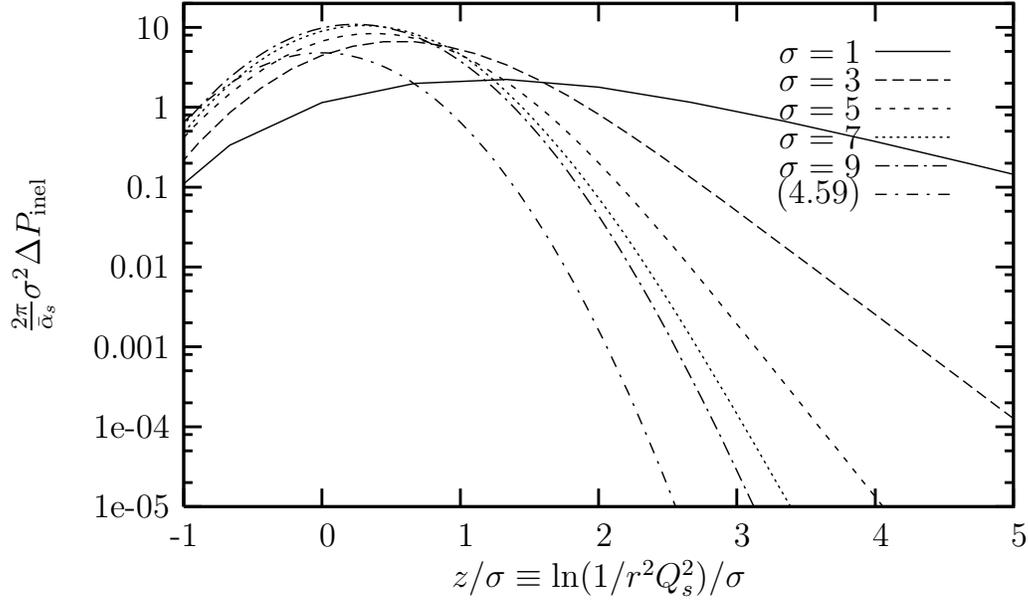}}
    \caption{\sl  The same as in Fig.~18 (multiplied by $\sim\sigma^2$) as a
function of $z/\sigma$.
    \label{DS_deltaP_inel}\vspace*{.5cm}}
    \end{figure}

The above result is valid for both small and large dipoles and
shows that $\Delta P_{\rm inel}(r)$  is an even function of
$z\equiv \ln(1/r^2 \langle Q_s^2 \rangle)$, which rapidly vanishes
when $|z|\gg \sigma$ and has a maximum at $r^2 \langle Q_s^2
\rangle=1$. This behaviour is compared to the corresponding mean
field behaviour  in Fig. \ref{deltaP_inel} which shows that, when
increasing the energy, the maximum around $r^2\sim 1/\langle Q_s^2
\rangle$ becomes flatter and flatter. The emergence of the
diffusive scaling with increasing $\sigma$ and the convergence
towards the asymptotic form (\ref{sigmaDfinal}) are illustrated in
Fig. \ref{DS_deltaP_inel}. (The various curves in  Figs.
\ref{deltaP_inel} and \ref{DS_deltaP_inel} are obtained by
numerically evaluating the r.h.s. of Eq.~(\ref{Pinels}) with the
dipole amplitude in Eq.~(\ref{Tave}) with $\gamma=1$, together
with a corresponding expression for the  two--dipole amplitude
$\lan T(\bm{s})T(\bm{r} -\bm{s})\ran$.)

By inserting the above result (\ref{sigmaDfinal}) into the r.h.s.
of Eq.~(\ref{dsigmaGdb}) and performing the convolutions with the
photon wavefunctions, we can compute the lowest--order
contribution to the differential diffractive cross--section at
high energy and $\beta\ll 1$. (As previously explained, this
quantity is the same as the the cross--section for gluon
production in diffractive DIS.) Via manipulations similar to those
encountered before (cf. Sect. \ref{SECT_HEDIS}), one finds
 \be\label{dsigmaGL}
 \frac{\rmd\sigma^{L}_{\rm  diff}}
 {\rmd^2 {b} \ \rmd \ln (1/\beta)}\,\simeq\,
    \atpi\, \frac{2 J F}{3}\, \frac{1}{\sigma^2}\,
    \exp\left(-\frac{2Z^2}{\sigma^2}\right),
    \ee
for the longitudinal piece and, respectively,
 \be\label{dsigmaGT}
 \frac{\rmd\sigma^{T}_{\rm  diff}}
 {\rmd^2 {b} \ \rmd \ln (1/\beta)}\,\simeq\,
 \atpi\, \frac{\sqrt{2\pi} J F}{6}\, \frac{1}{\sigma}\,
    \erfc\left(\frac{\sqrt{2}\,Z}{\sigma}\right),
    \eeq
for the transverse one. As usual, $F\equiv (N_c \alpha_{\rm em}/2
\pi^2) \sum_ f e_f^2$, and the above results are valid in the high
energy regime at $\sigma \gamma_0 \gg 1$ and $-\infty < Z \ll
\sigma^2$, up to corrections of $O(Z/\sigma^2)$.

Comparing Eqs.~(\ref{dsigmaGL}) and (\ref{dsigmaGT}) we see that
the transverse cross section dominates the longitudinal one
everywhere in the region of interest. Therefore,
    \beq\label{sigmagamma}
    \frac{\rmd\sigma^{\gamma}_{\rm  diff}}
 {\rmd^2 {b} \ \rmd \ln (1/\beta)}\,\simeq\,
  \atpi\, \frac{\sqrt{2\pi} J F}{6}\, \frac{1}{\sigma}\,
    \erfc\left(\frac{\sqrt{2}\,\ln
    (Q^2/\langle Q_s^2 \rangle)}{\sigma}\right)\qquad{\rm
    for}\qquad\beta\ll 1\,.
    \eeq
Recall that the average saturation scale $\langle Q_s^2 \rangle$
and the front dispersion $\sigma^2$ which enter this equation is
to be evaluated at a rapidity $Y_0=Y_{\rm gap}$. The
cross--section (\ref{sigmagamma}) is a monotonically decreasing
function of $Q^2$ (for fixed $Y$, that is, fixed $\sigma$ and
$\langle Q_s^2 \rangle$). Note that after multiplication by
$\sigma$, this cross--section exhibits diffusive scaling.

It might be interesting to compare this high--energy estimate to
the corresponding result at intermediate energies, as given by the
mean field approximation (say, by the solution to the BK
equation). In the high--$Q^2$ regime at $Q^2\gg Q_s^2$, the latter
can be estimated by using $\Delta P_{\rm inel}\sim \abar r^2
Q_s^2$, cf. Eq.~(\ref{sigmaDGBW}), which yields \cite{MS03}
 \beq\label{sigmaGGBW}
    \frac{\rmd\sigma^{\gamma}_{\rm  diff}}
 {\rmd^2 {b} \ \rmd \ln (1/\beta)}\,\sim\,
  \abar F\,
   \frac{Q_s^2}{Q^2}\, \ln
   \frac{Q^2}{Q_s^2}\,\qquad\mbox{for}\qquad Q^2\,\gg\, Q_s^2\,.
   \eeq
This mean--field result receives contributions from dipole sizes
$r$ uniformly distributed within the range $1/Q < r < 1/Q_s$. By
contrast, the high--energy estimate (\ref{sigmagamma}) is
dominated by $r\sim 1/Q$.

\subsubsection{The elastic diffraction}

In this subsection, we shall evaluate the increase $\Delta P_{\rm
el}$ in the probability for elastic scattering due to the emission
of one soft gluon (cf. Eq.~(\ref{PelQQG})). Our purpose is to
demonstrate that, in the fluctuation--dominated regime at high
energy and relatively large $Q^2$, cf. Eq.~(\ref{HEDIS}), this
increase is parametrically larger than the corresponding evolution
$\Delta P_{\rm inel}$ in the probability for inelastic
diffraction, that we have already computed.

Let us start by briefly discussing the case of the mean field
approximation, where the opposite situation holds: for $Q^2\gg
Q_s^2$ (or relatively small dipole sizes), the inelastic piece
dominates over the elastic one. Indeed, by using $\Delta P_{\rm
el} (r)\equiv 2 \avg{T(r)} \Delta {\cal A}(r)$ together with
$\Delta {\cal A} ={\del \lan T\ran} /{\del Y}$, cf.
Eqs.~(\ref{DAQQG})--(\ref{Tevol}), and the expression (\ref{TGBW})
for the mean--field amplitude, one immediately finds: $\Delta
P_{\rm el}\sim \abar\, (r^2 Q_s^2)^{2\gamma}$ for $r\ll 1/Q_s$.
This is a `higher--twist' effect, much smaller than the
corresponding inelastic contribution in Eq.~(\ref{sigmaDGBW}).

But at high energy, $\Delta P_{\rm el}$ turns out to be the
dominant contribution, down to very small dipole sizes, namely, so
long as $z \ll \sigma^2$. Moreover, its calculation turns out to
be quite subtle, as we explain now. On one hand, it is
straightforward to use $\Delta {\cal A} ={\del \lan T(r)\ran}
/{\del Y}$ together with Eq.~(\ref{DAHE}), to deduce :
 \be \label{deltaPel}
 \Delta P_{\rm el} (r)\,\simeq\,\lambda\abar
 \, \erfc\left(\frac{z}{\sigma}\right)\,
 \frac{1}{\sqrt{\pi}\sigma}\,
    \exp\left(-\frac{z^2}{\sigma^2}\right)\,,
    \ee
which is indeed {\em parametrically larger} (for any $z$) than the
corresponding inelastic piece, Eq.~(\ref{sigmaDfinal}). For
instance, for small dipole sizes, or large $z\gg \sigma$, this
elastic component behaves like
 \be \label{deltaPelI}
 \Delta P_{\rm el} (r)\,\simeq\,
   \frac{\abar}{\pi}\, \frac{\lambda}{z}\,
    \exp\left(-\frac{2z^2}{\sigma^2}\right)
 \qquad\mbox{for}\qquad \sigma \ll \, z\, \ll \sigma^2\,,\ee
and dominates over the inelastic piece by a large factor $\lambda
\sigma^2/z \gg 1$, as anticipated.
%(we assume $\lambda\sim O(1)$).

On the other hand, the above estimate for $\Delta P_{\rm el}$ has
the drawback to involve the average front velocity $\lambda$,
which is generally unknown. It is therefore tempting to try and
compute $\Delta {\cal A}(r)$ directly from its integral
representation (\ref{DAQQG}) together with
Eqs.~(\ref{Thighsigma})--(\ref{ThighsigmaN}) for the dipole
amplitudes. {\em A priori}, this calculation is independent of
$\lambda$ and should even enable us to {\em determine} $\lambda$,
by comparison with the expected result in Eq.~(\ref{DAHE}). But
this expectation appears to be naive: the coefficient of the
would--be dominant term at large $Y$ --- that term precisely which
should be identified as $\lambda/\sigma$
--- turns out to be zero. (This can be checked via manipulations
similar to those in Eqs.~(\ref{sigmaDtempb})--(\ref{Texpand}):
without taking the square of the amplitudes inside the integrand,
so like in Eq.~(\ref{sigmaDtempc}), the integral vanishes exactly
after the expansion (\ref{Texpand}).) This seems to imply the
unphysical result $\lambda=0$, but in reality it only means that
the approximations (\ref{Thighsigma})--(\ref{ThighsigmaN})
--- although the correct leading--order estimates for the average
amplitudes at high energy --- cannot be used to also evaluate the
right hand sides of the evolution equations. The latter are
sensitive to the subleading terms, which are not under control in
the present approximations.

\begin{figure}[t]
    \centerline{\hspace*{-.5cm}\epsfxsize=14cm\epsfbox{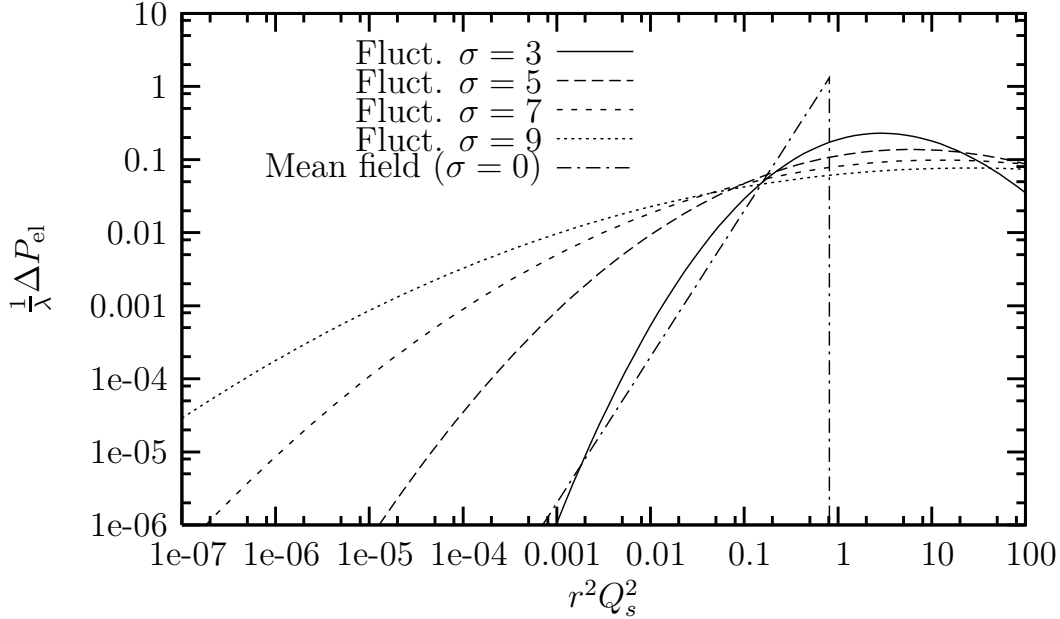}}
    \caption{\sl  The contribution of the $q\bar{q}g$ component of the onium
wavefunction to the elastic diffraction probability as a function
of $r^2\langle Q_s^2 \rangle$. The high--energy estimate
(\ref{deltaPel}) is shown for various values of $\sigma$ and
compared to the mean--field prediction obtained with
Eq.~(\ref{TGBW}).
    \label{deltaP_el}\bigskip}
    \end{figure}

To summarize, in the high--energy regime of interest, the
diffractive probability associated with the $q \bar q g$ component
is controlled by its elastic piece, $\Delta P_{\rm diff}\approx
\Delta P_{\rm el}$, and has the dominant behaviour displayed in
Eq.~(\ref{deltaPel}). Although not as symmetric as the inelastic
contribution (\ref{sigmaDfinal}), the function (\ref{deltaPel})
has still the properties to vanish for both very small and very
large dipole sizes, and to develop a rather flat maximum at $r\sim
1/\avg{Q_s}$. This behaviour is illustrated in Fig.
\ref{deltaP_el}, which also shows that, when increasing $\sigma$,
the elastic piece rises faster than the inelastic one (compare to
Fig. \ref{deltaP_inel}) and rapidly dominates over the latter at
all dipole sizes, in agreement with our previous estimates.

After replacing  $\Delta P_{\rm diff}\approx \Delta P_{\rm el}$ in
Eq.~(\ref{PdiffQQG}) and also using Eq.~(\ref{Pdiffqq}), it
becomes clear that the diffractive probability coincides with the
elastic probability, Eq.~(\ref{PelQQG}), which demonstrates
Eq.~(\ref{Pdiffel}) to the order of interest in $\abar\ln
(1/\beta)$. Although we have considered here only the simplest
evolution of the projectile (namely, the formation of the $q\bar q
g$ state), we expect the property (\ref{Pdiffel}) to hold also for
smaller values of $\beta$ (when higher Fock--space components come
into play), for the reasons explained after Eq.~(\ref{Pdiffel}).
Assuming this to be the case, we conclude that, within the
high--energy regime at $\sigma\gamma_0\gg 1$ and $Z\ll\sigma^2$,
the diffractive DIS cross--section is dominated by the elastic
scattering of the onium, and thus is independent of $\beta$ (for a
given value of $Y$), within the limits of the leading--logarithmic
approximation with respect to $\ln (1/\beta)$. To that accuracy,
the corresponding cross--section has already been evaluated in
Eq.~(\ref{sigmaqq}).

Let us conclude with a remark which applies to both the inclusive
and the diffractive cross--sections and refers to their behaviour
when increasing $Y$ in the high--energy regime: At the end of
Sect. \ref{SECT_PL}, we have mentioned the  `rigidity' of the
dipole amplitudes at high--energy, namely, their property to
increase only slowly when increasing $Y$ at fixed, but large
$Q^2$, in the weak scattering regime. Not surprisingly, the same
`rigidity' transmits to the DIS cross--section in the
diffusive--scaling window defined by Eq.~(\ref{HEDIS}). For
instance, the $Y$--derivative of the diffractive cross--section in
Eq.~(\ref{sigmaqq}) is estimated as
 \beq\label{sigmadiffQQGfin}
 \frac{\rmd\sigma^{\gamma}_{\rm diff}}
 {\rmd^2 {b} \ \rmd Y}
 \, \simeq\,
    \lambda \abar\,\frac{\pi F}{6}\,
    \erfc^2\left(\frac{\ln(Q^2/\langle Q_s^2
    \rangle)}{\sigma}\right),
    \eeq
which for $Q^2$ within the range indicated in Eq.~(\ref{HEDIS})
behaves like
 \beq\label{sigmacompHE}
 \frac{\rmd\sigma^{\gamma}_{\rm diff}}
 {\rmd^2 {b} \ \rmd Y}
 \,\sim\,\lambda
  \abar \,\frac{\sigma^2}{Z^2}\,
   \exp\left(-\frac{2Z^2}{\sigma^2}\right)
   \ \sim\ \lambda
  \abar\,
  \frac{Z}{\sigma^2}\ \frac{\rmd\sigma^{\gamma}_{\rm  diff}}
 {\rmd^2 {b}}\qquad\mbox{when}\quad \sigma\ll
Z\ll\sigma^2, \eeq that is, it involves an additional suppression
factor ${Z}/{\sigma^2}\ll 1$ as compared to the usual behaviour
expected in the context of the BFKL evolution.

\section{Conclusions}

In this paper, we have formulated the inclusive and diffractive
deep inelastic scattering of a virtual photon off a hadronic
target in the high--energy limit and in the large--$N_c$
approximation.  We have worked in a Lorentz frame in which the
high--energy limit is achieved by boosting the hadronic target,
which then evolves to very high gluon density, whereas the
rapidity of the projectile (the virtual photon) remains finite and
relatively small. Our main emphasis has been on a novel type of
universal behaviour which emerges at sufficiently high energy as a
consequence of saturation and gluon--number fluctuations in the
target wavefunction. The essential feature of this new regime is
the fact that the DIS cross--sections at fixed impact parameter
are dominated by the physics of saturation up to very large values
of $Q^2$, well above the (average) saturation momentum $\langle
Q_s^2 \rangle$ of the target (which itself increases exponentially
with $Y$). And the distinguished signature of this behaviour is a
new, {\em diffusive}, scaling law, which should be obeyed by all
the DIS cross--sections at sufficiently high energies, and which
reflects the Gaussian nature of the probability distribution for
the logarithm of the saturation momentum in the target.

To establish this behaviour, we have relied on two main types of
ingredients: \texttt{(i)} a set of factorization formul\ae\ which
relate the DIS cross--sections to dipole--target scattering
amplitudes, and \texttt{(ii)} the high--energy estimates for the
latter, as determined by the Pomeron loop equations via the
correspondence with statistical physics.

Concerning point  \texttt{(i)}, let us recall here that we have
restricted ourselves to diffractive processes in which the hadron
undergoes elastic scattering, and to leading--logarithmic
approximations for the high--energy evolutions of both the target
and the projectile. Besides, we have assumed the projectile to
remain dilute, and thus to obey BFKL evolution. Under these
assumptions, we have shown that a physically transparent
description of the diffractive processes, \`a la Good and Walker,
emerges when viewing the collision in the frame in which the
rapidity of the target is equal to the minimal rapidity gap. In
this frame, the final state interactions automatically cancel out,
and the projectile can be represented as a distribution of dipoles
which elastically scatter off the target.  The effect of the
collision is to destroy the coherence of the dipole superposition
and thus transform the incoming photon into a collection of
partons in the final state.

Concerning point  \texttt{(ii)}, we should emphasize that the
asymptotic functional form of the dipole amplitudes at high energy
follows from very general and elementary considerations --- the
average amplitudes are dominated by saturated gluon configurations,
for which the single--event amplitudes have reached their unitarity
limit $T=1$ ---\,, and as such it should be very robust: It is
likely that this dominant behaviour will not be modified by higher
order perturbative corrections, nor by corrections in $1/N_c$ (but
this remains to be investigated). On the other hand, the energy
dependencies of the average saturation momentum $\langle Q_s^2
\rangle$ and of the front dispersion $\sigma^2$ are sensitive to the
details of the evolution, and are presently unknown for realistic
values of $\alpha_s$. This is why, in this paper, we have
systematically measured $Q^2$ in terms of $\langle Q_s^2 \rangle$,
and $Y$ in terms of $\sigma^2$.

Perhaps the most interesting, qualitative, conclusion of our
analysis is the fact that, at sufficiently high energies and as a
result of fluctuations, the physics of gluon saturation starts to
be determinant at momenta $Q^2$ well above the saturation momentum
$\langle Q_s^2 \rangle$, within a logarithmic distance $\ln
(Q^2/\langle Q_s^2 \rangle)\sim \sigma^2$ which increases linearly
with $Y$. From previous studies of the mean field approximation,
we already knew that saturation can manifest itself at momenta
larger than $Q_s$, via the phenomenon of {\em geometric} scaling.
In the present analysis, we have shown that this phenomenon is
preserved by fluctuations so long as the front dispersion is not
too large (i.e., for $\sigma^2\ll 1$). What is however striking
about the high--energy regime at $\sigma^2 > 1$ is that the
physics above $\langle Q_s^2 \rangle$ is not only {\em
influenced}, but actually {\em dominated} by rare configurations
at saturation. This has tremendous physical consequences, for both
the dipole amplitudes and the DIS cross--sections :

\begin{itemize}
\item The {\em strong correlation} property $\langle T^2(r) \rangle \simeq
\langle T(r) \rangle$, which is natural at saturation where
$\langle T \rangle=1$, appears to be satisfied up to very large
$Q^2\equiv 1/r^2$, where $\langle T \rangle\ll 1$.

\item The amplitudes and the DIS cross--sections are very {\em rigid}, in
the sense that they rise unusually slowly when increasing $Y$ at
fixed $Q^2$ within the weak scattering regime.

\item At relatively large virtualities $Q^2\gg \langle Q_s^2 \rangle$,
cf. Eq.~\eqref{HEDIS}, the convolutions  with the photon
wavefunction yielding the  DIS cross--sections are dominated by
small sizes $r\sim 1/Q$ for the $q\bar q$ pair, as they select
those gluon configurations in the target wavefunction which are at
saturation on the resolution scale set by the virtual photon.

\item Similarly, the convolutions with the dipole probabilities
yielding the average over the BFKL (onium) wavefunction of the
projectile are dominated by dipole sizes of the order of their
parent dipole
--- the original $q\bar q$ pair --- and hence of order $1/Q$.

\item In the high--energy limit, the DIS diffractive cross--section
becomes purely elastic (at the level of the  $q\bar q$ pair), the
inelastic contribution being parametrically suppressed.

\item Within the high--energy regime defined by Eq.~\eqref{HER},
all the amplitudes or cross--sections show {\em diffusive}
scaling, that it, they depend upon $Q^2$ and $Y$ only via the
variable  $z/\sigma$ with $z\equiv \ln (Q^2/\langle Q_s^2
\rangle)$.

\item Within the diffusive scaling region alluded to above,
the `twist expansion' does not apply anymore: At large $Q^2\gg
\langle Q_s^2 \rangle$, the cross--sections decay as a Gaussian of
$z/\sigma$, and not as a power of $1/Q^2$.

\end{itemize}

These properties are in sharp contrast with the corresponding
predictions of the MFA (that we have also worked out in this
paper, at least qualitatively, for the sake of comparison), and
demonstrate the breakdown of the latter at sufficiently high
energy.

Note that the above conclusions depend upon the functional form of
the dipole amplitudes, but not upon the precise values of
coefficients $\lambda$ and $D_{\rm fr}$. But, of course, the
latter are necessary in order to translate our results into
physical units for $Y$ and $Q^2$. The small--$x$ phenomenology at
HERA suggests $\lambda \simeq 0.2 \div 0.3$ \cite{GBW99,IIM03},
but we refrain ourselves from suggesting a similar, experimental,
measure of the diffusion coefficient $D_{\rm fr}$, as we do not
believe that the high--energy regime that we discuss here has been
approached at HERA. (For instance, the diffractive data at HERA
indicate the dominance of inelastic diffraction at small $\beta$
\cite{MS03}, at variance with the high--energy behaviour that we
predict here.) The conceptually proper way to compute $\lambda$
and $D_{\rm fr}$ for realistic values of $\alpha_s$ (and within
the present approximations: high energy, leading--order
perturbative QCD, and large $N_c$) would be by solving the Pomeron
loop equations of Ref. \cite{IT05}. This would also allow one to
explicitly test our present, asymptotic, predictions, and to study
the preasymptotic behaviour, like the dependence upon the initial
conditions or upon the impact parameter.

Such more detailed calculations should allow one to determine the
physical region in which this high--energy behaviour should start
to manifest itself and, especially, whether this region might be
accessible at the LHC.

\section*{Acknowledgments}

We are grateful to A.~Mueller and R.~Peschanski  for  discussions
and for useful comments. Y.~H.~ and D.~T.~ thank SPhT Saclay for
hospitality.  This research was supported in part by RIKEN BNL
Research Center and the U.~S.~Department of Energy  [Contract No.
DE-AC02-98CH10886]. G.S. is funded by the National Funds for
Scientific Research (Belgium).

\appendix
\section{The virtual photon wavefunction}

In the dipole frame we have adopted throughout this paper, the DIS
cross-sections are factorized as the product of the photon
wavefunction ($\gamma^* \to q\bar q$) with the quark-antiquark
interaction amplitude. In this appendix, we give the $q\bar q$
dissociation probabilities for the virtual photon. This well-known
process can be computed in perturbative QED (to lowest order in
$\alpha^{}_{\rm em}$) and reads \be \label{PsiT} \vert
\psi^{\gamma}_{T}(v,r;Q)\vert^2 &=&\frac{\alpha^{}_{\rm
em}N_c}{2\pi^2}
                       \sum_f e_f^2 \, \left\{
                       \big(v^2 +(1-v)^2\big) \bar Q_f^2  \rmK_1^2(\bar Q_f r)
                        + m_f^2 \rmK_0^2 (\bar Q_f r)\right\},\\
\label{PsiL} \vert \psi^{\gamma}_L(v,r;Q)\vert^2
&=&\frac{\alpha^{}_{\rm em}N_c}{2\pi^2}
                       \sum_f e_f^2 \, \left\{
                      4 Q^2 v^2 (1-v)^2 \rmK_0^2(\bar Q_f
                      r)\right\}\,,
                      \ee
where $\bar Q_f^2 \equiv v (1-v)Q^2 + m_f^2$, $m_f$ and $e_f$ is
the mass and the electric charge of the quark with flavor $f$, and
$\rmK_0$ and $\rmK_1$ are the modified Bessel functions.

In order to get estimates of the DIS cross-sections, {\em i.e.} to
estimate the convolution of this wavefunction with the $q\bar q$
interaction amplitude, we shall ignore the quark masses (these are
important only for relatively small $Q^2 \sim m_f^2$) and use
simple approximations for the Bessel functions, as quite common in
the literature (see, e.g., \cite{GBW99} for similar
manipulations). Specifically, since $\rmK_\nu(x)$ decreases
exponentially at large $x$, one can restrict the integrations in
Eqs.~(\ref{sigmatot})--(\ref{sigmadiff}) to values $v$ and $r$
such that $\bar Q_f r < 1$. Besides, for $x\ll 1$, $\rmK_1(x)\sim
{1}/{x}$ and $\rmK_0(x)\sim \ln (1/x)$, so the Bessel functions
can be approximated as
 \be
 \rmK_0(x)\sim \Theta(1-x) ,\qquad \rmK_1(x)\sim \frac{1}{x}\,
 \Theta(1-x),\label{K0_K1} \ee
where we ignore the overall normalization as well as the
logarithmic singularity of $\rmK_0$ as $x\to 0$ (the latter is
innocuous within the relevant convolutions).  Another
simplification arises when performing the integral over $v$ : the
constraint $v (1-v)Q^2 <1$ together with the fact that $v(1-v)\le
1/4\, $ for $\,0\le v\le 1$ makes it natural to distinguish
between

\texttt{(a)} {\it symmetric} configurations, for which $Q^2r^2 <4$
(``small dipoles''), and

\texttt{(b)} {\it aligned jet} configurations, for which $4 <
Q^2r^2$ (``large dipoles'').

In the first case, there is no restriction on the $v$ integral,
which is then dominated by symmetric values $v\sim 1/2$. In the
second case, the dominant contributions correspond to the
situation in which one of the two dipoles carries most of the
total longitudinal momentum, that is, $v$ is either close to zero,
or close to one, which allows us to further simplify the
respective integrand. As a result of such simplifications, the
convolutions involving the photon wavefunctions can be finally
estimated as shown in Eqs.~(\ref{sigmaTfin})--(\ref{sigmaLfin}).

\section{The diffractive cross--section near $\beta= 1$}

In this Appendix we present the result for the differential
diffractive cross--section per unit rapidity for $\beta\simeq 1$
as obtained when relaxing the leading--logarithmic approximation
in $\ln(1/\beta)$. As compared to the formul\ae\ in the main text,
this quantity represents the integrand which would give the
elastic piece in the r.h.s. of Eqs.~\eqref{sigmadiffint} if that
quantity was computed beyond the leading--log approximation w.r.t.
$\ln(1/\beta)$. Notice that, within the limits of that
approximation, the elastic scattering implies $\beta=1$, hence the
corresponding differential distribution is just a
$\delta$--function $\sim \delta(\beta-1)$. But this distribution
gets smeared after more accurately taking into account the
kinematics for the diffractive production of the $q\bar q$ pair
into which the virtual photon has fluctuated.

In the leading--logarithmic approximation with respect to
$\ln(1/x),$ one finds \cite{BP96,W97}
 \be
 \f{\rmd\sigma_{\rm diff}^{q\bar q}}{\rmd^2b\ \rmd\ln(1/\beta)}
 =\frac{Q^2}{4\pi\beta}\sum_f \int \rmd^2 {\bm r}\rmd^2 {\bm r}'
 \rmd v \, v(1-v)\, \Theta({\bm \kappa}_f^2)\
 {\rme}^{i{\bm \kappa}_f\cdot({\bm
 r}'-{\bm r})}
 \nonumber\\
 \sum_{\alpha=L,T}\Phi^f_\alpha(v,{\bm r},{\bm r}') \langle
 T({\bm{r},\bm{b}}) \rangle_{Y_{\rm gap}} \langle T({\bm{r}',\bm{b}})
 \rangle_{Y_{\rm gap}} \label{sigmaqq1}\ee
where ${\bm \kappa}_f^2=v(1-v)M_X^2-m_f^2.$ The photon
wavefunctions $\Phi^f_T$ and $\Phi^f_L$ are given by
 \be \Phi^f_T(v,{\bm r},{\bm r}')=\,
\frac{\alpha_{\rm em}N_c}{2\pi^2} e_f^2 \Big((v^2+(1-v)^2)\bar
Q_f^2 \,\frac{{\bm r}\cdot{\bm r}'}{r\ r'}\, \rmK_1(\bar Q_f r)\,
\rmK_1(\bar Q_f r') \nonumber\\ +\, m_f^2 \, \rmK_0(\bar Q_f
r)\rmK_0(\bar Q_f
 r')\Big) \ee
(we use the same notations as in Appendix A) and respectively
 \be \Phi^f_L(v,{\bm r},{\bm r}')=\,
\frac{\alpha_{\rm em}N_c}{2\pi^2}\, e_f^2\, 4Q^2 v^2(1-v)^2\,
 \rmK_0(\bar Q_f r)\, \rmK_0(\bar Q_f r') \ee
with $r=|\bm{r}|$ and $r'=|\bm{r}'|.$ Note that these functions
are such that \be \sum_f \Phi^f_\alpha(v,{\bm r},{\bm
r})=|\psi^\gamma_\alpha(v,r;Q)|^2 \ee is the photon wavefunction
squared discussed in Appendix A. Note furthermore that the r.h.s.
of  \eqref{sigmaqq1} depends upon $\beta$, via ${\bm \kappa}_f^2$.
(Recall that $\beta=Q^2/(Q^2+ M_X^2)$.)

The cross--section $({\rmd\sigma_{\rm diff}^{q\bar q}}/{\rmd^2b})$
is obtained by integrating \eqref{sigmaqq1} from $\ln(1/\beta)=0$
up to $Y-Y^{\rm min}_{\rm gap}$. If this upper limit is relatively
large, one recovers the $q\bar q$ part of our expression in the
main text, cf. Eq.~(\ref{sigmadiffHE}) (with $Y^{\rm min}_{\rm
gap}$ renoted as $Y_{\rm gap}$, for simplicity) :
 \be \f{\rmd\sigma_{\rm
diff}^{q\bar q}}{\rmd^2b} (Y,Y_{\rm gap},Q^2) =\int \rmd^2 {\bm
r}\int \rmd v \sum_{\alpha=L,T}\,|\psi^\gamma_\alpha(v,r;Q)|^2\,
\langle T({\bm{r},\bm{b}})\rangle_{Y_{\rm gap}}^2\ .
\label{qqllb}\ee This can be seen by changing the integration
variable to $\kappa_f^2:$ the typical values of $r$ and $r'$ are
set by the photon wavefunctions, and thus are of order $1/\bar
Q_f.$ On the other hand, the integral over $\kappa_f^2$ fixes the
difference $r-r'$ to be of order $1/\kappa_f^{\rm max}$, with
$\kappa_f^{\rm max}$ determined by the maximal diffractive mass
$M_X^2$ (the one corresponding to the minimal rapidity gap $Y_{\rm
gap}$). When $\kappa^{\rm max}_f\gg \bar Q_f $ (in practice, this
condition amounts to $M_X^2\gg Q^2$), the difference $r-r'$ is
small as compared to both $r$ and $r'$, so one can approximate the
integral $\int \rmd^2\kappa_f \,e^{i{\bm \kappa}_f\cdot({\bm
r}'-{\bm r})}$ by $\delta^{(2)}({\bm r}'-{\bm r}),$ which then
yields \eqref{qqllb}.

Note finally that the differential cross--section is proportional
to the diffractive structure function $F_2^{D,3}(x,Q^2,\beta),$
which is measured at HERA: \be
F_2^{D,3}(x,Q^2,\beta)=\f{Q^2\beta}{4\pi^2\alpha_{\rm em}x}\,
\f{\rmd\sigma_{\rm diff}^\gamma}{\rmd\ln(1/\beta)} \ . \ee

%\bibliographystyle{unsrt}
%\bibliography{myrefs}
\end{document}